\documentclass[10pt,aps,prd,notitlepage,superscriptaddress,nofootinbib,amsmath,amssymb,twocolumn,showpacs]{revtex4-1}
\usepackage[T1]{fontenc} 
\usepackage{ulem}

\pdfoutput=1

\usepackage[outline]{contour}
\usepackage{xcolor,graphicx}
\usepackage{caption}
\usepackage{subcaption}
\usepackage{bm}

\usepackage{color,marvosym}
\usepackage{graphicx} 
\usepackage{comment}
\usepackage{breakurl}

\definecolor{darkred}{rgb}{0.5,0,0}
\definecolor{darkgreen}{rgb}{0,0.5,0}
\definecolor{darkblue}{rgb}{0,0,0.5}

\usepackage{hyperref}
\hypersetup{colorlinks,
linkcolor=darkblue,
filecolor=darkgreen,
urlcolor=darkblue,
citecolor=darkblue }
\newcommand{\obs}{_{{\cal O}}}

\newcommand{\UD}[2]{\ensuremath{^{#1}_{\phantom{#1} #2}}}

\newcommand{\calO}{\ensuremath{\mathcal{O}}}

\newcommand{\calV}{{\ensuremath{\cal V}}}

\newcommand{\dd}{\ensuremath{\textrm{d}}}

\newcommand{\beq}{\begin{equation}}
\newcommand{\eeq}{\end{equation}}
\newcommand{\bea}{\begin{eqnarray}}
\newcommand{\eea}{\end{eqnarray}}
\newcommand{\bit}{\begin{itemize}}
\newcommand{\eit}{\end{itemize}}
\newcommand{\bfi}{\begin{figure}}
\newcommand{\efi}{\end{figure}}
\newcommand{\bfic}{\begin{figure*}}
\newcommand{\efic}{\end{figure*}}
\newcommand{\bce}{\begin{center}}
\newcommand{\ece}{\end{center}}
\newcommand{\bt}{\begin{table}}
\newcommand{\et}{\end{table}}
\newcommand{\btb}{\begin{tabular}}
\newcommand{\etb}{\end{tabular}}

\newcommand{\calD}{\ensuremath{\mathcal{D}}}

\newcommand{\calE}{\ensuremath{\mathcal{E}}}
\newcommand{\calW}{\ensuremath{\mathcal{W}}}

\newcommand{\Dang}{\ensuremath{D_{\textrm{ang}}}}

\newcommand{\Dpar}{\ensuremath{D_{\textrm{par}}}}

\newcommand{\Dlum}{\ensuremath{D_{\textrm{lum}}}}

\newcommand{\Omegamzero}{\ensuremath{\Omega_{{\rm m}_0}}}
\newcommand{\Omegakzero}{\ensuremath{\Omega_{{k}_0}}}
\newcommand{\OmegaLambdazero}{\ensuremath{\Omega_{{\Lambda}_0}}}

\newcommand{\qed}{\nobreak \ifvmode \relax \else
      \ifdim\lastskip<1.5em \hskip-\lastskip
      \hskip1.5em plus0em minus0.5em \fi \nobreak
      \vrule height0.75em width0.5em depth0.25em\fi}

\makeatother

\begin{document}

\title{Geometric optics in relativistic cosmology: new formulation and a new observable}

\author{Miko\l{}aj Korzy\' nski}

\author{Eleonora Villa}
\email{Corresponding author: villa@cft.edu.pl}
\affiliation{Center for Theoretical Physics, Polish Academy of Science, \\ Al. Lotnik\' ow 32/46, 02-668 Warszawa, Poland}

\date{\today}

\begin{abstract} 
We discuss a new formalism for light propagation which can be used within the regime of validity of geometric optics, but with no limitation on the scales of interest: from inside the Galaxy to the ultra-large scales of cosmology. One of our main results is that within this framework it is possible to calculate all relevant observables (image magnification, parallax, position drift or proper motion) by simply differentiating the photon trajectory with respect to the initial data. We then focus on a new observable, which we name the distance slip: it is defined as the relative difference between the angular diameter distance and the parallax distance.  Its peculiarity lies in the fact that its value is independent of the momentary motions of both the source and the observer and that for short distances it shows a tomographic property, being proportional  to the amount of matter along the line of sight.
After describing further its properties and methods of measurement, we specialize our study of the distance slip to cosmology. We show that it does not depend on the Hubble constant $H_0$ and that its dependence on the other cosmological parameters is different from other distance indicators. This suggests that the distance slip may contain new information.
\end{abstract}

\pacs{}

\maketitle

\section{Introduction}
\label{sec:intro}
In general relativity the spacetime geometry, related via Einstein equations to the matter and energy content, leaves an imprint on the light beams received by the observer, affecting this way all the observable quantities, e.g. the magnification of distant objects, their redshift, but also on the parallax and proper motions. This is the physical foundation of most - if not all - the methods we use to extract information about our Universe by measuring electromagnetic radiation and gravitational waves emitted from distant sources.

Recently a new theoretical formulation of the problem of light propagation in curved spacetimes within the geometric optics approximation has been introduced in \cite{Grasso:2018mei}. It provides a new, covariant, frame-independent and unified framework to calculate all the optical observables one can construct from comparing the properties of neighbouring geodesic through the spacetime from the source to the observer. It also extends the standard Sachs formalism (see \cite{sachs, seitzschneiderehlers, perlick-lrr} and \cite{ehlers-jordan-sachs}, the last one translated and reprinted in \cite{Jordan2013reprint}) by considering the view of distant objects from various observation points, displaced in both space and time, instead of a single observer at a fixed spacetime event. It is therefore particularly suited for
calculating the parallax effects as well as the time variations, also called the drifts, of the values of optical observables registered by an observer \cite{Quercellini:2010zr, Korzynski:2017nas, Marcori:2018cwn, Hellaby:2017soj}. 

In the literature different methods are proposed for various observables, and for some observable more than one method has been used (see e.g. \cite{Yoo:2016vne} for a comparison of four different approaches for the calculation of the luminosity distance in the cosmological context).
On the other hand the main result of the new formulation of \cite{Grasso:2018mei} emphasizes the advantage of having a unified framework: all the observables - the parallax, the magnification, the position drift, the angular diameter distance etc. -
registered by a given observer are expressed in terms of one key quantity only, the so-called \emph{bilocal geodesic operators} (BGOs),
and the kinematical variables characterizing the momentary positions and motions of  the source and the observer with respect to their local inertial frames. 
In addition the BGOs can be written as (non-local and non-linear) functionals of the curvature tensor along the line of sight, given by solutions of certain matrix ODEs \cite{Grasso:2018mei, Korzynski:2017nas}. Therefore, in their turn, the observables can be expressed in terms of the curvature along the line of sight and the momentary 4-velocities and 4-accelerations of both the observer and the source.
Finally, we remark that the bilocal formulation provides a simple and transparent way to investigate the dependence of the observables on the choice of the frame by just changing the 4-velocities we plug into the appropriate expressions. This is especially important for the drift effects, which depend on the momentary motions of the sources and the observer via a number of effects, including the relative transverse motion, the 
aberration effect, the Shapiro delay of light signals etc.

The results presented in \cite{Grasso:2018mei} are completely general.
However the case of spacetimes such that the null geodesic equation can be integrated exactly up to quadratures turns out to be particularly interesting.
As we show here, this property provides a shortcut for calculating the bilocal geodesic operators between any two points connected by a null geodesic without solving any additional (non-linear) ODE, besides the geodesic equation. Indeed, we show that it is possible to obtain the components of the BGOs directly by simply differentiating the
null geodesic curve with respect to the initial data. Within the bilocal formulation for geometric optics, this means also that the general solution of the null geodesic equation is the only quantity we need to obtain expressions for observables like the angular diameter distance, the parallax, the parallax distance and the position drift for \emph{any} pair of source and observer, located at \emph{any} two points connected by a null geodesic. In this paper we describe this method, which we call ``the variation method'', and we specialize our result for the observables to cosmology and in particular to the Friedmann-Lema\^{i}tre-Robertson-Walker (FLRW) spacetimes flat, open and closed. Note that a part of the results derived for the FLRW metric here has already been published in \cite{Fleury:2013sna}: the authors derived there the expressions for the transverse, spatial components of two BGOs, called in their terminology the Jacobi and scale matrices. In \cite{Fleury:2013sna} they were derived as an intermediate result when discussing the Hubble diagram for an inhomogeneous swiss-cheese Universe model.

The main topic of our paper is a new observable introduced in \cite{Grasso:2018mei}: it is defined as the relative difference between the parallax distance and the angular diameter distance. We name it the \emph{distance slip} and it can be expressed as $\mu= 1- \Dang^2/\Dpar^2$ (in the absence of strong lensing\footnote{More precisely, the definition of the distance slip is $\mu= 1- \sigma\Dang^2/\Dpar^2$, where $\sigma=\pm 1$. We may have $\sigma =-1$ in some situations, but only for strongly lensed objects. For more details see section~\ref{sec:mu}.}). It is an interesting observable in astrophysics for three reasons. Firstly, it can in principle be measured using purely astrometric methods, by combining the parallax distance - measured via parallax effects - with the angular diameter distance - measured via the angular size of the image.  Secondly, it is a direct signature of the spacetime curvature. This can be seen as follows: in a flat spacetime the results of both distance measurements must coincide. On the other hand, if curvature is present between the source and the observer, it affects both methods of distance determination and, as it turns out, it affects each of them differently. Therefore the relative difference between the two distances may serve as a direct measure of the spacetime curvature 
along the line of sight.
In this respect, one can also prove that for short distances $\mu$ is directly related to an integral of the stress-energy tensor along the line of sight, giving this way a new, tomography-like method to map the dark and ordinary matter content of the spacetime.
Thirdly, the value of the distance slip is completely independent of the momentary motions of both the observer and the source, eliminating this way any possible measurement systematics or noise due to the peculiar motions.

The distance slip seems fairly challenging to measure, because for sources located at short distances its value is quite small, and therefore very precise astrometric measurements are needed to determine its value. 
However, as we show in this work, $\mu$ attains significant values (of the order of 1) on cosmological distances. The difficult task on these scales is to measure both the parallax and the angular diameter distance of the same object. Distant quasars seem very promising candidates for such a measurement. Although parallax measurements on extragalactic scales seem currently beyond the reach of available instruments, in the near future, a realistic possibility of observing the cosmic parallax of distant quasars is offered by the Gaia mission, see  \cite{Rasanen:2014mca, Quercellini:2010zr} for recent discussions.
In addition, the measurements of the angular diameter distance or the closely related luminosity distance, also required for measuring the distance slip, have either been recently proven possible, \cite{2018Natur.563..657G}, or are already under way: in \cite{Risaliti:2018reu} the authors present a new measurement of the expansion rate of the Universe based on a Hubble diagram of quasars up to redshift $z\sim 6$. The use of this kind of sources offers new possibilities to test the $\Lambda$CDM concordance model in a redshift range which is yet poorly explored, between the farthest observed Supernov\ae\,\,Ia and the cosmic microwave background radiation (CMB).  

On the one hand, exploiting new probes at our disposal, e.g. using other sources for the very same observation, as in  \cite{Risaliti:2018reu}, is one crucial way to take advantage of the huge progress in observational cosmology: it has been evolving rapidly during the last century and now it is considered a precision science, offering an unprecedented opportunity to test gravity on ultra-large scales and/or high redshift.
However, the success of precision cosmology depends not only on accurate observations, but also on the theoretical modelling, which must be understood to at least to the same level of accuracy. Therefore, on the other hand, the contribution from the theory side is also important: theoretical studies have to be targeted to a better interpretation of the cosmological observations and potentially to provide new, clean probes.
In this respect, a unified and comprehensive approach valid for all observables, as the one proposed in \cite{Grasso:2018mei}, would be particularly useful because it may help to better understand and to keep track of different approximations/assumptions that are commonly used in the literature but that we may eventually want to relax. In our work we specialize the machinery of \cite{Grasso:2018mei} to the FLRW spacetime and we focus our study on the new observable $\mu$ with the aim to investigate its potential use as a new cosmological probe.

The paper is organized as follows: in section~\ref{sec:formulation} we review the formulation of geometric optics in terms of the BGOs and we show how to calculate them using the variation of the null geodesic curve with respect to initial data. In section~\ref{sec:final} we derive the expressions of the Jacobi map, the magnification matrix, the angular diameter distance, the parallax distance and the position drift  in terms of the BGOs.
Section~\ref{sec:mu} is dedicated to the distance slip $\mu$: we discuss its general properties and some issues related to its measurement. In section~\ref{sec:cosmo} we focus on cosmology: after discussing the possibility of its measurement on cosmological scales, we begin by reporting the expression for $\mu$ in any FLRW metric, flat and curved, of which we give a detailed derivation with our new variation method in Appendix~\ref{app:0orderW}. We investigate the dependence of this new observable on the cosmological parameters in the redshift range accessible to the observations and we also give and comment its expansion at low redshift. We collect our final remarks in Section~\ref{sec:concl}.

\paragraph{Notation:} Greek indices ($\alpha, \beta, ...$) run from 0 to 3, while Latin indices ($i,j, ...$) run from 1 to 3 and refer to spatial coordinates only. Latin indices ($A,B, ...$) run from 1 to 2.
Boldface indices denote tensors and bitensors expressed in a semi-null frame(s) (as opposed to a coordinate frame(s)), namely the Greek boldface ($\bm \alpha, \bm \beta, \dots$) run again from
0 to 3, Latin boldface indices ($\bm i, \bm j, \dots$) from 1 to 3 and capital boldface Latin $(\bm A, \bm B, \dots)$ again from 1 to 2. 
Dot denotes derivative with respect to conformal time. The subscript $\mathcal{O}$ denotes quantities evaluated at the observer position, i.e. $f(\lambda\obs) \equiv f\obs \equiv f(\mathcal{O})$, $\lambda$ being the affine parameter along the null geodesic connecting observer and source. We will use $f\obs$ or $f(\mathcal{O})$ depending on notational convenience. Analogously, subscript $\calE$ denotes the point of emission by the source. We use the unit system in which $c = 1$.

\section{Formulation}
\label{sec:formulation}

We begin by a short review of the bitensorial formalism applied to geometric optics, for a longer discussion see \cite{Grasso:2018mei}. Let $\gamma_0$ be a null geodesic segment connecting the observation point $x_\calO$, corresponding to the value $\lambda_\calO$ of the affine parameter, with the emission point $x_\calE$, corresponding to an arbitrary value $\lambda$. We fix a coordinate system  which covers the neighbourhoods of both geodesic endpoints. The geodesic curve $x^\mu(x_\calO^\nu, \ell_\calO^\nu,\lambda)$ is function of the 
initial point $x_\calO^\mu$ and the initial tangent vector $\ell_\calO^\mu$ at the observer's position, and of the
affine parameter value $\lambda$.

Consider a perturbation of the initial data for the geodesic at  $\lambda_\calO$, namely the variation position and the tangent vector at the observer $\left(x^\mu_{\cal O}, \ell^\mu_{\cal O} \right)$. Then the deviation at the other endpoint for a fixed value $\lambda_\calE$ of the affine parameter at linear order takes the form
\bea
		\delta x^\mu = {\calW_{XX}}\UD{\mu}{\nu}\,\delta x_\calO^\nu + {\calW_{XL}}\UD{\mu}{\nu}\,\Delta \ell_\calO^\nu, \label{eq:endpointvariation0}
\eea
where $\delta x_\calO^\mu$ and $\delta x^\mu$ are the displacements at $\lambda_\calO$ and $\lambda$ respectively, and $\Delta \ell_\calO^\mu$ is the covariantly defined deviation of the initial tangent vector, given by
\bea
		\Delta \ell_\calO^\mu = \delta \ell_\calO^\mu + \Gamma\UD{\mu}{\alpha\beta}(\calO)\,\ell_\calO^\alpha\, \delta x_\calO^\beta. \label{eq:covariantvariation}
\eea
${\calW_{XX}}{}\UD{\mu}{\nu}$  and ${\calW_{XL}}{}\UD{\mu}{\nu}$ are bitensors, mapping tangent vectors from $\calO$ to $\calE$, called the bilocal geodesic operators, or BGOs (transport operators in differential geometry literature or bundle transfer matrices in non-relativistic geometric optics \cite{Uzun:2018yes}). They can be expressed as solutions of matrix ODEs along the fiducial geodesic $\gamma_0$ involving the Riemann tensor \cite{Fleury:2013sna, Grasso:2018mei}. Namely, it follows from the 1st order GDE that in a parallel-propagated frame $\calW_{XX}{}\UD{\bm \mu}{\bm \nu}$ and $\calW_{XL}{}\UD{\bm \mu}{\bm \nu}$ solve the equations
\bea
\frac{\dd^2}{\dd \lambda^2}\,{\calW_{XX}}\UD{\bm \mu}{\bm \nu} - R\UD{\bm \mu}{\bm \alpha \bm \beta \bm \sigma}\,\ell^{\bm \alpha}\,\ell^{\bm \beta} \,{\calW_{XX}}\UD{\bm \sigma}{\bm \nu} &=& 0 \label{eq:WfromR1}\\
\frac{\dd^2}{\dd \lambda^2}\,{\calW_{XL}}\UD{\bm \mu}{\bm \nu} - R\UD{\bm \mu}{\bm \alpha \bm \beta \bm \sigma}\,\ell^{\bm \alpha}\,\ell^{\bm \beta} \,{\calW_{XL}}\UD{\bm \sigma}{\bm \nu} &=& 0 \label{eq:WfromR2}
\eea
with the initial data
\bea
{\calW_{XX}}\UD{\bm \mu}{\bm \nu} \Big|_{\lambda=\lambda_\calO} &=& \delta\UD{\bm \mu}{\bm \nu} \label{eq:WfromR3} \\
\frac{\dd}{\dd \lambda}{\calW_{XX}}\UD{\bm \mu}{\bm \nu}\Big|_{\lambda=\lambda_\calO} &=& 0 \label{eq:WfromR4}  \\
{\calW_{XL}}\UD{\bm \mu}{\bm \nu}\Big|_{\lambda=\lambda_\calO} &=& 0 \label{eq:WfromR5} \\
\frac{\dd}{\dd \lambda}{\calW_{XL}}\UD{\bm \mu}{\bm \nu}\Big|_{\lambda=\lambda_\calO} &=& \delta\UD{\bm \mu}{\bm \nu} \label{eq:WfromR6}  . 
\eea

\subsection{Bilocal geodesic operators from the variations of the general solution of the geodesic equation.} \label{deviationmethod}

Equations (\ref{eq:WfromR1})-(\ref{eq:WfromR6}) relate the bitensors $\calW_{XX}$ and $\calW_{XL}$ directly to the curvature along $\gamma_0$, but  they are not all that useful    in the cosmological setting. We present therefore another way to evaluate them, based on direct differentiation of the general solution of the geodesic equation.

Consider the general solution of the geodesic equation $x^\mu(x_\calO^\nu, \ell_\calO^\nu, \lambda)$, given in a particular coordinate system, depending on the 
initial point $x_\calO^\mu$ at the observer's position, the initial tangent vector $\ell_\calO^\mu$ at the observer's position, and on the
affine parameter $\lambda$ at the emission point. The idea is to express the BGOs, decomposed in the coordinate frames, by the derivatives of this general solution with respect to the initial data and by other geometric objects, such as the Christoffel symbols. This is an entirely new method and, to our knowledge, it hasn't been comprehensively 
discussed in the literature so far. It can be applied whenever we know the general solution of the general geodesic equation  or just the null geodesic equation. The solution can be perturbative or exact, possibly even given by implicit relations and quadratures.
We will now sketch this method briefly.

Note first that if we allow additionally for a variation of the affine parameter  $\lambda$ at the endpoint $\calE$ of the geodesic segment, instead of~\eqref{eq:endpointvariation0} we obtain
\bea
		\delta x^\mu = {\calW_{XX}}\UD{\mu}{\nu}\,\delta x_\calO^\nu + {\calW_{XL}}\UD{\mu}{\nu}\,\Delta \ell_\calO^\nu + \ell_\calE^\mu\,\delta \lambda,
		\label{eq:endpointvariation1}
\eea
where $\ell_\calE^\mu$ is the tangent vector to $\gamma_0$ at $\calE$. This is 
because for fixed initial data (i.e. $\delta x_\calO^\mu =0$ and $\Delta l_\calO^\mu = 0$) a small variation of the final value of $\lambda$ produces a shift of the endpoint proportional to the tangent vector at the endpoint. We can interpret relation~(\ref{eq:endpointvariation1}) as follows: the total variation of the geodesic endpoint with respect to the initial data and the affine parameter, obtained by differentiating the general solution $x^\mu(x_\calO^\nu, \ell_\calO^\nu, \lambda)$ and expressed in the basis
given by the variations $\delta x_\calO^\mu$, $\Delta \ell_\calO^\mu$ and $\delta \lambda$, yields the components of the bilocal geodesic operators $\calW_{XX}$, $\calW_{XL}$, as well as the tangent vector $\ell_\calE$ in the appropriate
coordinate basis. We can therefore regard the 4 functions $x^\mu(x_\calO^\nu, \ell_\calO^\nu, \lambda)$, representing the general solution of the general geodesic equation, as 
analogs of the thermodynamical potentials: their total derivatives give physically interesting quantities as expansion coefficients (components) \emph{when expressed in the correct basis of differentials}.  Keep in mind that it is important that we take the variations in \emph{all} components of the initial data as well as the affine parameter, and that the basis of expansion is precisely the one described above, i.e. $(\delta x_\calO^\mu, \Delta \ell_\calO^\mu, \delta\lambda)$ in the chosen coordinate system.

For the practical purpose of a convenient calculation of the bilocal geodesic operators,  assume  we are given the functions $x^\mu(x_\calO^\nu, \ell_\calO^\nu, \lambda)$ in a coordinate system. Then we calculate  their total variation with respect to all variables\footnote{We use here the notation borrowed from thermodynamics, where $\left(\frac{\partial F}{\partial x}\right)_{y,z}$ means the partial derivative of $F$ with respect to $x$ with $y$ and $z$ kept fixed.}
\bea
 \delta x^\mu &=& \left(\frac{\partial x^\mu}{\partial x^\nu_\calO}\right)_{\ell_\calO, \lambda}\,\delta x^\nu_\calO + 
 	\left(\frac{\partial x^\mu}{\partial \ell^\nu_\calO}\right)_{x_\calO, \lambda}\,\delta \ell^\nu_\calO  \nonumber \\ 
 	&&+ \left(\frac{\partial x^\mu}{\partial\lambda}\right)_{x_\calO, \ell_\calO}\,\delta \lambda\,.
\eea
We can now make use of (\ref{eq:covariantvariation}) to change the basis of variations from $(\delta x_\calO^\mu, \delta \ell_\calO^\mu, \delta\lambda)$ to
$(\delta x_\calO^\mu, \Delta \ell_\calO^\mu, \delta\lambda)$ and compare the result with (\ref{eq:endpointvariation1}). We obtain the following relations:
\bea
{\calW_{XX}}\UD{\mu}{\nu} &=&  - 
\left(\frac{\partial x^\mu}{\partial \ell_\calO^\sigma}\right)_{x_{\cal O},\lambda} \Gamma\UD{\sigma}{\alpha\nu}(\calO)\,\ell_\calO^\alpha \nonumber\\
&& + \left(\frac{\partial x^\mu}{\partial x_\calO^\nu}\right)_{\ell_{\cal O},\lambda} \label{eq:relation1}\\
{\calW_{XL}}\UD{\mu}{\nu} &=& \left(\frac{\partial x^\mu}{\partial \ell_\calO^\nu}\right)_{x_{\cal O},\lambda} \label{eq:relation2}\\
\ell_\calE^\mu &=& \left(\frac{\partial x^\mu}{\partial \lambda}\right)_{x_\calO,l_\calO}. \label{eq:relation3}
\eea
They express the two geodesic bitensors (and the tangent vector at $\calE$) explicitly in terms of the partial derivatives of the solution of the geodesic equation. In the next section we will demonstrate how these bitensors can then be used directly to calculate the magnification matrix, the parallax and the position drifts for any observers and sources located at $\calO$ and $\calE$ respectively. Therefore the method of endpoint variations sketched here allows for calculating all those three optical effects for any observer-source pair with one calculation.
 
Now, in many physically interesting cases, including the FLRW metric, we do not have a simple, closed form of the general solution of the geodesic equation, but rather the 
general solution for \emph{null} 
geodesics. This restricts the type of variations of the initial tangent vector we may consider, and thus restricts the components of $\calW_{XL}$ we may
obtain by the variational method.
Note that it should nevertheless be possible to recover the \emph{optical} properties of the spacetime just from that limited information. While the variational method sketched above requires the knowledge of all geodesics in the neighbourhood of a given one, we may modify it a little bit to make it work even if only the general solution for null geodesics is available.

The requirement for the perturbed geodesics to remain null at linear order is equivalent to a constraint on the admissible initial deviation vector:
\bea
\Delta \ell_\calO^\sigma \,\ell_{\calO\,\sigma} = 0, \label{eq:admissible}
\eea
or
\bea
\Delta \ell_\calO^0 = -\frac{\ell_{\calO\,i}}{\ell_{\calO\,0}}\,\Delta\ell_{\calO}^i  \label{eq:admissible2}
\eea

Assume we are just given the solution for past-directed null geodesics, parametrized by the initial point and the three spatial components of the initial tangent vector
$x^\mu(x_\calO^\mu,\ell_\calO^i,\lambda)$. The number of independent variables is thus reduced by one and the total variation reads
\bea
 \delta x^\mu &=& \left(\frac{\partial x^\mu}{\partial x^\nu_\calO}\right)_{\ell_\calO, \lambda}\,\delta x^\nu_\calO + 
 	\left(\frac{\partial x^\mu}{\partial \ell^i_\calO}\right)_{x_\calO, \lambda}\,\delta \ell^i_\calO \nonumber \\ &&+ \left(\frac{\partial x^\mu}{\partial\lambda}\right)_{x_\calO, \ell_\calO}\,\delta \lambda\,, \label{eq:nullvariations}
\eea
where $i$ runs from 1 to 3.
This formula needs to be related to (\ref{eq:endpointvariation1}) in order to obtain the relation between the partial derivatives and the bilocal operators. 
Note that admissible deviation vectors $\Delta \ell_\calO^\mu$, satisfying (\ref{eq:admissible}), can be parametrized just by the spatial components $\Delta \ell_\calO^i$.

Let us introduce the notation ${\calV_{XL}}\UD{\mu}{i}$ for the ${\calW_{XX}}$ operator acting on admissible vectors, and expressed in terms of their spatial components, i.e. let
\bea
 {\calV_{XL}}\UD{\mu}{i}\,\Delta \ell_{\calO}^i ={\calW_{XL}}\UD{\mu}{\nu}\,\Delta \ell_{\calO}^\nu  \label{eq:VwsW}
\eea
for all  vectors $\Delta \ell_\calO^\mu$ satisfying (\ref{eq:admissible}). From (\ref{eq:admissible2}) we can get an exact relation to the components of $\calW_{XL}$:
\bea
 {\calV_{XL}}\UD{\mu}{i} = {\calW_{XL}}\UD{\mu}{i} -  {\calW_{XL}}\UD{\mu}{0}\,\frac{\ell_{\calO\,i}}{\ell_{\calO\,0}}.
\eea
Briefly speaking, $\calV_{XL}$ is the $\calW_{XL}$ operator restricted to variations of directions respecting the null conditions, and expressed in a convenient, purely spatial
parametrization. On the other hand, its components constitute  precisely those combinations of components of $\calW_{XL}$ which we can be extracted from the variations of the initial data restricted to null geodesics, i.e. those variations to which we have access via the relation (\ref{eq:nullvariations}). 

The reader may now check that for the restricted variations we have
\bea
 \delta x^\mu = {\calW_{XX}}\UD{\mu}{\nu}\,\delta x_\calO^\nu + {\calV_{XL}}\UD{\mu}{i}\,\Delta \ell_\calO^i+ \ell_\calE^\mu\,\delta \lambda.
		\label{eq:endpointvariation1null}
\eea
Applying the identity $\Delta \ell_\calO^i =  \delta \ell_\calO^i + \Gamma\UD{i}{\alpha\beta}(\calO)\,\ell_\calO^\alpha\, \delta x_\calO^\beta$ and comparing with
(\ref{eq:nullvariations}) we get the analog of relations (\ref{eq:relation1})-(\ref{eq:relation3}) for null geodesics
\bea
{\calW_{XX}}\UD{\mu}{\nu} &=&  - 
\left(\frac{\partial x^\mu}{\partial \ell_\calO^i}\right)_{x_{\cal O},\lambda} \Gamma\UD{i}{\alpha\nu}(\calO)\,\ell_\calO^\alpha\nonumber\\ &&+  \left(\frac{\partial x^\mu}{\partial x_\calO^\nu}\right)_{\ell_{\cal O},\lambda}\label{eq:relation11}\\
{\calV_{XL}}\UD{\mu}{i} &=& \left(\frac{\partial x^\mu}{\partial \ell_\calO^i}\right)_{x_{\cal O},\lambda} \label{eq:relation22}\\
\ell_\calE^\mu &=& \left(\frac{\partial x^\mu}{\partial \lambda}\right)_{x_\calO,l_\calO}. 
\eea
 The equations above allow to calculate the \emph{optical part} of the two geodesic bitensors in terms of partial derivatives of the general solution of the \emph{null} geodesic equation. They constitute the first important result of this article. 
We shall use them throughout the rest of the paper to calculate $\calW_{XX}$ and $\calV_{XL}$ for the unperturbed FLRW solution.

\section{Optical observables from the bilocal geodesic operators}
\label{sec:final}

The main advantage of the BGOs lies in the fact that we can express a number of observables of interest in a unified framework via $\calW_{XX}$ and $\calV_{XL}$ (or $\calW_{XL}$) and the kinematical variables 
describing the momentary motions of the source and the observer in the moments of light emission and observation respectively \cite{Grasso:2018mei, Korzynski:2017nas}.
The observables  in question 
are the angular diameter distance $\Dang$, the luminosity distance $\Dlum$, the magnification matrix $M\UD{\bm A}{\bm B}$, the parallax and the position drift (or proper motion) $\delta_\calO r^{\bm A}$.
We can therefore consider not only observers and sources comoving with the cosmic flow or defined in a particular gauge, but also consider situations in which both are boosted
with respect to the large-scale flow, for example due to the  small-scale non-linearities.

We first note that the Jacobi map can be expressed using $\calW_{XL}$ or $\calV_{XL}$. 
Let 
$e_{\bm A}$ denote a parallel-propagated Sachs basis of two vectors orthogonal to $\ell^\mu$ along $\gamma_0$. 
Recall that the Jacobi map $\calD$ relates the initial direction deviation with the displacement along a null geodesic for vectors orthogonal to $\ell^\mu$:
\bea
 \xi^{\bm A}(\lambda) = \calD\UD{\bm A}{\bm B}(\lambda)\,\Delta l^{\bm B}_\calO.
\eea

Adding two more vectors, a parallel-propagated, normalized timelike vector $u^\mu$ and the null tangent $\ell^\mu$, we obtain the parallel-propagated semi-null frame (SNF) $(u^\mu, e_{\bm A}^\mu, \ell^\mu)$.
In this frame the components of the Jacobi map $\calD$ simply coincide with the transverse components of $\calV_{XL}$ and $\calW_{XL}$:
\bea \label{eq:jacobi}
 \calD\UD{\bm A}{\bm B} &=& \calV_{XL}{}\UD{\bm A}{\bm B} = \calW_{XL}{}\UD{\bm A}{\bm B}.
 \eea
 This allows us to write all the observables derived from the Jacobi map in terms of the transverse components of the BGOs. Note that we may use either $\calW_{XL}$ or $\calV_{XL}$ for this purpose since their transverse components always coincide. Substituting $\calV_{XL}$ by $\calW_{XL}$ is also possible for the other observables discussed below, since they only make use of the transverse components of $\calV_{XL}$.
It is also noteworthy that the values of $\calD\UD{\bm A}{\bm B} $ do not depend on the choice of the timelike vector, making this way the formalism observer frame-invariant \cite{Korzynski:2017nas, Grasso:2018mei}.
 
The Jacobi matrix is directly related to the magnification matrix $M\UD{\bm A}{\bm B}$, which in turn relates  the transverse displacements along the null geodesics to the the angles on the observer's sky:
\bea
 \delta \theta^{\bm A} = M\UD{\bm A}{\bm B}\,\delta x_\calE^{\bm B}. \nonumber
\eea
Namely, for an observer with 4-velocity $u_\calO$ we have
\bea
M\UD{\bm A}{\bm B} &=& \left(l_{\calO\,\mu}\,u_\calO^\mu\right)^{-1}\,\calD^{-1}{}\UD{\bm A}{\bm B} \nonumber \\ &=&  \left(l_{\calO\,\mu}\,u_\calO^\mu\right)^{-1}\,\left(\calV_{XL}{}\UD{\bm A}{\bm B}\right)^{-1}, \label{eq:definitionofM}
\eea
where $\left(\calV_{XL}{}\UD{\bm A}{\bm B}\right)^{-1}$ denotes the inverse of the transverse submatrix of $\calV_{XL}$. 

The angular diameter distance to an object is formally defined as the square root of the ratio between the cross-sectional area of a luminous object and the solid angle taken by its image in the observer's
celestial sphere \cite{perlick-lrr}. It can be expressed as the determinant of the magnification map in a Sachs frame:
\bea
\Dang &=& \left| \det M\UD{\bm A}{\bm B} \right|^{-1/2} = \left(l_{\calO\,\mu}\,u_\calO^\mu\right)\,\left|\det \calD\UD{\bm A}{\bm B}\right|^{1/2} \nonumber \\
&=&\left(l_{\calO\,\mu}\,u_\calO^\mu\right)\,\left|\det \calV_{XL}{}\UD{\bm A}{\bm B}\right|^{1/2}. \label{eq:definitionofDang}
\eea
The prefactor $\left(l_{\calO\,\mu}\,u_\calO^\mu\right)$ in (\ref{eq:definitionofM}) and (\ref{eq:definitionofDang}) represents the relativistic light aberration effect: the same objects appear larger or smaller 
for observers passing through the same point $\calO$ with different 4-velocities. The difference in the apparent size is related to the difference of 4-velocities and to the direction of observation, defined  by the null tangent vector $\ell_\calO^\mu$.

The emitter-observer asymmetry operator $m\UD{\bm A}{\bm \alpha}$ determines how the effect of displacements on one end of the null geodesics differ from the displacements on the other one
\cite{Grasso:2018mei}. It was introduced first in \cite{Korzynski:2017nas} and it appears in the expressions for the parallax and position drifts, or proper motions. It can be read out from $\calW_{XX}$ expressed in a parallel-propagated semi-null frame: 
\bea
 m\UD{\bm A}{\bm \alpha} = \calW_{XX}{}\UD{\bm A}{\bm \alpha} - \delta\UD{\bm A}{\bm \alpha} \label{eq:mfromWXX}
\eea
(recall that the boldface indices are used for geometric objects decomposed in the semi-null frame: capital Latin indices $\bm A, \bm B, \dots$ run from 1 to 2, while Greek indices $\bm \mu, \bm \nu, \dots$ run from 0 to 3).
Consider now the parallax matrix $\Pi\UD{\bm A }{\bm B}$, relating the displacement of the position of observation in a transverse direction $\delta x^{\bm A}_\calO$ with the apparent shift of the source's position $\delta \theta^{\bm A}$
on the observer's sky, defined with respect to parallel propagated directions on the celestial sphere\footnote{One can also prove that $\Pi\UD{\bm A}{\bm B}$ is always a symmetric matrix, but this is irrelevant for our purposes.}:
\bea
 \delta \theta^{\bm A} = -\Pi\UD{\bm A}{\bm B}\,\delta x_\calO^{\bm B}. \label{eq:parallaxmatrixdef}
\eea
It can also be expressed using the BGOs. Namely, in \cite{Grasso:2018mei} the following relation has been derived:
\bea
\Pi\UD{\bm A}{\bm B} &=& \left(l_{\calO\,\mu}\,u_\calO^\mu\right)^{-1}\,\calD^{-1}{}\UD{\bm A}{\bm C}\,\left(\delta\UD{\bm C}{\bm B} + m_\perp{}\UD{\bm C}{\bm B} \right)\label{eq:definitionofPi}
\eea
It follows then that
\bea
\Pi\UD{\bm A}{\bm B} &=& \left(l_{\calO\,\mu}\,u_\calO^\mu\right)^{-1}\,\left(\calV_{XL}{}\UD{\bm A}{\bm C}\right)^{-1}\,\calW_{XX}{}\UD{\bm C}{\bm B}. \label{eq:PiviaBGOs}
\eea
The parallax effect is used in astronomy to measure distances to luminous sources in an astrometric technique  as the trigonometric parallax.
The theoretical justification of this method is based on the flat spacetime analysis of the geometry of light rays and obviously requires a modification if we want to include the curvature effects. 
The parallax distance in a general, curved spacetime can be defined in many ways \cite{rasanen, Grasso:2018mei},
the differences coming from different methods of averaging over the baseline orientation. In this paper we use the one based on the determinant of the parallax matrix, fully analogous to 
(\ref{eq:definitionofDang}):
\bea
\Dpar  &=& \left| \det \Pi\UD{\bm A}{\bm B} \right|^{-1/2} \\\nonumber
&=& \left(l_{\calO\,\mu}\,u_\calO^\mu\right)\,\left| \det{ \calD{}\UD{\bm A}{\bm B}}\right|^{1/2}  \, \left| \det \left(
\delta\UD{\bm A}{\bm B} + m_\perp{}\UD{\bm A}{\bm B}\right) \right|^{-1/2} . \label{eq:definitionofDpar}
\eea
In terms of the BGOs $\Dpar$  is then given by
\bea
\Dpar  &=& \left(l_{\calO\,\mu}\,u_\calO^\mu\right)\,\left| \det \calV_{XL}{}\UD{\bm A}{\bm B}\right|^{1/2}  \nonumber\\
&&\cdot \left| \det \calW_{XX}{}\UD{\bm A}{\bm B} \right|^{-1/2} . \label{eq:definitionofDpar2}
\eea

Finally we may consider the proper motions or position drifts, i.e. the rate of change  of the sources' positions on the observer's celestial sphere in the observer's proper time. 
The position change is defined here  with respect to the fixed spatial directions given by a Fermi-Walker transported frame. For a source with momentary 4-velocity $u_\calE^\mu$ at $\calE$ and an observer with momentary 4-velocity $u_\calO^\mu$ and 4-acceleration $w_\calO^\mu$ at $\calO$ we have \cite{Korzynski:2017nas}:
\begin{widetext}
\bea
\delta_\calO r^{\bm A} = \left(l_{\calO\,\mu}\,u_\calO^\mu\right)^{-1}\,\calD^{-1}{}\UD{\bm A}{\bm B}\,\left(\left(\frac{1}{1+z}\,u_\calE - \hat u_\calO\right)^{\bm B} - m\UD{\bm A}{\bm \mu}\,u_\calO^{\bm \mu}\right)
 + w_\calO^{\bm A},
\eea
where $\delta_\calO r^{\bm A}$ is the position drift rate in radians per a unit of the observer's proper time, $z$ is the redshift measured by the observer and $\hat u_\calO$ is the parallel  transport of $u_\calO$ from $\calO$ to $\calE$.
Again this quantity can be expressed directly using BGOs in parallel propagated SNF:
\bea
\delta_\calO r^{\bm A} = \left(l_{\calO\,\mu}\,u_\calO^\mu\right)^{-1}\,\left(\calV_{XL}{}\UD{\bm A}{\bm B}\right)^{-1}\,\left(\frac{1}{1+z}\,u_\calE^{\bm B} - \calW_{XX}{}\UD{\bm B}{\bm \nu}\,u_{\calO}^{\bm \nu}\right)
 + w_\calO^{\bm A}.
\eea
\end{widetext}
We stress  that in order to calculate \emph{any} of these observables, measured by \emph{any} observer $u_\calO$, comoving or not, and with respect to \emph{any} source $u_\calE$, we only need to evaluate
the BGOs $\calV_{XL}$ or $\calW_{XL}$ as well as $\calW_{XX}$ between two points connected by a null geodesic. As we have shown above, this can be done by varying the functional form of the null geodesic, obtained exactly or perturbatively.

\section{A new observable: the distance slip}
\label{sec:mu}
\subsection{Definition and properties}
\label{sec:muprop}
In \cite{Grasso:2018mei} the following dimensionless quantity has been defined
\bea \label{eq:defofmu}
\mu = 1- \frac{\det \Pi\UD{\bm A}{\bm B}}{\det M\UD{\bm A}{\bm B}}. 
\eea
We can rewrite with the help of Eqs.~\eqref{eq:definitionofM} and \eqref{eq:definitionofPi} in terms of the emitter-observer asymmetry operator, and thus in term of the spacetime curvature:
\bea \label{eq:defofmu2}
\mu =  1- \det \left(\delta\UD{\bm A}{\bm B} + {m_\perp}\UD{\bm A}{\bm B}\right),
\eea
${m_\perp}\UD{\bm A}{\bm B}$ denoting here the transverse components of the full operator $m\UD{\bm A}{\bm \mu}$. 
On the other hand, using Eqs.~\eqref{eq:definitionofDpar} and~\eqref{eq:definitionofDang}, it can be expressed in terms of the parallax distance and the angular diameter distance from the observation point to a single object far way:
\bea
 \mu = 1 - \sigma\,\frac{\Dang^2}{\Dpar^2}, \label{eq:mufromDangDpar2}
\eea
where $\sigma = \pm 1$ defines the sign and depends on the parity of the magnification matrix and the parallax matrix, i.e. $\sigma = \textrm{sgn} \left(\det M\,\UD{\bm A}{\bm B}\right)\,\textrm{sgn}\left( \det \Pi\UD{\bm A}{\bm B} \right)$. Note that for most objects observed in the Universe we detect simple images, i.e. $\det M\UD{\bm A}{\bm B} > 0$ (inverted images may appear only for strongly lensed objects, which are relatively rare) and 
the dependence of the parallax on the displacement is not inverted either (except, again, strongly lensed images), i.e.
$\det \Pi\UD{\bm A}{\bm B} > 0$. This means that for most objects we have simply
\bea
 \mu = 1 - \frac{\Dang^2}{\Dpar^2}. \label{eq:mufromDangDpar}
\eea
In other words, for a given observer and a given distant source $\mu$ measures 
 the relative difference between the results of two methods of distance determination: by the source's parallax and by  its angular size. We will therefore call $\mu$ the \emph{distance slip}. 
 
 Since the angular diameter distance in related to the luminosity distance $\Dlum$ and the redshift by the Etherington's reciprocity relation $\Dlum = \Dang(1+z)^2$ \cite{etherington, etherington2, perlick-lrr}, we can also express
 $\mu$ using $\Dlum$ and $z$:
\bea
\mu = 1 - (1+z)^{-4}\frac{\Dlum^2}{\Dpar^2} \label{eq:mufromDlumDpar}
 \eea
The distance slip as an observable  has a number of peculiar properties, not shared by the standard observables like the redshift or the luminosity distance, which we will now  briefly summarize. These properties hold for any spacetime as long as we may use the first order geodesic deviation equation approximation and the distant observer approximation. For proofs and longer discussion see \cite{Grasso:2018mei}.

\paragraph{ Independence from momentary motions of the observer and the emitter. } Consider a spacetime with fixed emission and observation points $\calE$ and $\calO$, connected by a null geodesic. The parallax distance and the angular diameter distance depend on both the spacetime geometry as well as 
the momentary 4-velocity of the observer $u_\calO^\mu$ at the moment of observation:
\bea
\Dang &\equiv & \Dang\left[ g_{\mu\nu}, u_\calO^\mu\right] \\
\Dpar &\equiv & \Dpar\left[g_{\mu\nu} , u_\calO^\mu\right]\,,
\eea
where $g_{\mu\nu}$ denotes here the spacetime geometry.
Note that they do not depend on the momentary 4-velocity of the emitter in the moment of light emission $u_\calE^\mu$, or any other quantities describing the motions of both the emitter and observer,
such as the momentary 4-accelerations. The independence of $\Dang$ from the emitter's rest frame is a standard result (see \cite{perlick-lrr}), which can be seen 
as a consequence of the Sachs shadow theorem \cite{sachs}. The independence of $\Dpar$ of the emitter's motion on the other hand is a fairly straightforward consequence of the  
relativistic parallax definition as given by a momentary measurement, using light emitted in a single moment along the source's worldline, see \cite{Grasso:2018mei}.
The remaining dependence of both distances on $u_\calO^\mu$ is due to the standard light aberration effect, described by special relativity: small regions of the sky appear larger or smaller depending on
the observer's 4-velocity. 
This dependence appears in (\ref{eq:definitionofDang}) and (\ref{eq:definitionofDpar}) as the common prefactor $\ell_{\calO\,\mu}\,u_{\calO}^{\mu}$. The reader may check, however, that  $\mu$ does not depend on \emph{any} kinematical variables describing the momentary motions of the source and the observer, because in the ratio $\Dang^2 / \Dpar^2$, appearing in its definition (\ref{eq:mufromDangDpar2}), the $u_\calO^\mu$-dependent prefactors cancel out. The remaining expression is a functional of the spacetime geometry only:
\bea
\mu  \equiv \mu\left[ g_{\mu\nu} \right] .
\eea
In other words, for a given spacetime and two events $\calE$ and $\calO$, connected by a null geodesic, we can be sure that \emph{any}
emitter-observer pair  will measure the same value of $\mu$ when passing through $\calE$ and $\calO$ respectively.

\paragraph{ Distance slip as a functional of the curvature along the line of sight. } Let $\gamma_0$ denote the null geodesic connecting $\calE$ and $\calO$, $\lambda$ be its affine parameter and $\ell^\mu$ its tangent vector. One can show that the distance slip
$\mu$ can be expressed as a non-linear functional of the curvature tensor along $\gamma_0$. Namely, let $e_{\bm A}^\mu$ denote the Sachs basis, i.e. two parallel propagated, normalized and orthogonal spatial vectors, perpendicular  to $\ell^\mu$. Then we can define the matrix ${m_\perp}\UD{\bm A}{\bm B}$ of the transverse emitter-observer symmetry operator as the solution of the following ODE in that basis:
\bea
\frac{\dd^2}{\dd\lambda^2} {m_\perp}\UD{\bm A}{\bm B} - R\UD{\bm A}{\alpha\beta\bm C}\,\ell^\alpha\,\ell^\beta\,{m_\perp}\UD{\bm C}{\bm B} &=&  R\UD{\bm A}{\alpha\beta\bm B}\,\ell^\alpha\,\ell^\beta \quad \label{eq:mperpODE}
\eea
with the initial data at $\calO$:
\bea
\frac{\dd}{\dd\lambda} {m_\perp}\UD{\bm A}{\bm B} \big|_{\lambda_\calO}  &=&  0  \label{eq:mperpODEID1}\\
{m_\perp}\UD{\bm A}{\bm B}\big|_{\lambda_\calO} &=& 0. \label{eq:mperpODEID2}
\eea
${m_\perp}\UD{\bm A}{\bm B}$ will now denote the solution at $\calE$. Then we can apply (\ref{eq:defofmu2}):
\bea
\mu = 1 - \det\left(\delta\UD{\bm A}{\bm B} + {m_\perp}\UD{\bm A}{\bm B}\right). \label{eq:mufrommperp}
\eea
We see therefore that  $\mu$ depends on the spacetime geometry  via the Riemann tensor along $\gamma_0$ or, more precisely, via the transverse components of the optical tidal tensor
$R\UD{\mu}{\nu\rho\sigma}\,\ell^\nu\,\ell^\rho$:
\bea
 \mu \equiv \mu\left[ R\UD{\bm A}{\alpha\beta\bm B}\,\ell^\alpha\,\ell^\beta \Big|_{\gamma_0} \right].
\eea
The reader may check that $\mu$ given by (\ref{eq:mufrommperp}) is independent of the choice of the parallel-transported Sachs frame.

\paragraph{ Distance slip as a curvature detector. } In a flat space we have $\mu = 0$ along any null geodesic. This can be seen directly from equations (\ref{eq:mperpODE})-(\ref{eq:mufrommperp}) if we substitute $R\UD{\mu}{\nu\rho\sigma} = 0 $. Alternatively, we note that in a flat spacetime both methods of distance determination must give the same result for the same object, i.e. $\Dang = \Dpar = D$, where $D$ is the spatial distance  between $\calO$ and $\calE$, calculated on the 3D hypersurface of the observer's rest frame. Since all images are simple in a flat spacetime and the parallax map is not inverted we have $\mu = 0$ from (\ref{eq:mufromDangDpar}).
Conversely, any deviation of $\mu$ from 0 means that the spacetime must be curved somewhere along $\gamma_0$ between points $\calO$ and $\calE$. Note that this property makes $\mu$ 
somewhat similar to the 
angle deficit of a geodesic triangle in 2-dimensional non-Euclidean geometry. Namely, the measurement of the angle deficit probes curvature within a finite region of the manifold, defined by the  interior of a geodesic triangle, and
the same way the measurement of $\mu$ probes the curvature along the fiducial null geodesic $\gamma_0$ in the segment between the emission and the observation points.

\paragraph{Tomographic property for short distances. } One can prove that for short distances or weak curvature $\mu$ can be expanded as a series in the powers of the curvature tensor:  
\bea
 \mu &=& \int_{\lambda_{\calO}}^{\lambda_{\calE}} R_{\mu\nu}\,\ell^\mu\,\ell^\nu \, (\lambda_\calE - \lambda) \,\dd\lambda + O\left(\mathbf{R}^2\right) \label{eq:mufromRicciT}
  \\  &=& 8\pi G \int_{\lambda_{\calO}}^{\lambda_{\calE}} T_{\mu\nu}\,\ell^\mu\,\ell^\nu \, (\lambda_\calE - \lambda) \,\dd\lambda + O\left(\mathbf{R}^2\right), \label{eq:mufromT}
\eea
where $O\left(\mathbf{R}^2\right)$ denotes terms involving quadratic and higher powers of the Riemann tensor and for the second equality we make use of Einstein equations.
The leading order linear term should be sufficient whenever the impact of curvature on light propagation is small. This is always true if the distance between $\calO$ and $\calE$ is
short in comparison to the characteristic scale of the curvature of the spacetime. In the cosmological setting this condition means that the distance is small with respect to the Hubble radius and  that the null geodesics $\gamma_0$ does not stay for too long in  strongly overdense regions. 

We note from (\ref{eq:mufromT}) that in the leading, linear order the Weyl tensor  drops out of the integral, leaving only the stress-energy tensor contracted twice with the null tangent. The cosmological constant drops out as well,
since the term $\Lambda\,g_{\mu\nu}$, contracted twice with  null vector $\ell^\mu$, vanishes too. In the end we are left in
(\ref{eq:mufromRicciT})  with just the integral of the stress-energy tensor of the matter  (dark and baryonic). 
Therefore $\mu$ \emph{depends in the leading order only on the gravitating matter content, both dark and ordinary, along the line of sight}. The linear kernel $\lambda_\calE - \lambda$ in the integral
makes the result more sensitive to the matter distribution closer to the observer than far away.

Note also that mass concentrations located off the
optical axis may easily
influence the exact position of the emitter's image on the observer's sky due to gravitational light bending and at the same time cause a sizable image distortion due to tidal forces. However, as we can see from (\ref{eq:mufromRicciT}), they cannot directly influence  $\mu$ in the leading order, unless they happen to be positioned exactly between the observer and the emitter along $\gamma_0$. Thus $\mu$ yields  a weighted integral of the matter density  located precisely between the source and the emitter, reminiscent of tomography.

\subsection{Methods of measurement}
 
One of the main advantages of $\mu$ as an observable is that it can be measured using purely astrometric methods, by comparing the parallax distance with the angular diameter distance to the same object. As we already noted, the latter can be also measured indirectly,
by measuring the luminosity distance and the redshift, see Equation (\ref{eq:mufromDlumDpar}). Therefore the objects we use for measurements must be  standard rulers or standard candles for which we can additionally  measure  the parallax effect. In this subsection we will  discuss the methods of measurement as well as some of the issues connected with calibration.

\paragraph{Parallax distance determination using the annual parallax in a curved spacetime. }The definition of the parallax distance (\ref{eq:definitionofDpar}), introduced in \cite{Grasso:2018mei}, requires the determination of the exact position of the object simultaneously from at least 3 points of view, by three comoving observers (the classic parallax in the terminology of \cite{rasanen, Grasso:2018mei}). The measurements
must be performed at the moment the observers cross  the future light cone centered at a single point on the source's worldline. This way all observers register the light emitted by the source at the same  moment. In the distant observer approximation this can be achieved by appropriate timing of observation using an appropriate null time coordinate, see \cite{Grasso:2018mei}.
 This kind of simultaneous measurement from many points of view is not feasible in astronomy, and the standard trigonometric parallax measurements actually use the time variations of the apparent positions due to the annual Earth's motion, see \cite{klioner2003}. In a flat spacetime this is easy to justify, because for sufficiently short time scales the apparent position on the sky (i.e. the single-worldline parallax defined in \cite{Grasso:2018mei}) for Earth-based observers varies with time according to the formula\footnote{We neglect here the contribution from the aberration, since it is commonly subtracted from the parallax measurements.}
\bea
 \delta \theta^{\bm A}(t_b) = v^{\bm A}\cdot t_{b} - \Dpar^{-1}\,\delta x_{\calO}^{\bm A}(t_b), \label{eq:parallaxtime0}
\eea 
$t_b$ being the appropriate null time coordinate related to the barycentric time, $\delta x_{\calO}^{\bm A}(t_b)$ the momentary position of the Earth wrt the Solar System barycenter. The first term corresponds to the peculiar motion with constant angular velocity $v^{\bm A}$ and the second one is the  ``pure" parallax effect we want to measure. 
Note that both terms are easy to separate since the first one is linear, while the second one is periodic with the period of one year corresponding the the Earth's orbit.
This decomposition is the cornerstone of all practical parallax measurements, including those performed from the space observatory Gaia, \cite{klioner2003}. It is currently feasible only for objects
at galactic distances, with the record distance of around $20\textrm{kpc}$ obtained to a water maser source by the Very Long Baseline Array (VLBA) observatory \cite{2017Sci...358..227S}.
 
Fortunately, it turns out that it is possible to determine the parallax matrix and the parallax distance in a curved spacetime, with all relativistic corrections, in a very similar way assuming that the gravitational field does not vary very much on short scales. More precisely, as was also shown  in \cite{Grasso:2018mei}, for a source which is in free fall and for the observer in a gravitationally bound system, 
undergoing a periodic motion around a free falling barycenter, the variation of the apparent position is given by the peculiar motion, i.e. drift across the sky
with constant angular velocity $v^{\bm A}$, and a periodic signal  proportional to the observer's transverse displacement with respect to the barycenter. The result is that the apparent position variation for short times is given by a relation with the same structure of the one in flat spacetime, Eq.~\eqref{eq:parallaxtime0}, namely\footnote{Here we neglect again the aberration effects and also the light bending effects from the Solar System bodies, which are under control and subtracted from the parallax measurements.}
\bea
 \delta \theta^{\bm A}(t_b) = v^{\bm A}\cdot t_{b} - \Pi\UD{\bm A}{\bm B}\,\delta x_{\calO}^{\bm B}(t_b), \label{eq:parallaxtime}
\eea
where the angular velocity of the proper motion $v^{\bm A}$ and the parallax matrix $\Pi\UD{\bm A}{\bm B}$ are again constant, $t_{b}$ is a null time coordinate related to the barycenter time  and $\delta x_{\calO}$ is the momentary displacement of the observer with respect to the barycenter. 
Just like in the flat case the first term grows linear in time, while the second one has the annual periodicity of the Earth's orbital motion. Moreover,
we see that the periodic component of the signal is given by the product of the constant parallax matrix $\Pi\UD{\bm A}{\bm B}$ and the transverse components of the observer's position. 
 Both terms should therefore be easily separable in the observational data if the measurement is made over many orbital periods and the components of $\Pi\UD{\bm A}{\bm B}$ should be possible to determine  after removing the linear drift from the data.

The  result above  holds for \emph{any} curved spacetime as long as the curvature scale is much larger than the size of the object we observe and of the Solar System. 
Therefore, under the assumptions above, the standard method of parallax determination by  decomposing the apparent motion of the source into the constant proper motion and periodic parallax should work well even if we take into account  all relativistic corrections (gravitational light bending, Shapiro delays) to the light propagation due to the curved spacetime\footnote{Note, however, that the corrections due to the non-flat geometry \emph{within the Solar System}, i.e. light bending and Shapiro delays due to the Sun and large planets, need to be taken into account separately \cite{klioner2003}.}. The only small modification we need to introduce in non-flat geometry is that we cannot a priori assume that the 
parallax angle's direction is exactly opposite to  the transverse displacement of the observer, as in (\ref{eq:parallaxtime0}). This
proportionality of vectors holds if and only if the parallax matrix itself is proportional to the unit matrix, i.e. $\Pi\UD{\bm A}{\bm B} = \Dpar^{-1}\,\delta\UD{\bm A}{\bm B}$. This may happen for example if the geometry is rotationally symmetric with respect to the optical axis. However, if the light between the source and the observer undergoes shear due to tidal forces the parallax matrix can in principle be any symmetric matrix. 
The data analysis should therefore assume a more general form of the periodic term, i.e. a linear relation between the momentary position of the Earth and the apparent position on the barycentric celestial sphere, given by a symmetric matrix, as assumed in (\ref{eq:parallaxtime}).

The need to consider $\Pi\UD{\bm A}{\bm B}$ as a linear mapping in two dimensions rather than a rescaling should pose no problem for sources located sufficiently far from the ecliptic. For these sources the projection of the 
Earth's orbit on the transverse plane is an ellipse with semiaxes of comparable size, $\delta x_\calO^{\bm A}$ probes both transverse dimensions and therefore we can obtain all components of $\Pi\UD{\bm A}{\bm B}$ from the measurement. However, for sources close to the ecliptic   the projected Earth's orbit degenerates to a line or an extremely elongated ellipse. In that case only one baseline direction
is probed by the Earth's motion and we may obtain only 2 out of 3 independent components of $\Pi\UD{\bm A}{\bm B}$. 
This is sufficient  if we for some reason may also assume that the shear effects are negligible.

Another issue we would like bring up  is connected with the problem of the fixed reference frame. Recall that the parallax is currently measured using the position variation with respect to the non-rotating frame given by  a set of distant ``fixed quasars'' \cite{rasanen, 2018A&A...616A..14G}. On the other hand,
strictly speaking, the definition of parallax in (\ref{eq:parallaxmatrixdef}) calls
for the comparison of the apparent positions using the parallel transport between the observation events. Physically this means that we should use  the local inertial frame, determined by the inertial effects within the Solar System, to define the notion non-rotating directions with respect to the barycenter. The results of these two
measurements are in general different, the difference being due to a possible slow, secular rotation of one frame with respect to the other, caused for example by the peculiar motions of the quasars.  
For precise measurements this difference, as well as the variability and
individual motions of the ``fixed quasars'', need to be taken into account, \cite{rasanen}.

Finally we note here also one important subtlety regarding the simultaneous measurements of $\Dlum$ and $\Dpar$: 
recall that the standard methods of measurement for the luminosity  distance, either using the period-luminosity relation in variable stars (RR Lyrae, Cepheids) or the Type Ia supernov\ae, require calibration 
on short distances. This is achieved with the help other methods available in the distance ladder for sufficiently close objects. The methods of calibration 
for variable stars make use of various astrometric techniques of distance determination  \cite{Humphreys_2013, 2013Nature}, including the trigonometric parallax distance measurements for stars contained within the Milky Way  \cite{2007AJ....133.1810B, 10.1111/j.1365-2966.2007.11972.x, Casertano_2016, refId0, 2018ApJ...855..136R}. 
Therefore, in order to avoid a vicious circle in the distance ladder calibration and the data analysis
we need  to separate clearly the \emph{local} measurements of parallax and luminosity distance, for which we neglect the distance slip and which 
we then may use for calibration purposes only, and the measurements made at larger distances, which we use for the determination of $\mu$ using the calibration obtained from the short-distance data.

\section{Cosmological applications}
\label{sec:cosmo}

In this section we will show that the properties of $\mu$ make it a particularly interesting observable in the cosmological context. Before that, however, we must note that 
its measurement is much more difficult that on shorter distances.   The measurement of the distance slip, as we mentioned, requires the determination of both the angular diameter distance, or equivalently the luminosity distance and the redshift, as well as the parallax distance. As for the former two, we need to note that different types of standard rulers or candles are available on extragalactic distances than on the galactic scales. Obviously the need for a simultaneous measurement of the parallax together with the luminosity or angular diameter distance strongly restricts the type of sources that may be used for measurements on cosmological scales.  We will now briefly discuss the problems of determination of each of the quantities involved and go through the possible sources, as they appear in the recent literature. 

Let us discuss first the measurement of the parallax.
On extragalactic or cosmological scales the 1 AU baseline provided by the Earth's motion may be too small for an effective measurement of the parallax. It was therefore suggested to use
the motion of the whole Solar System with respect to the CMB frame for the measurement, \cite{kardashev}, which provides the baseline of around 78 AU yearly, with the signal growing secularly over the years. We refer the reader to Ref~\cite{rasanen} and Refs. therein for the first studies on the cosmic parallax and to Ref.~\cite{Marcori:2018cwn} for a more detailed discussion of the methods and feasibility of the measurements. Here we just note that separating out of the parallax effects due to the observer's motion from the drifts due to the peculiar motions of the sources is more difficult in this case.  In \cite{PhysRevLett.121.021101,  Marcori:2018cwn} the authors propose to  average the component of the drift aligned with the direction of the CMB dipole over many sources. The uncorrelated peculiar motions of the sources should then average out, leaving this way the  signal due to the motion of the local group with respect to the CMB frame. This signal has been estimated in \cite{Marcori:2018cwn}  to be around
$0.3 \mu\textrm{as}/\textrm{yr}$ for objects at $z = 0.1$ and $0.06 \mu \textrm{as}/\textrm{yr}$ for $z = 1.48$, for short distances dominating over the aberration drift.\footnote{In \cite{Marcori:2018cwn} the authors use a different terminology, separating the parallax effect as we define it in this paper into  the aberration drift and ``pure" parallax drift. This splitting is done using the standard coordinates of the background FLRW metric.} In \cite{rasanen} a similar order-of-magnitude estimate of  $10^{-2} \mu \textrm{as}/\textrm{yr}$ has been obtained for sources on cosmological distances, although without the contribution of the local environment  or the peculiar motions of the sources. These values are much smaller than the precision of standard astrometric measurements of an \emph{individual} source, but given  sufficiently many sources the cosmic parallax can be measurable for the first time by the Gaia satellite\footnote{http://www.cosmos.esa.int/web/gaia.} launched in December 2013. In 5 years is expected the parallax measurement of at least $N \sim 5 \times 10^5$ quasars in the redshift range $z\in [0,5]$ with an average precision for a single measurement of $100\,\mu {\rm as}$ which will be reduced of a factor of $1/ \sqrt{2 N}$ for the full duration of the mission. Therefore it is expected that the cosmic parallax signal is within the range of sensitivity of Gaia, see \cite{Ding:2009xs} and \cite{Quartin:2009xr} for a more detailed analysis of the uncertainties, and \cite{Quercellini:2010zr} and Refs. therein for further details. 

The luminosity distance determination on the other hand requires sources whose absolute luminosity can be determined from optical observations alone. The  most important standard candles on cosmological distances are the Type Ia supernov\ae\,\,, see e.g. \cite{1996ApJ...473...88R, Riess_1998, 1999ApJ...517..565P}. Note that supernov\ae\,\,are luminous but also transient sources, lasting less than a year, while the parallax determination requires position measurements extending over many years or even decades. Supernov\ae \,\,Ia events may therefore only be suitable if the host galaxy is identified as well. The same problem arises if we try to use the gravitational wave signal  provided by binary black hole or neutron star mergers as standard sirens 
 \cite{1986Natur.323..310S, 2005ApJ...629...15H}: the transient nature of the signal and problems with precise pointing of the source precludes the secular position variation measurement.

The most promising sources to measure the distant slip  are therefore quasars: their positions can be determined with fairly high precision and they are suitable for long-term position variation measurements. We will now briefly review the recent developments in the field of the angular diameter distance and the luminosity distance measurements to quasars. In Ref.~\cite{Risaliti:2018reu} the authors obtained an Hubble diagram by measuring the luminosity distance from a sample of $\sim 1600$ quasars using a relation between UV and X ray emission that makes quasars standard candles. The advantage with respect to the same measurement from the luminosity distance of Type Ia supernov\ae\,\,is that it is possible to probe a larger redshift range: in Ref.~\cite{Risaliti:2018reu} the redshift range is $0.05 < z < 5.5$ whereas the farthest supernov\ae \,\,are observed at $z\lesssim 2$. 
Another method to make quasars standard candle is related to the so-called reverberation-mapping technique. It consists in the measure of time-delay response between the continuum and the Broad Emission Line Region (BELR) of a quasar: the time delay is directly related with the physical size of the BELR which in turn is related to the continuum luminosity of the source, via the well-known Radius-Luminosity relation from which the luminosity distance follows by its very definition. The values of the $\Lambda$CDM parameters determined this way are in agreement with other cosmological probes at $2\sigma$ level. In the near future the constraints will improve significantly: the redshift range of quasars detectable by the Large Synoptic Survey Telescope\footnote{https://www.lsst.org} is $0<z<7$ and the quasar counts will raise enormously, with an estimation of $\sim 3000$ reverberation-mapped AGNs, thus providing a much better statistics for this type of signal for cosmological purposes.
Finally the authors of Ref.~\cite{Elvis:2002ja} suggested to use the reverberation-mapping technique to make quasars standard rulers: according to their proposal, having estimated physical size of the BERL by accurately measuring the time delay, in principle it is possible to resolve angular size of the BELR region of the quasar by using interferometric methods. The GRAVITY collaboration has recently succeeded in applying  this method to a quasar \cite{2018Natur.563..657G}. For a recent review on the reverberation-mapping technique applied to quasar for cosmological purposes we remind the reader to Ref.~\cite{Panda:2019cvs} and Refs. therein. 

In the rest of this section we simply assume that the distance slip, i.e. $\Dpar$ together with $\Dang$ (or with the redshift $z$ and $\Dlum$), is measured for a sufficiently large sample of sources on cosmological scales and we discuss what kind of information can be obtained from the results. We specialize the calculation of the distance slip $\mu$ to the FLRW spacetime, i.e. to a homogeneous and isotropic matter distribution. We 
consider comoving observer and emitter, although we note that the distance slip is in the end independent from the motions of both. 

\subsection{Distance slip in an unperturbed FLRW Universe. }
 We start by considering the FLRW line element written in the form 
\bea
\dd s^2 = -\dd t^2 + a(t)^2\left(\dd\chi^2 + S_k(\chi)^2\,\dd\Omega^2\right)
\eea
if cosmic time $t$ is used as time variable and
\bea
\dd s^2 = a(\eta)^2\left[-\dd \eta^2 + \left(\dd\chi^2 + S_k(\chi)^2\,\dd\Omega^2\right)\right]
\eea
if we use conformal time $\eta$. 
The two time variables are linked by $\dd t = a \,\dd \eta$. In the above expressions for the metric $\dd\Omega^2$ is the infinitesimal solid angle and the specific form of the function $S_k(\chi)$ depends on the curvature of the spatial hypersurface. We have
\bea
S_k(\chi) &=& \left\{\begin{array}{ll} \label{eq:Sk}
                    \frac{1}{\sqrt{k}}\,\sin(\sqrt{k}\,\chi) & \textrm{if }k > 0 \\
                    \chi & \textrm{if }k = 0 \\
                    \frac{1}{\sqrt{|k|}}\,\sinh(\sqrt{|k|}\,\chi) & \textrm{if }k < 0 \,,
                   \end{array}\right.
                   \eea
where $\chi$ plays the role of the radial coordinate. 
We also define the derivative of  $S_k(\chi)$ which will be useful in the following
                   \bea
C_k(\chi) &\equiv& \frac{\dd S_k}{\dd\chi} = \left\{\begin{array}{ll} \label{eq:Ck}
                    \cos(\sqrt{k}\,\chi) & \textrm{if }k > 0 \\
                    1 & \textrm{if }k = 0 \\
                    \cosh(\sqrt{|k|}\,\chi) & \textrm{if }k < 0 \,.
                   \end{array}\right.
\eea
Consider now an observer placed  at the origin $\chi = 0$ at the present moment, corresponding to 
$z = 0$. By convention we assume that the scale factor $a_\calO$ at present is set to 1.
The $\chi$ coordinate of  a light source observed with redshift $z$ defines the comoving distance to the source and is given by the integral
\begin{equation}\label{Dcom}
\chi(z) = \int _0^z \frac{{\rm d} \hat{z} }{(1+\hat{z}) {\cal H}(\hat{z})}=  \int _0^z \frac{{\rm d} \hat{z}}{H(\hat{z})} .
\end{equation}
We  normalize the FLRW photon geodesics such that the time component of the tangent vector is equal to unity at the observer position and the affine parameter increases toward the source, i.e. $\ell^0\obs = -1$. The two Hubble parameters in~\eqref{Dcom}, $H \equiv (\dd a / \dd t)/ a$ and $\mathcal{H} \equiv (\dd a / \dd \eta)/ a$, are related by $\mathcal{H}(1+z)= H$.
For any spatial curvature we consider a universe containing ordinary and dark matter and a cosmological constant $\Lambda$. The Hubble parameter in terms of the redshift then reads
\bea
 H(z)^2 = H_0^2\,\left( \Omegamzero(1+z)^3 + \Omegakzero(1+z)^2 + \OmegaLambdazero\right)\, 
\eea
where $H_0$ denotes the today value and $\Omegamzero, \Omegakzero$ and $\OmegaLambdazero$ are respectively the matter, curvature and cosmological constant parameters at present. It is also useful to consider the dimension-less comoving distance defined as
\bea
E(z) = H_0\chi(z)&=& \int _0^z {\rm d} \hat{z}\left(\Omegamzero(1+\hat z)^3 + \Omegakzero(1+\hat z)^2 + \right.\nonumber\\
 &&\left. + \OmegaLambdazero \right)^{-1/2},\label{eq:Ez}
\eea
which is independent of $H_0$.

In the following we report explicitly the results for the distance slip $\mu$ and the angular diameter and parallax distance in an FLRW background with arbitrary curvature, namely flat, open or closed. For a detailed derivation with the help of the machinery we have introduced in Section~\ref{sec:formulation} we refer the reader to Appendix~\ref{app:0orderW}.

From its definition in Eq.~\eqref{eq:defofmu2} and the result in FLRW in Eq.~\eqref{eq:mformulageneral2}, expressed in terms of the redshift $z$ of the source, the distance slip $\mu$ is then
\begin{equation}\label{eq:muFLRW}
 \mu = 1 - \left[\frac{1}{1+z}\Big(C_k(\chi) +  H_0S_k(\chi)\Big)\right]^2. 
\end{equation}
In flat FLRW this reads
\begin{equation}\label{eq:muflat}
\mu(z) = 1 - \left[\frac{1 +  E(z)}{1+z} \right]^2
\end{equation}
with $\Omegakzero=0$ in~\eqref{eq:Ez}, whereas for the curved cases, by expressing the constant curvature of the spatial hypersurfaces $k$ in terms of the curvature parameter at present $\Omega_{k_0}$ as $k= -\Omega_{k_0} H^2_0$, we get 
\begin{widetext}
\begin{equation}\label{eq:muclosed}
 \mu(z) = 1 - \left\{\frac{1}{1+z}\Big[\cos \Big(  \sqrt{-\Omega_{k_0}}E(z)\Big) + \frac{1}{\sqrt{-\Omega_{k_0}}} \sin \Big(\sqrt{-\Omega_{k_0}} E(z) \Big) \Big]\right\}^2
\end{equation}
for a closed universe with $\Omega_{k_0} < 0$ and we get
\begin{equation}\label{eq:muopen}
 \mu(z) = 1 - \left\{\frac{1}{1+z}\Big[\cosh \Big(  \sqrt{\Omega_{k_0}}E(z)\Big) + \frac{1}{\sqrt{\Omega_{k_0}}} \sinh \Big(\sqrt{\Omega_{k_0}} E(z) \Big) \Big]\right\}^2
\end{equation}
for an open universe with $\Omega_{k_0} > 0$, according to~\eqref{eq:Sk} and~\eqref{eq:Ck}. 
\end{widetext}

Finally, we report the expressions for the angular diameter distance
\begin{equation}\label{eq:DangFLRW}
\Dang = \frac{S_k(\chi)}{1+z}
\end{equation}
and the parallax distance 
\begin{equation} \label{eq:DparFLRW}
\Dpar  =  \frac{S_k(\chi)}{C_k(\chi) +  H_0\,S_k(\chi)}\,,
\end{equation}
both calculated using our approach (see Eqs.~\eqref{eq:definitionofDang} and~\eqref{eq:definitionofDpar} and Appendix~\ref{app:0orderW}).
We checked that our results coincide with previous ones in the literature, see e.g. \cite{DiDio:2016ykq} for the angular diameter distance and \cite{rasanen} for the parallax distance.

\begin{figure}[h]
    \includegraphics[width=8.4cm]{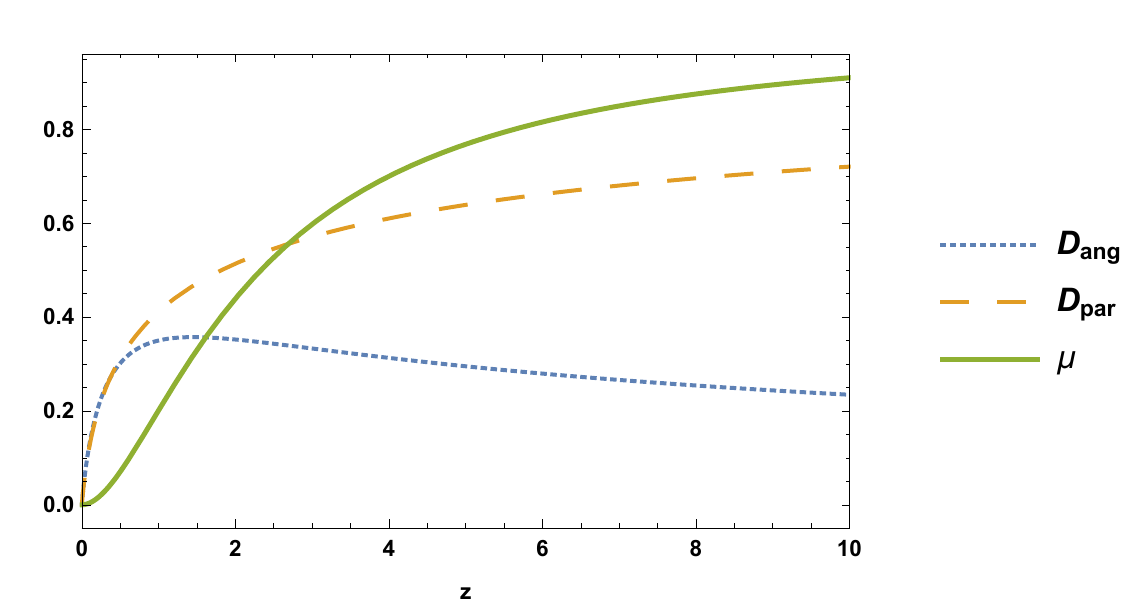}
     \caption{The angular diameter distance $\Dang$, the parallax distance $\Dpar$ and the distance slip $\mu$ for flat FLRW. Here $\Dang$ and $\Dpar$ are plotted dimension-less in terms of the Hubble radius $R_{\text{H}}= 1/H_0$. We set the value for the cosmological parameters from \cite{Aghanim:2018eyx}.}
   \label{fig:dist}
   \end{figure}

We plot in Fig.~\ref{fig:dist} the distance slip $\mu$, $\Dang$ and $\Dpar$ in the FLRW spacetime as function of the redshift, with the cosmological parameters fixed to the values as measured from Planck. Note that while the angular diameter distance decreases with redshift, the distance slip and the parallax distance increase, meaning that the parallax does not become arbitrarily small but approaches a constant value. However the difficulty to measure the parallax also increases with redshift, due to the decreasing of the apparent luminosity of the source.

\paragraph{Remark.} The reader may check that for the de Sitter space ($\Omegamzero = 0$, $\OmegaLambdazero > 0$) and the anti-de Sitter space
($\Omegamzero = 0$, $\OmegaLambdazero < 0$) we have $\mu \equiv 0$ everywhere. 
This can be seen easily from the expression for $\mu$ in \eqref{eq:mufrommperp} and from equation \eqref{eq:mperpODE} if we note that the components of the optical tidal tensor appearing in \eqref{eq:mperpODE}  vanish for spacetimes with only the cosmological constant present in the curvature tensor.
In particular $\Dang = D_{\rm par}$ in both spacetimes along any null geodesic, just like in the flat space.

\subsection{Dependence on the cosmological parameters}
\label{sec:depOs}
\begin{figure}[h!]
  \begin{subfigure}[h!]{0.36\textwidth} 
    \includegraphics[width=\textwidth]{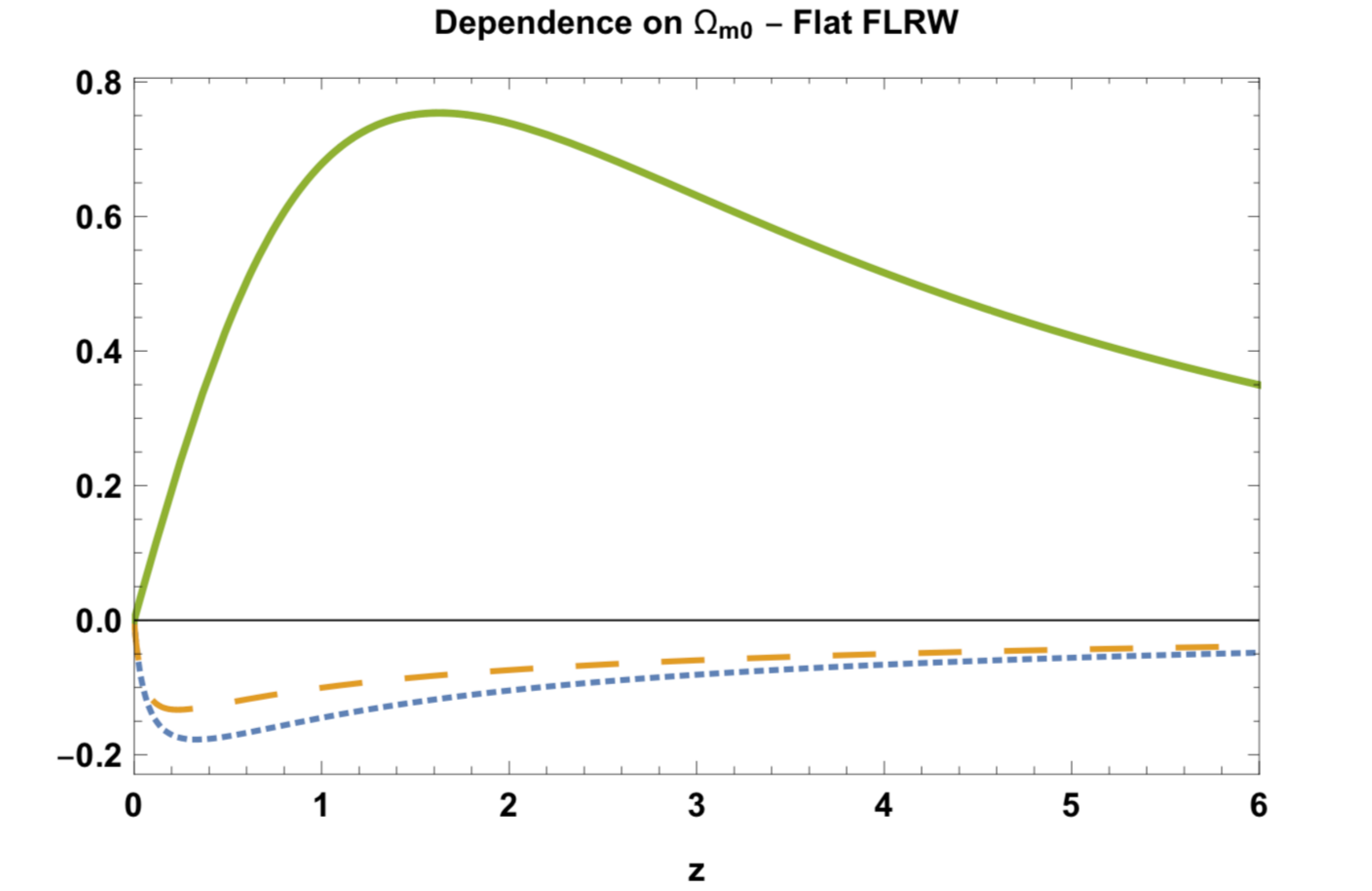}
  \end{subfigure}
  \hfill
  \begin{subfigure}[h!]{0.45\textwidth}
    \includegraphics[width=\textwidth]{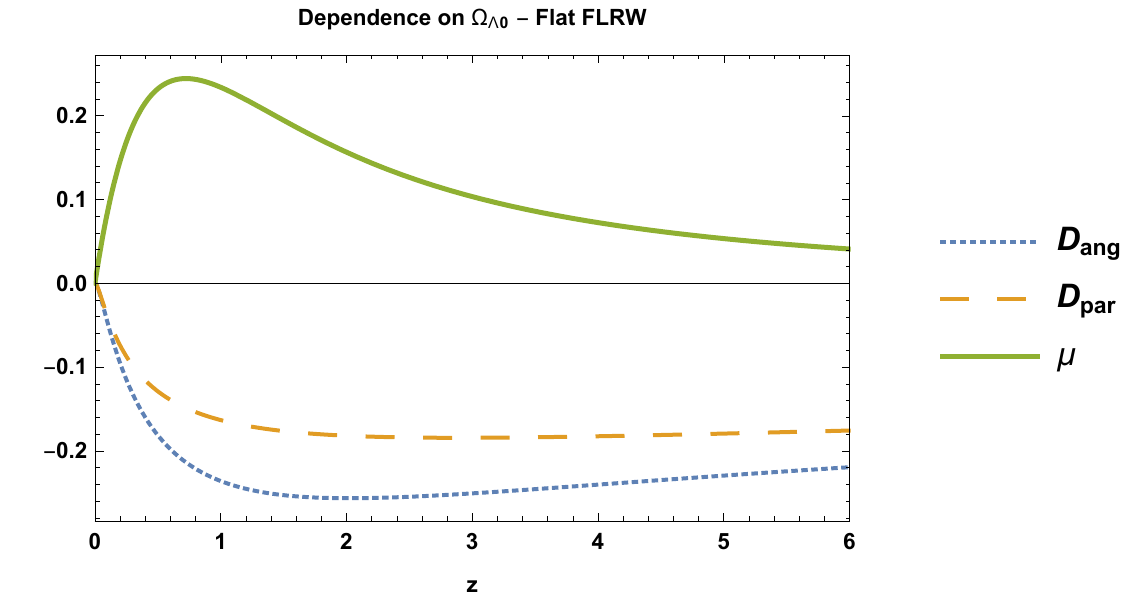}
    \end{subfigure}
    \caption{Dependence of the dimension-less angular diameter distance and parallax distance and the distance slip on the cosmological parameters for the flat FLRW model. The plots show the derivatives with respect to $\Omega_{{\rm m}_0}$ and $\Omega_{\Lambda_0}$. We set $\Omega_{{\rm m}_0}= 0.266018$  as fiducial value, \cite{Aghanim:2018eyx}, and we obtain $\Omega_{\Lambda_0}=0.733982$ from the closure condition.}
     \label{fig:depflat}
   \end{figure}

\begin{figure}[h!]
  \begin{subfigure}[h!]{0.36\textwidth}
    \includegraphics[width=\textwidth]{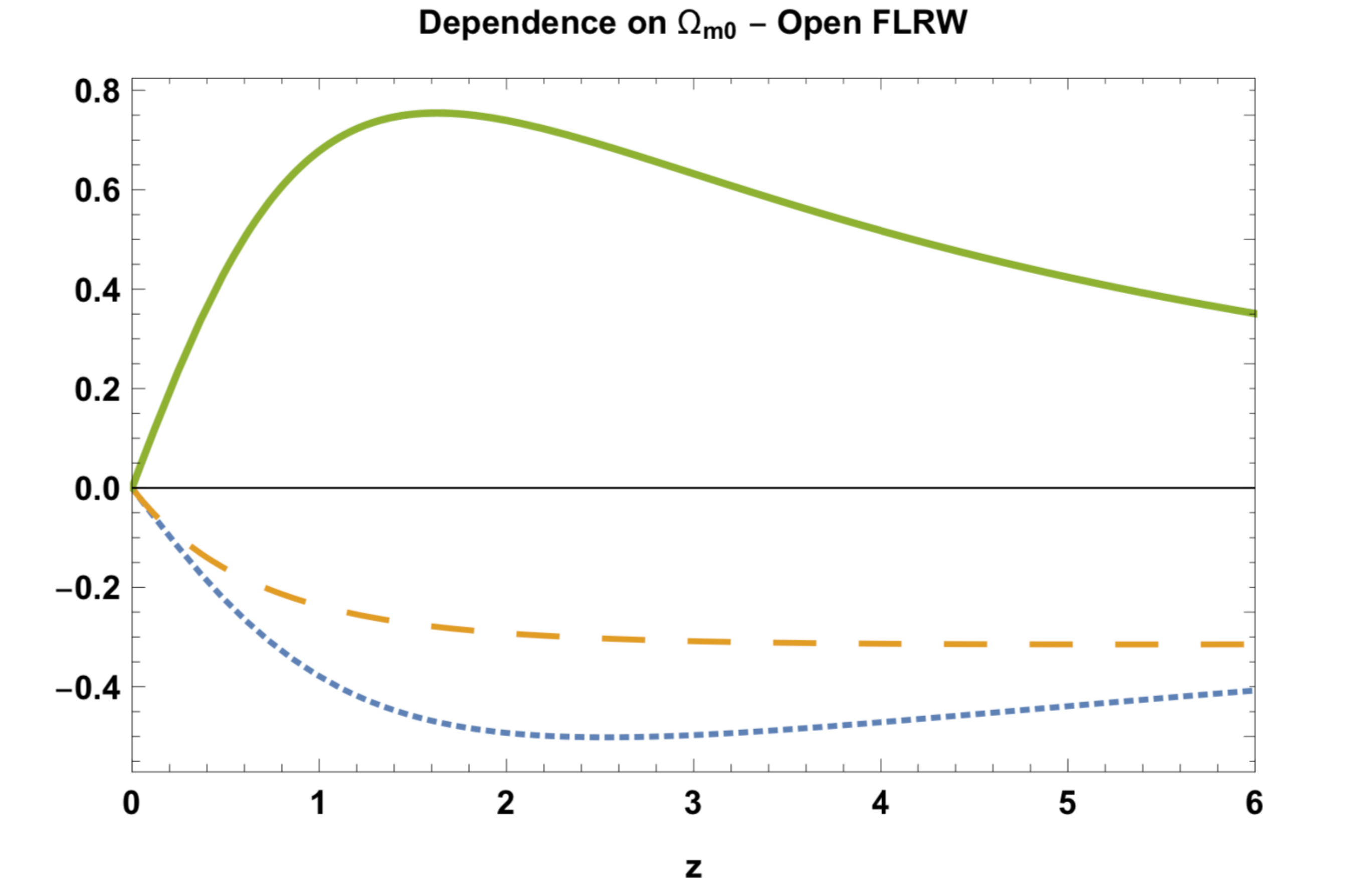}
  \end{subfigure}
  \hfill
  \begin{subfigure}[h!]{0.45\textwidth}
    \includegraphics[width=\textwidth]{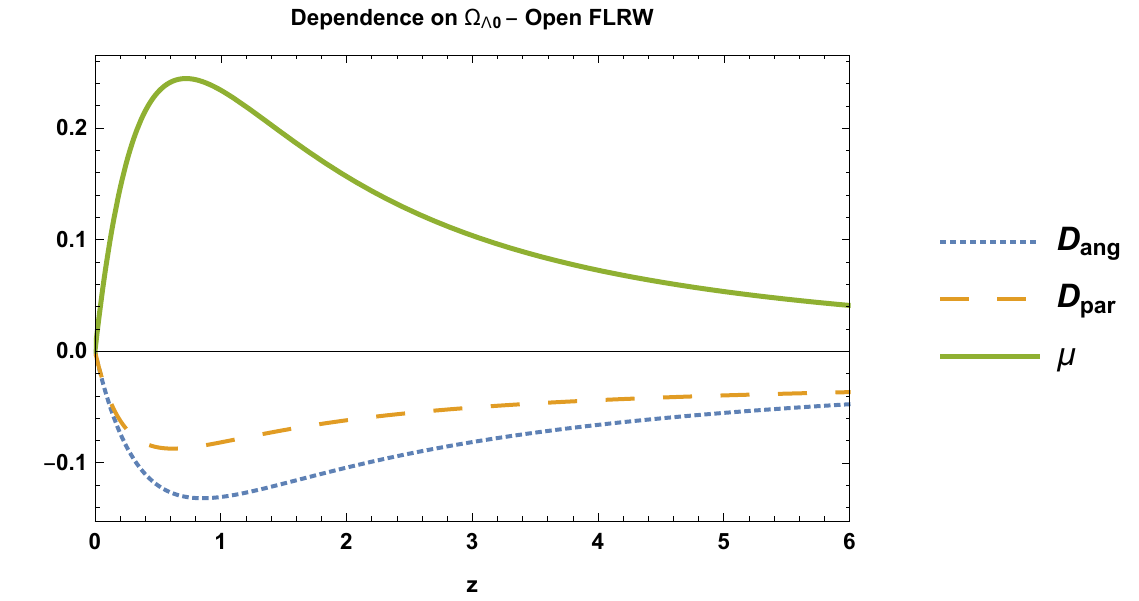}
    \end{subfigure}
    \caption{Dependence of the dimension-less angular diameter distance and parallax distance and the distance slip on the cosmological parameters for an open FLRW model. The plots show the derivatives with respect to $\Omega_{{\rm m}_0}$ and $\Omega_{\Lambda_0}$. We set $\Omega_{{\rm m}_0}= 0.266018$ and  $\Omegakzero=0.0010$ as fiducial values, \cite{Aghanim:2018eyx}, and we obtain $\Omega_{\Lambda_0}=0.732982$ from the closure condition.}
      \label{fig:depopen}
   \end{figure}

\begin{figure}[h!]
  \begin{subfigure}[h!]{0.36\textwidth}
    \includegraphics[width=\textwidth]{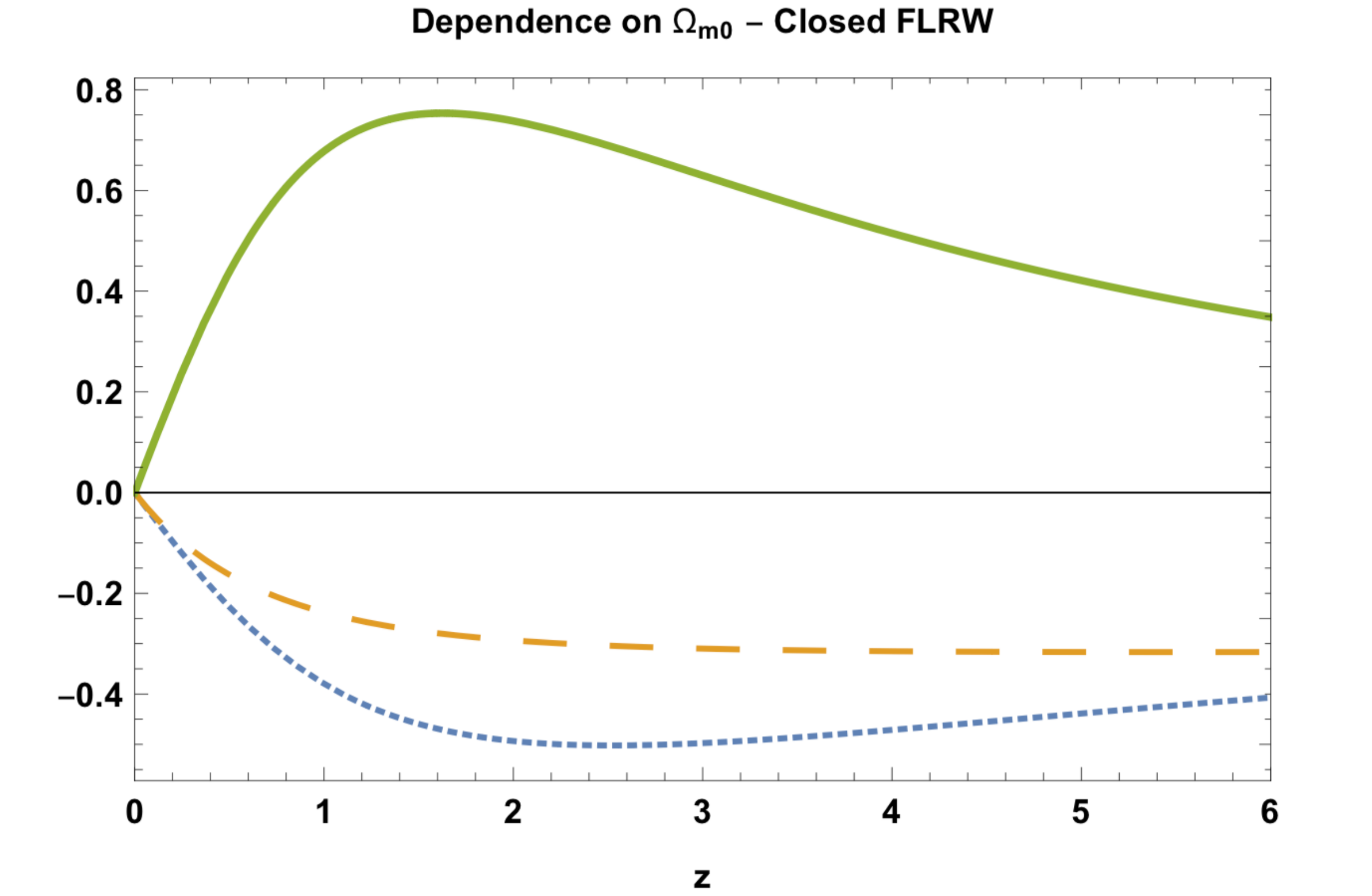}
  \end{subfigure}
  \hfill
  \begin{subfigure}[h!]{0.45\textwidth}
    \includegraphics[width=\textwidth]{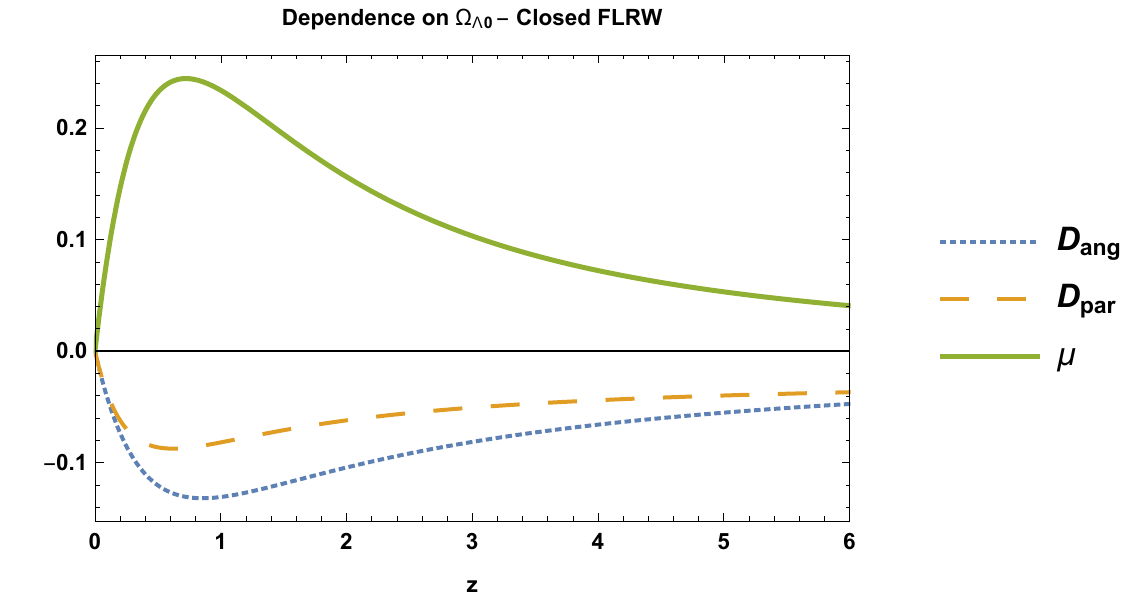}
    \end{subfigure}
    \caption{Dependence of the dimension-less angular diameter distance and parallax distance and the distance slip on the cosmological parameters for a closed FLRW model. The plots show the derivatives with respect to $\Omega_{{\rm m}_0}$ and $\Omega_{\Lambda_0}$.We set $\Omega_{{\rm m}_0}= 0.266018$ and  $\Omegakzero=-0.0010$  as fiducial values, \cite{Aghanim:2018eyx}, and we obtain $\Omega_{\Lambda_0}=0.734982$ from the closure condition.}
     \label{fig:depclosed}
   \end{figure}

\begin{figure}[h!]
  \begin{subfigure}[h!]{0.36\textwidth}
    \includegraphics[width=\textwidth]{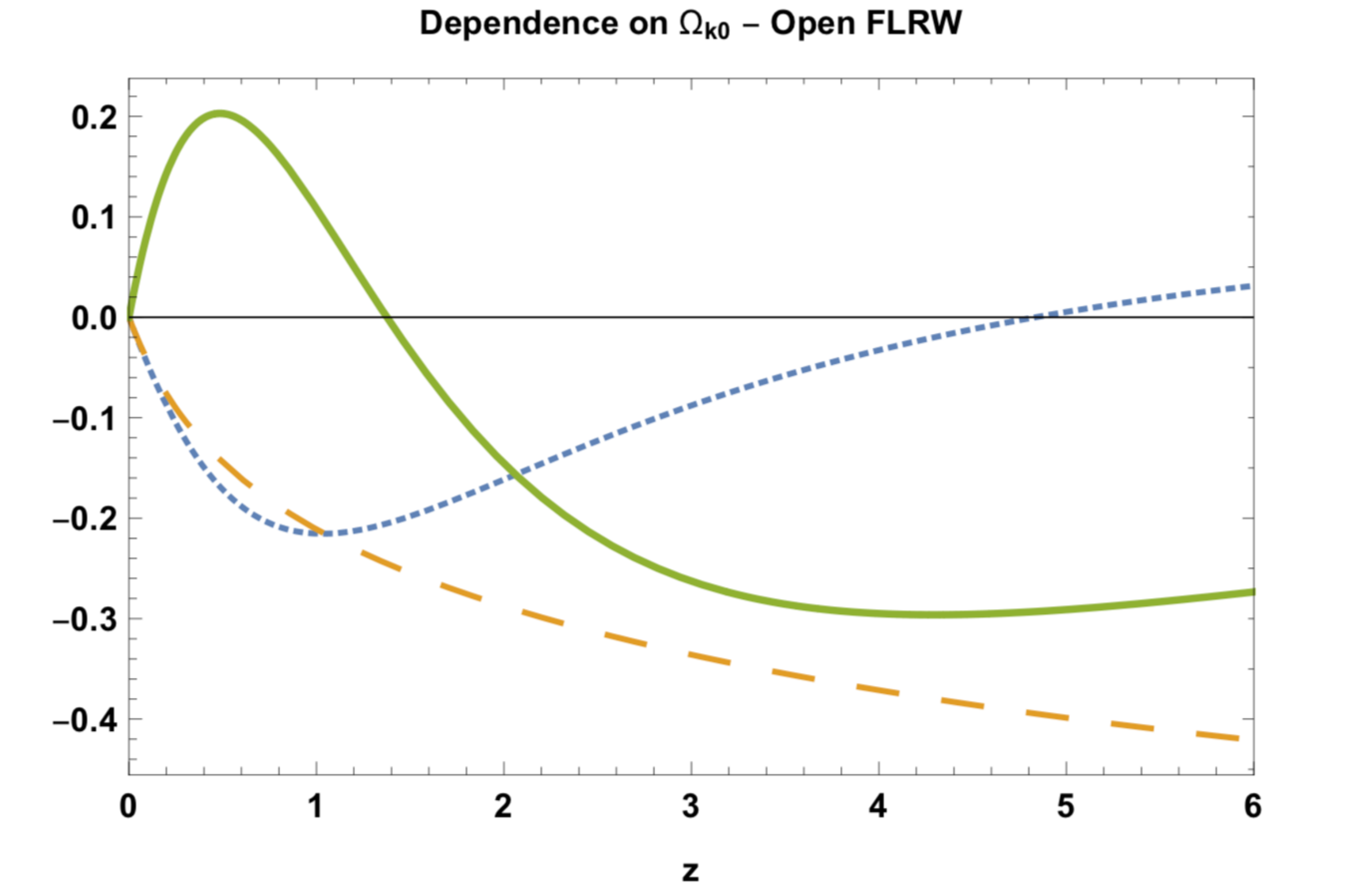}
  \end{subfigure}
  \hfill
  \begin{subfigure}[h!]{0.45\textwidth}
    \includegraphics[width=\textwidth]{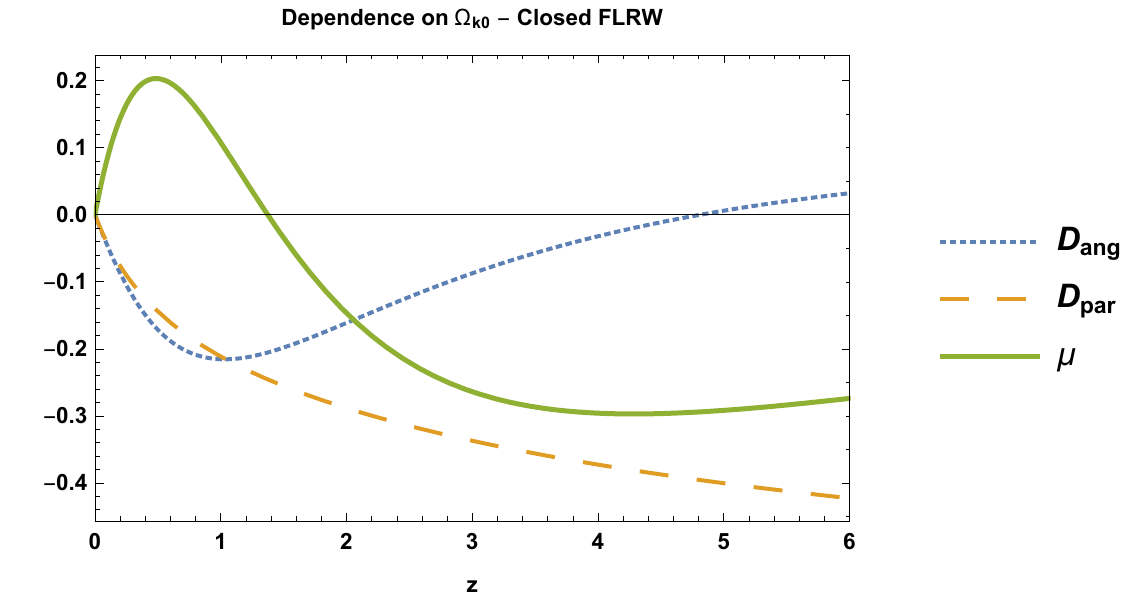}
    \end{subfigure}
    \caption{Dependence of the dimension-less angular diameter distance and parallax distance and the distance slip on the curvature parameter for a closed and open FLRW models. The plots show the derivatives with respect to $\Omega_{{\rm k}_0}$. All the other cosmological parameters are fixed to fiducial values from \cite{Aghanim:2018eyx}.}
     \label{fig:depOk}
   \end{figure}

In this section we explore the dependence of the distance slip $\mu$ on the cosmological parameters $H_0$, $\Omegamzero$, $\Omegakzero$, $\OmegaLambdazero$  in comparison with the angular diameter and the parallax distance.
First of all, by simply looking at their expressions in \eqref{eq:muflat}-\eqref{eq:muopen}, \eqref{eq:DangFLRW} and \eqref{eq:DparFLRW}, we notice that $\Dang$ and $\Dpar$ individually depend on $H_0$\footnote{The dimension-less expressions of $\Dang$ and $\Dpar$ in terms of the Hubble radius, which are simply obtained multiplying \eqref{eq:DangFLRW} and \eqref{eq:DparFLRW} by $R_H^{-1}=H_0/c$, of course do not depend on $H_0$. But in this case it is implicitly assumed that $H_0$ is known from other measurements.} whereas their ratio, i.e. $\mu$, does not. A measurement of the distance slip would therefore have the advantage to determine the cosmological parameters fully independently from $H_0$. This holds of course for any curvature of the FLRW spacetime.  An accurate estimation of the constraints on the cosmological parameters that can be obtained from a measurement of the distance slip from e.g. the simultaneous measurements of $\Dang$ and $\Dpar$ of quasars is beyond the scope of this paper and would require precise estimations of the uncertainties in the measured values of the two distances from this kind of sources at different redshifts. Here we just investigate the dependence on the cosmological parameters of the distance slip compared with that of the parallax and angular diameter distance alone, in order to understand if it contains additional and potentially useful information as a new probe in cosmology. In Fig.~\ref{fig:depflat} - Fig.~\ref{fig:depOk} we plot the derivative of the three observables (the dimension-less expressions of $\Dang$ and $\Dpar$, and $\mu$) with respect to one parameter at a time, the other being fixed to their fiducial values from \cite{Aghanim:2018eyx}, in function of the redshift. Fig.~\ref{fig:depflat}, Fig.~\ref{fig:depopen}, and Fig.~\ref{fig:depclosed} show the dependence on $\Omegamzero$ and $\OmegaLambdazero$ for the flat, open and closed FLRW universe, respectively. Fig.~\ref{fig:depOk} is dedicated to the dependence on the curvature parameter $\Omegakzero$ for the open (left panel) and closed case (right panel). These derivatives represent the dependence coming from theory only, i.e. from the functional form of the observable at hand. They are those appearing in the Fisher matrix which, together with the specifications about each measurement, is used to forecast e.g. the constraints on the model parameters achievable with a specific instrument. 
We note that in all cases the dependence of the distance slip is quite different from that of the parallax and the angular diameter distance. In particular, from Fig.~\ref{fig:depflat}, Fig.~\ref{fig:depopen}, and Fig.~\ref{fig:depclosed} we note that the the dependence of $\mu$ on $\Omegamzero$ and $\OmegaLambdazero$ is very different from that of $\Dang$ and $\Dpar$, which are very similar to each other.
This may suggest that potentially new information is contained in $\mu$ and that the parameter degeneracies may be different from that in the measurements of $\Dang$ and $\Dpar$ separately.
We finally note from Fig.~\ref{fig:depOk} that each of the three observables depends on the curvature in a peculiar way.

\subsection{Low-redshift expansions}\label{sec:lowz}
By expanding Eq.~\eqref{eq:muFLRW} for small redshift we find for the $\Lambda$CDM model up to third order
\begin{equation}\label{eq:mulowz}
\mu(z) = \frac{3}{2}\,\Omega_{{\rm m}_0}z^2 + \left(-  \frac{1}{2}\,\Omega_{{\rm m}_0} - \frac{3}{2}\Omegamzero\,\Omegakzero - \frac{9}{4}\Omegamzero ^2  \right) z^3 \,,
\end{equation}
where we used the closure condition $\OmegaLambdazero = 1 -\Omegamzero-\Omegakzero$ to get rid of $\OmegaLambdazero$ and the above expansion is valid for the open, the closed and the flat FLRW model, with $\Omegakzero = 0$ for the latter\footnote{If we use the closure condition to get rid of $\Omegakzero $ we obtain instead $$
\mu(z) = \frac{3}{2}\,\Omega_{{\rm m}_0}z^2 + \left(-  2\,\Omega_{{\rm m}_0} + \frac{3}{2}\Omegamzero\,\OmegaLambdazero - \frac{3}{4}\Omegamzero ^2  \right) z^3\,.$$}.
We see that only $\Omegamzero$ appears in the leading, quadratic, term. This is in perfect agreement with the general result of Eqs.~\eqref{eq:mufromRicciT} and~\eqref{eq:mufromT}, showing that for short distances the distance slip depends only on the dark and baryonic matter content: here this is just specialized to any FLRW spacetime with matter and a cosmological constant.
The dependence of $\mu$ on the curvature $\Omegakzero$ appears only at the third order in the redshift.  Let us also recall that there is no dependence on $H_0$ at all orders, as we have noticed in section~\ref{sec:depOs}. A straightforward consequence is that measurements of $\mu$ for very small redshifts offer a simple way to determine the value of $\Omegamzero$ locally, bypassing the uncertainties of the determination of $H_0$ or $\Omegakzero$.

\begin{figure}[h!]
    \includegraphics[width=8cm]{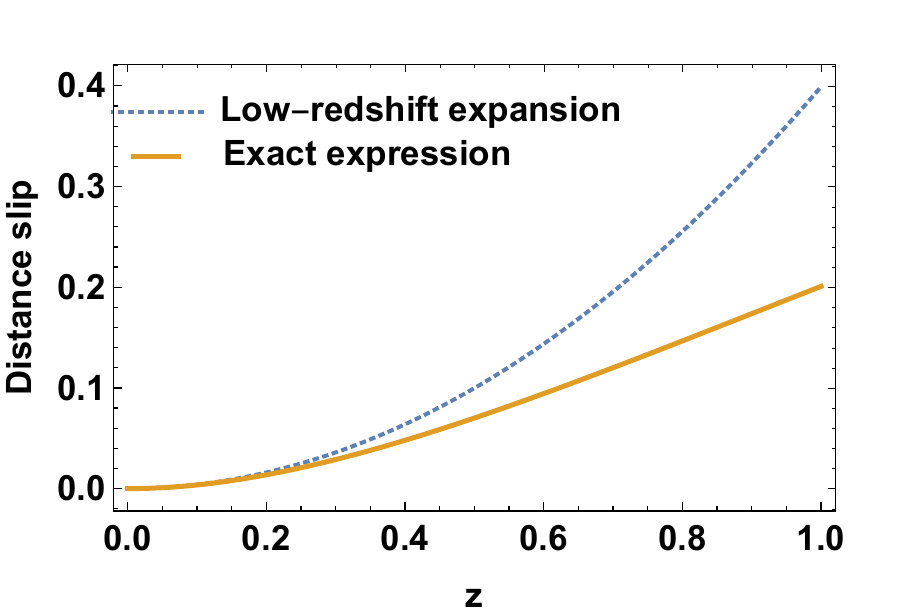}
     \caption{Comparison between the low-redshift expansion of the distance slip as in Eq.~\eqref{eq:mulowz} and its exact expression for flat FLRW, Eq.~\eqref{eq:muflat}. The low-redsfhit expansion truncated at the first term $\propto z^2$ is very accurate for redshift $z\lesssim 0.2$. We set the value for the cosmological parameters from \cite{Aghanim:2018eyx}.}
   \label{fig:lowz}
   \end{figure}

In Fig.~\ref{fig:lowz} we plot the low-redshift expansion of the distance slip in Eq.~\eqref{eq:mulowz} versus its exact expression for flat FLRW, Eq.~\eqref{eq:muflat}. They start to differ at $z\gtrsim0.05$. At $z=0.1$ the difference is $\sim 10\%$ and it increases monotonically with the redshift.
Let us remark that, although the distance slip is very small at low redshift (being quadratic in $z$) and thus its measurement would be difficult, it would be also lead to a measure of $\Omega_{{\rm m}_0}$  independent of any other cosmological parameters, i.e. $H_0$, $\OmegaLambdazero$, and $\Omega_{{\rm k}_0}$.

  \paragraph{Distance slip - angular diameter distance relation.} Beside the dependence of the distances $\Dang$ and $\Dpar$ and the distance slip $\mu$ on the redshift we can also  consider directly the relations between these quantities, bypassing this way the redshift as observable.
   As an example we discuss here the relation between $\mu$ and $\Dang$ for short distances. Note that since \emph{both} quantities in question do not depend on the 
   motions of the sources all results of measurements derived from this relation are free from any systematics or noise due to the peculiar motions of the sources, unlike the redshift-based measurements\footnote{The residual dependence of the value of the angular 
   diameter distance on the motion of the observer can be fixed for example by boosting the measurement results to the CMB frame defined by the CMB dipole.}. This may be important for short-distance measurements where
   peculiar motions may constitute a significant part of the error budget.
   
Up to third order the relation between $\mu$ and $\Dang$ reads
 \begin{eqnarray}
\mu(\Dang) &=&  \frac{3}{2}\Omegamzero\,H_0^2\,\Dang^2+\frac{5}{2}\Omega_{{\rm m}_0} H_0^3\,\Dang^3\,. \label{eq:muofDang}
\end{eqnarray}
It follows that fitting the results of the measurements of $\mu$ and $\Dang$ for a sample of relatively close sources (meaning $\Dang$ much smaller than the Hubble distance) to (\ref{eq:muofDang}) yields
 the local value of the combination $\Omegamzero\,H_0^2$ as the coefficient in the quadratic term and, if the data allow, also the value of $\Omegamzero\,H_0^3$ as the next order coefficient. 
Let us note that the leading order term $\Omegamzero\,H_0^2 \propto \rho_{\textrm{m}_0}$  is another evidence of the tomographic property of $\mu$ for short distances, mentioned in section \ref{sec:muprop}.

\subsection{Dynamical dark energy} \label{sec:dynDE}
\begin{figure}[h!]
  \begin{subfigure}[h!]{0.36\textwidth} 
    \includegraphics[width=\textwidth]{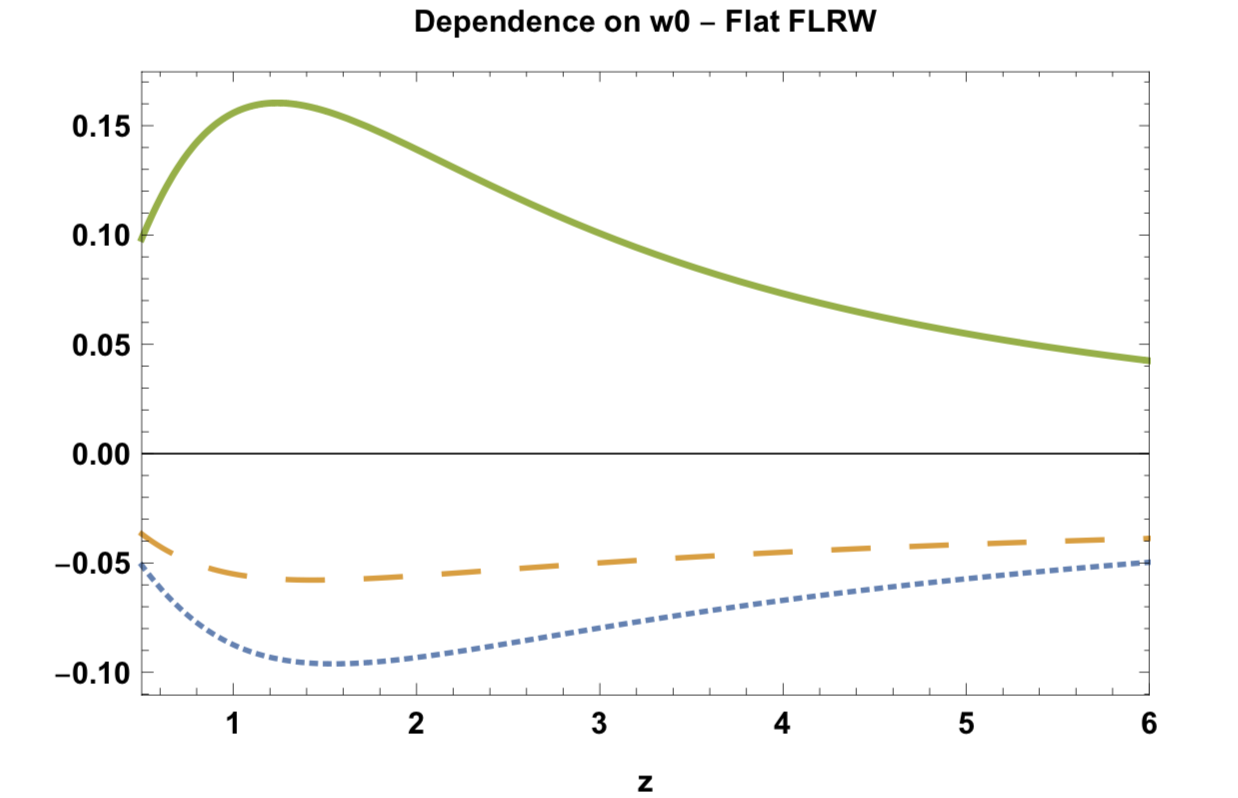}
  \end{subfigure}
  \hfill
  \begin{subfigure}[h!]{0.45\textwidth}
    \includegraphics[width=\textwidth]{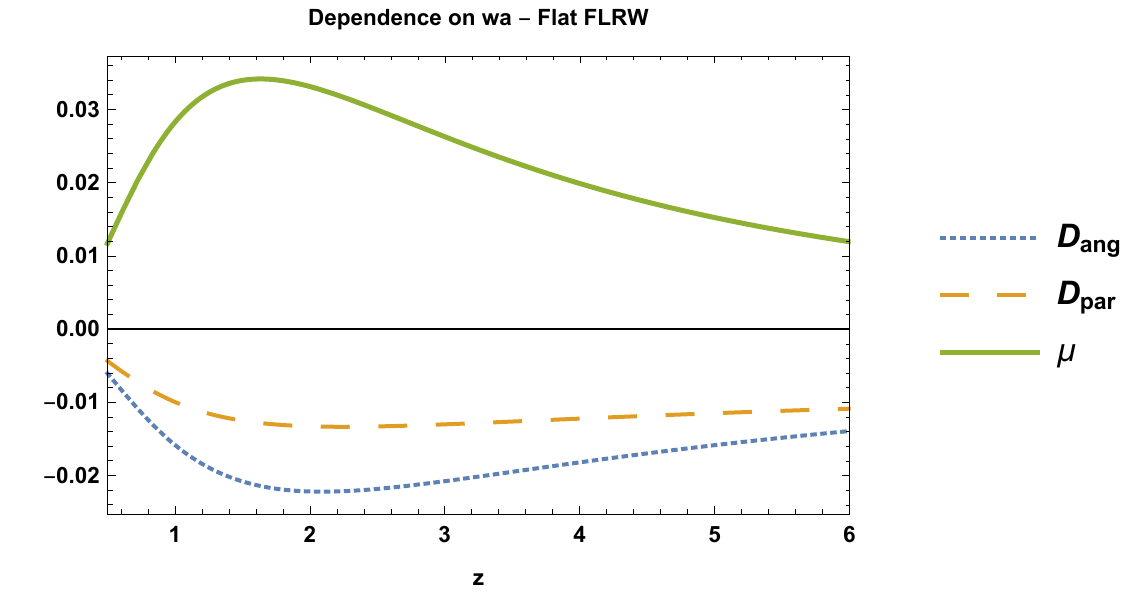}
    \end{subfigure}
    \caption{Derivatives of the dimension-less angular diameter distance and parallax distance and the distance slip with respect to the parameters of the dynamical dark energy model in \eqref{eq:w0wa} in the flat FLRW case. The plots show the derivatives with respect to $w_0$ and $w_a$. We set the fiducial values of the other cosmological parameters from \cite{Aghanim:2018eyx}.}
     \label{fig:depwflat}
   \end{figure}

We consider here a simple modification of the $\Lambda$CDM in which the equation of state $w=p/\rho$ for dark energy is not constant in time as it is for the cosmological constant $\Lambda$.
We follow the usual parametrization for the equation of state varying with time which was introduced in \cite{Chevallier:2000qy} and \cite{Linder:2002et}
\begin{equation}\label{eq:w0wa}
w(z)= w_0 + \frac{z}{1+z} w_a\,,
\end{equation}
where $w_0$ is the value of $w$ today and $w_a$ governs the time dependence. For the $\Lambda$CDM model $w_0 = -1$ and $w_a =0$. The expression for the angular diameter distance, the parallax distance and the distance slip for dynamical dark energy are formally the same as for $\Lambda$CDM, i.e. \eqref{eq:DangFLRW}, \eqref{eq:DparFLRW} and \eqref{eq:muFLRW}, where however the Hubble parameter is modified as
\begin{eqnarray}
 H(z)^2 &=& H_0^2\,\left( \Omegamzero(1+z)^3 + \Omegakzero(1+z)^2 +\right. \\\nonumber
 &&+\left.\Omega_{\rm DE} e^{-3 w_a \frac{z}{1+z}} (1+z) ^{3(1+w_0 +w_a)}\right).
\end{eqnarray}
We explore the dependence of $\Dang$, $\Dpar$ and the distance slip on the two parameters of the modification of the $\Lambda$CDM in \eqref{eq:w0wa}. Our results for the flat geometry are shown in Fig.~\ref{fig:depwflat}: we note again that $\mu$ shows a different behaviour from those of $\Dang$ and $\Dpar$, which are in turn very similar, as for the parameters of the standard $\Lambda$CDM, see section~\ref{sec:depOs}

For small redshift $\mu$ takes the form
\begin{equation}
\mu(z)= \frac{3}{2}\Big[ 1+ w_0 (1-\Omegamzero) - \Omegakzero(w_0 +1) \Big]z^2 +O(z^3)\,
\end{equation}
for the three geometries, and $\Omegakzero =0$ in the above equation gives the result for the flat case.
As expected, there is a dependence on the equation of state of dynamical dark energy: at the leading order we find that $\mu$ depends on $w_0$ but not on $w_a$, because the effect of time variation appears at higher order in redshift.

\section{Conclusions}
\label{sec:concl}
In this paper we have discussed the new approach for the study of light propagation in the geometric optics regime presented in \cite{Grasso:2018mei}, which is based on the bilocal geodesic operators, BGOs, a new fundamental tool to fully characterize light propagation in a given spacetime and on all the scales of interest. In section~\ref{sec:final} we provide the relations between the BGOs and all the important quantities and observables already present in the literature like the Jacobi map, the magnification matrix, the angular diameter distance, the parallax distance and the position drift. The novelty of our results lies in the fact that all of them can be obtained within a unified framework and from one key quantity only, the BGOs. In addition, we show in section~\ref{deviationmethod} that in spacetimes where  an analytic expression for the null geodesic curve - physically representing the photon trajectory - is known one can avoid to solve the ODEs for the BGOs and simply calculate all the observables of interests by differentiating the expression of the photon geodesic with respect to initial data.
This new method is applicable to the cases where an exact solution of the Einstein equations allows for a solution of the geodesic equation and also in presence of perturbations around it. 

The main topic of our work is the study a new observable, the distance slip $\mu$,  introduced for the first time in~\cite{Grasso:2018mei}. It is a (dimensionless) combination of known observables, the parallax distance and the angular diameter distance or, alternatively, the parallax distance, the redshift and  the luminosity distance, and is defined  by relations (\ref{eq:mufromDangDpar2})-(\ref{eq:mufromDlumDpar}). Its usefulness stems from its peculiar properties, not shared by the known distance measures themselves: in any spacetime it is invariant with respect to the boosts of the observer and the source  which would make its measurement highly resistant to (ideally independent of) the noise and systematics due to peculiar motions.
Moreover, the distance slip can always be expressed as a nonlocal functional of the spacetime curvature along the line of sight, see equations (\ref{eq:mufrommperp}) and (\ref{eq:mperpODE})-(\ref{eq:mperpODEID2}). In particular, we also show that for short distances its value is simply proportional to a weighted integral of the matter density (equation (\ref{eq:mufromT})), reminiscent of tomography. This makes distance slip is a convenient tool for determining the geometry of the spacetime and its matter content.

We specialize our study of the peculiar properties of the distance slip, focusing on cosmology and on the differences between this new observable and those it is constructed from.
First of all, as it is immediately evident from the expression (equations \eqref{eq:muflat}-\eqref{eq:muopen}, \eqref{eq:DangFLRW} and \eqref{eq:DparFLRW}), the distance slip is independent of the Hubble parameter today $H_0$, unlike the angular diameter distance, the parallax distance and the luminosity distance.
We then go further and investigate the dependence of $\mu$ as opposed to $\Dang$ and $\Dpar$ on the other cosmological parameters, considering the curved and flat FLRW models for a universe containing cold dark matter and a cosmological constant (section \ref{sec:depOs}) as well as cold dark matter and dynamical dark energy (section \ref{sec:dynDE}). It is well known that the angular diameter distance $\Dang$ and the luminosity distance $\Dlum$ are related by a simple algebraic relation, namely the Etherington's duality formula $\Dlum = \Dang \,(1+z)^2$, 
\cite{etherington, etherington2}. Therefore the 
relations $\Dlum(z)$ and $\Dang(z)$, measured for a sample of sources, contain exactly the same information about the spacetime geometry \cite{perlick-lrr}. 
On the other hand, this does not hold for the parallax distance: $\Dpar(z)$ is known to contain independent information about the spacetime geometry, which were investigated in the FLRW metric case \cite{kermack_mccrea_whittaker_1934, weinberg-letter, kasai, rosquist, rasanen}. This very fact is particularly evident for the curvature parameter $\Omegakzero$, as we show here in Fig.~\ref{fig:depOk}. However, regarding the other cosmological parameters, in our case $\Omegamzero$, $\OmegaLambdazero$ and $w_0$ and $w_a$ the dependences as a function of the redshift z of the two distances display similar behaviour whereas that of $\mu$ is completely different. Although performing a detailed estimation of the constraining power of the distance slip is beyond the aim of our work, these results indicate that it may contain useful new information. We have also proposed to consider directly the relation $\mu(\Dang)$, without taking into account $z$ as an observable,
since it is strictly invariant with respect to the boosts of the sources, and therefore highly resistant to the noise due to peculiar motions.

The measurements of the distance slip are difficult for a fundamental reason: for sources located at short distances $\mu$ is very small, and thus its determination 
requires very precise astrometric measurements. The distance slip becomes significant only at cosmological distances (for instance  $\mu=0.22$ at $z = 1$), but at those distances
any parallax measurements are challenging. Nevertheless recent publications suggest that with the advances in astrometric techniques the parallax effects can be measured even at cosmological distances \cite{Quercellini:2010zr}, at least for $z < 1$, where the signal due to the Solar System's motion with respect to the CMB frame is expected to be larger than 
the effects of perturbations \cite{rasanen} or the aberration drift due to the motions within the local group \cite{Marcori:2018cwn}. In a subsequent paper \cite{KorzynskiVilla2} we will 
discuss this problem in detail and investigate the effects of local inhomogeneities on the distance slip measurements.

\section*{ACKNOWLEDGMENTS}
This work was supported by the National Science Centre, Poland (NCN) via 
the SONATA BIS programme, grant No~2016/22/E/ST9/00578 for the project
\emph{``Local relativistic perturbative framework in hydrodynamics and 
general relativity and its application to cosmology''}.

We thank Enea Di Dio, Giuseppe Fanizza, Andrzej Krasi\'n{}ski, Wojciech Hellwing and Michal Zaja\v{c}ek for stimulating discussions.

\appendix

\section{Bilocal geodesic operators in any unperturbed FLRW metric}
\label{app:0orderW}

We present the derivation of the Jacobi map $\calD\UD{\bm A}{\bm B}$ and the emitter-observer asymmetry operator $m\UD{\bm A}{\bm \mu}$ using the methods introduced in
Section \ref{sec:formulation}. The transverse part of the $\calW_{XX}$ and $\calW_{XL}$ matrices has already been derived in \cite{Fleury:2013sna}, but here we extend the result to the non-transverse part of $m\UD{\bm A}{\bm \mu}$, important for the position drift effects.

We derive the optical operators by the means the standard conformal trick. We first define the conformal time variable $\eta$ given by
\bea
\dd \eta = a^{-1}\,\dd t. \label{eq:conformaltime}
\eea 
 The unperturbed physical, expanding metric takes now the
form of
\bea
 g &=& a(\eta)^2(-\dd \eta^2 + \dd \chi^2 + S_k(\chi)^2 \dd \Omega^2)  \nonumber\\
 &=& a(\eta)^2\,\tilde g,
\eea
where $S_k(\chi)$ is defined by (\ref{eq:Sk}) and $\dd \Omega^2= \dd \theta^2 + \sin^2\theta\,\dd \varphi^2$ is the infinitesimal solid angle.
We have introduced the conformal metric $\tilde g$:
\bea
\tilde g = -\dd \eta^2 + \dd \chi^2 + S_k(\chi)^2 \dd \Omega^2. \label{eq:conformalg}
\eea
Note that in this derivation we do not assume a priori that the scale factor at the observation moment is equal to 1, unlike in Sec.~\ref{sec:cosmo}, i.e. we have $a\left(\eta_\calO\right) \equiv a_\calO \neq 1$ in general. 
This is because in the derivation we need to vary the observation moment, and therefore also the value of the scale factor $a_\calO$. 

The null geodesics of $g$ are the same as for $\tilde g$, except for the affine parametrization. Namely, let $\tilde x^\mu(x_{\calO}^\nu,\ell_\calO^\nu, \tilde\lambda)$ denote a null geodesic in $\tilde g$, with initial data $\tilde x^\mu(\lambda_\calO) = x_\calO^\mu$, $\ell^\mu(\lambda_\calO) = \ell_\calO^\mu$.
 It is a standard result that the null geodesic of $ g$, $ x^\mu(x_\calO^\nu, \ell_\calO^\nu, \lambda)$,  with the same initial data
can be obtained by simple reparametrization of the conformal one, i.e.
\bea
  x^\mu(x_\calO^\nu, \ell_\calO^\nu, \lambda) = \tilde x^\mu\left(x_{\calO}^\nu,\ell_\calO^\nu, \tilde\lambda(x_{\calO}^\nu,\ell_\calO^\nu,\lambda)\right), \label{eq:confo2}
\eea
where the function $ \tilde\lambda(x_{\calO}^\nu,\ell_\calO^\nu,\lambda)$ gives the initial data-dependent reparametrization.
We show right below that this reparametrization function can be obtained by solving the ODE
\bea
\frac{\dd \tilde \lambda}{\dd  \lambda} = \frac{a_\calO^2}{a^2}, \label{eq:ODEparametrizations}
\eea
with the initial data of the form $\tilde\lambda\left(x_{\calO}^\nu,\ell_\calO^\nu, \lambda_\calO\right) = \lambda_\calO$. 

We can prove  (\ref{eq:ODEparametrizations}) by comparing directly the tangent vectors $\tilde \ell^\mu$ and $\ell^\mu$ at each point of $\gamma_0$. First note that the component 0 of $\tilde\ell^\mu$ (associated with the conformal time $\eta$) scales according to
\bea
\tilde\ell^0 = \tilde\ell_\calO^0 \left(\frac{a_\calO}{a}\right)^2. \label{eq:l0scaling}
\eea
This can be seen in the following way: 
the $t$ component of $\ell^\mu$ in the $(t,\chi,\theta,\varphi)$ coordinate system must scale according to the redshift law, i.e. $\ell^t = (1+z)^{-1}\,\ell^t_\calO = \frac{a_\calO}{a}\,\ell^t_\calO$. On the other hand we have
$\ell^t = a\, \ell^0$ from the definition of the conformal time $\eta$ (\ref{eq:conformaltime}), so (\ref{eq:l0scaling}) follows immediately. 
We also know that $\tilde \ell^0 = -\tilde g_{\mu\nu} \,(\partial_\eta)^\mu \, \tilde \ell^\nu$ must remain constant because $\partial_\eta$ is a Killing vector of $\tilde g$.
Thus the $0$ components of both tangent vectors are related by $\ell^0 = \tilde\ell^0 \left(\frac{a_\calO}{a}\right)^2$ and consequently
the whole tangent vectors must be related by the scaling $\ell^\mu = \tilde \ell^\mu \left(\frac{a_\calO}{a}\right)^2$. The relation (\ref{eq:ODEparametrizations}) between the two parametrizations follows immediately.

The derivation of the optical operators $\calD$ and $m$ proceeds now in three steps. We first obtain the bilocal geodesic operators (BGOs)
$\tilde \calW_{XX}$ and $\tilde \calW_{XL}$ in the conformal spacetime, by solving the geodesic deviation equation (GDE) around the geodesics of the conformal metric $\tilde g$, from equations (\ref{eq:WfromR1})-(\ref{eq:WfromR2}).
Then we relate them to the operators $\calW_{XX}$ and $\calV_{XL}$ 
between the same two points on the same geodesic, but with respect to the metric $g$. This second part of the calculation is derived using the variational formulas (\ref{eq:relation11})-(\ref{eq:relation22}). Finally
we obtain the general expressions in the expanding spacetime for $\calD$ and $m$ by expressing $\calW_{XX}$ and $\calV_{XL}$ in the semi-null frame (SNF) of $ g$.

\paragraph{BGOs in the conformal spacetime. }

Consider the radial null geodesics $\gamma_0$ of the metric  $\tilde g$, passing through the observation point of coordinates $\eta = \eta_\calO$, $\chi = \chi_\calO$, $\theta = \frac{\pi}{2}$, $\varphi = 0$,  with the initial tangent vector $\ell_\calO^\mu = (\ell_\calO^0, \ell_\calO^0, 0, 0)$, $\ell_\calO^0 < 0$. Note that in the derivation we do not assume the observation point to be at the center of the 
spatial coordinate system, i.e. $\chi_\calO \neq 0$ or that at the observation time $a_\calO = a(\eta_\calO) = 1$, as it is assumed in Section \ref{sec:cosmo}.

The reader may check that the general solution reads
\bea
\tilde x^\mu(\lambda) = (\eta_\calO + \ell_\calO^0\,\tilde\lambda, \chi_\calO - \ell_\calO^0\,\tilde\lambda, \frac{\pi}{2}, 0), \label{eq:tildegeodesic}
\eea
where we have assumed for simplicity that $\tilde\lambda = \lambda_\calO = 0$ at the observation point. 
The tangent vector in the coordinate frame, which reads from~\eqref{eq:tildegeodesic}
\bea
\tilde \ell^\mu = \left(\ell_\calO^0, -\ell_\calO^0, 0, 0\right)\,, \label{eq:conformaltangent}
\eea
remains constant along the null geodesic.

We now report all the quantities necessary for the GDE (\ref{eq:WfromR1})-(\ref{eq:WfromR2}) and thus to obtain the BGOs. We begin with the SNF along $\gamma_0$, namely the  frame which is parallel-propagated
along the null geodesic with respect to the connection of the conformal metric $\tilde g$. It is given by
\bea
\tilde e_{\bm 0} &=& \partial_\eta\label{eq:SNF0} \\ 
\tilde e_{\bm 1} &=& S_k(\chi)^{-1}\,\partial_\theta \label{eq:SNF1} \\ 
\tilde e_{\bm 2} &=& \left(S_k(\chi)\,\sin\theta\right)^{-1}\,\partial_\varphi \label{eq:SNF2}\\
\tilde e_{\bm 3} &= &\ell_\calO^0\,\left(\partial_\eta - \partial_\chi\right) \label{eq:SNF3} \,,
\eea
where we note from (\ref{eq:conformaltangent}) that the last vector $\tilde e_{\bm 3}$ is simply equal to the tangent vector $\tilde \ell^\mu$. 

Then we need to calculate the Riemann  tensor $\tilde R\UD{\mu}{\nu\alpha\beta}$ of the conformal metric (\ref{eq:conformalg}), contract it twice with $\tilde \ell^
\mu$ from (\ref{eq:conformaltangent}) to obtain the optical tidal tensor and express it in the SNF (\ref{eq:SNF0})-(\ref{eq:SNF3}). The result is
\bea
\tilde R\UD{\bm\mu}{\bm\nu\bm\rho\bm\sigma} \, \tilde \ell^{\bm\nu}\, \tilde \ell^{\bm\rho} = \left(\ell_\calO^0\right)^2\,\left( \begin{array}{rrrr}
 0 & 0 & 0 & 0 \\ 0 & k & 0 & 0 \\ 0 & 0 & k & 0 \\ 0 & 0 & 0 & 0 \end{array}\right) \,,
\eea
which shows that in the SNF frame the optical tidal tensor turns out to have
constant coefficients

The operators $\tilde \calW_{XX}$ and $\tilde \calW_{XL}$ in the SNF can be now obtained easily from the matrix equations (\ref{eq:WfromR1})-(\ref{eq:WfromR2}).
We have
\bea
{{\tilde  \calW}_{XX}}{}\UD{\bm\mu}{\bm\nu} &=& \left(\begin{array}{cccc} 1 & 0 & 0 & 0 \\
0 & C_k\left(\Delta\chi\right) & 0 & 0 \\
0 & 0 & C_k\left(\Delta\chi\right) & 0 \\
0 & 0 & 0 & 1 \end{array} \right) \label{eq:tildeWXX}
\eea
and
\bea
{{\tilde \calW}_{XL}}{}\UD{\bm\mu}{\bm\nu} &=& \left(\begin{array}{cccc} \tilde \lambda & 0 & 0 & 0 \\
0 & -\frac{S_k\left(\Delta\chi\right)}{ \ell_\calO^0}& 0 & 0 \\
0 & 0 & -\frac{S_k\left(\Delta\chi\right)}{\ell_\calO^0} & 0 \\
0 & 0 & 0 & \tilde \lambda \end{array} \right) \label{eq:tildeWXL}
\eea
in the SNF $\tilde e_{\bm \mu}$ of~\eqref{eq:SNF0}-\eqref{eq:SNF3}. The functions $S_k$ and $C_k$ are given by (\ref{eq:Sk})-(\ref{eq:Ck}) and here as well as in the rest of this section their argument is the coordinate distance $\Delta\chi$ between the emission and observation point in the $\chi$ coordinate. Namely, from (\ref{eq:tildegeodesic}) we have
 \bea
\Delta\chi \equiv \chi_\calE- \chi_\calO = -\ell_\calO^0\,\tilde\lambda, \qquad \Delta\chi > 0\,.
 \eea
 From now on the argument is intended to be $\Delta\chi$, unless stated otherwise, and we drop it for notational convenience\footnote{Note that to switch from $\Delta\chi$ to $\ell_\calO^0\,\tilde\lambda$ we have a sign flip: $S_k\left(\ell_\calO^0\,\tilde \lambda\right)=-S_k(\Delta\chi)$ and $C_k\left(\ell_\calO^0\,\tilde \lambda\right)=C_k(\Delta\chi)$.}, i.e. $C_k \equiv C_k(\Delta \chi)$ and $S_k \equiv S_k(\Delta\chi)$.
Let us finally remark that, although we have chosen a special, radial null geodesic for the derivation, the results above hold for any null geodesic in the conformal space, because all null geodesics in the conformal metric are equivalent  due to the large isometry group of $\tilde g$.

\paragraph{BGOs in the expanding spacetime. }
In the second step we will obtain the operators $\calW_{XX}$ and $\calV_{XL}$, related to the expanding metric, from the conformal ones we have just obtained above. To do so, we will find the direct relation between them (see equations (\ref{eq:WviatildeW1a})-(\ref{eq:WviatildeW2a}) below) by using our variation method for the calculation of the BGOs, given by the relations (\ref{eq:relation11})-(\ref{eq:relation22}), written here in the common coordinate system $(\eta,\chi,\theta,\varphi)$ and by exploiting the fact that the null geodesics of both metric coincide up to a reparametrization. The BGOs are the coefficients of the total variation of the null geodesic curve with respect to initial data. This relation reads in the conformal spacetime
\bea
 \delta \tilde x^\mu = {\tilde \calW_{XX}}{}\UD{\mu}{\nu}\,\delta x_\calO^\nu + {{\tilde \calV}_{XL}}{}\UD{\mu}{i}\,\tilde\Delta \ell_\calO^i+ \tilde\ell_\calE^\mu\,\delta \tilde\lambda.
		\label{eq:endpointvariation11null}
\eea
with the operators $ {\tilde \calW_{XX}}{}\UD{\mu}{\nu}$, ${{\tilde \calV}_{XL}}{}\UD{\mu}{i}$ and the vector $\tilde \ell_\calE^\mu$ already known and $\tilde \Delta \ell_\calO^i$ denoting
the covariant direction deviation with the conformal Christoffel symbols $\tilde \Gamma(\calO)\UD{\mu}{\nu\sigma}$. The equation above holds for admissible variations $(\delta x_\calO^
\mu,\delta l_\calO^\mu)$ of the initial data, i.e. those
satisfying $\tilde\Delta \ell_\calO^\mu \, \ell_{\calO}^\nu\,\tilde g_{\mu\nu} = 0$. The reader may check that such variations are automatically admissible in the 
expanding metric, i.e. $\Delta \ell_\calO^\mu \, \ell_{\calO}^{\nu}\, g_{\mu\nu} = 0$ holds for them as well, and vice versa. The underlying reason is that a null tangent vector with respect
to $\tilde g$ is also null with respect to $ g$.  
We can therefore write down the same relation for the admissible null geodesic in the expanding metric:
\bea
 \delta  x^\mu = \calW_{XX}{}\UD{\mu}{\nu}\,\delta x_\calO^\nu + {{\calV}_{XL}}{}\UD{\mu}{i}\,\Delta \ell_\calO^i+ \ell_\calE^\mu\,\delta\lambda. \label{eq:endpointvariation2null}
\eea
Note that because of (\ref{eq:confo2}) the variations on the left hand sides of both equations must be equal for the same admissible variations of the initial data ($\delta x_\calO^\mu$, $\delta  l_\calO^\mu$) provided that  the variations of the affine parameters $\lambda$ and $\tilde \lambda$ are related appropriately, by the means of the variation of the relation (\ref{eq:parametrizationsint}). 
Therefore, the procedure to pass from one set of operators to the other one is fairly straightforward: we simply need to re-express the basis of differentials $(\delta x_\calO^\mu, \tilde \Delta \ell_\calO^i, \delta \tilde \lambda)$ 
in terms of the basis given by $(\delta x_\calO^\mu,  \Delta \ell^i_\calO, \delta  \lambda)$ and equate the right hand sides of (\ref{eq:endpointvariation11null}) and (\ref{eq:endpointvariation2null}). We will do so by deriving step by step the conformal transformations from the conformal basis to the basis in the expanding spacetime.

Let us begin with the covariant differentials of the spatial components of the tangent vector, i.e. the second term of the basis. We have to recall that the covariant derivatives of two conformally related metrics are in turn related by
\bea
\tilde \nabla_\mu \xi^\nu &=& \nabla_\mu \xi^\nu + C\UD{\mu}{\nu\alpha}\,\xi^\alpha ,
\eea
where in our case $C\UD{\mu}{\nu\alpha}$ is a tensor given by the derivative of the scale factor $a$:
\bea
C\UD{\mu}{\nu\alpha} &=& \frac{a_{,\kappa}}{a} \,\tilde g^{\kappa\mu}\,\tilde g_{\nu\alpha} - \frac{a_{,\alpha}}{a}\,\delta\UD{\mu}{\nu} - \frac{a_{,\nu}}{a} \,\delta\UD{\mu}{\alpha} .
\label{eq:Cformda}
\eea
The differentials $\tilde \Delta \ell_\calO^i$ and $\Delta \ell_\calO^i$ are therefore related by
\bea
 \tilde \Delta \ell_\calO^i = \Delta \ell_\calO^i + C\UD{i}{\mu\nu}\,\ell_\calO^\mu\,\delta x_\calO^\nu. \label{eq:covariantdervrelation}
\eea 
The relation between the variations of the affine parameters $\tilde \lambda$ and $\lambda$ is a bit more complicated, because we first need the relation between the two affine parameters. It is obtained  by solving for $\lambda$ the ODE in~\eqref{eq:ODEparametrizations}:
\bea
\frac{\dd \lambda}{\dd \tilde \lambda} = \frac{a^2}{a_\calO^2}. \label{eq:ODEparametrizations2}
\eea
Integrating (\ref{eq:ODEparametrizations2}) with the initial condition $\lambda = \tilde \lambda = 0$ at $\calO$ leads to:
\bea
 \lambda(\eta_\calO,\ell_\calO^0,\tilde \lambda) = \int_0^{\tilde\lambda} \frac{a(\eta_\calO + \ell_\calO^0\,\tilde \lambda')^2}{a(\eta_\calO)^2}\,\dd \tilde\lambda'. \label{eq:parametrizationsint}
\eea
Once this reparametrization and the conformal null geodesic are known (see~\eqref{eq:tildegeodesic} above), the null geodesic of the expanding metric simply follows from (\ref{eq:confo2}).
The conformal transformation of the differentials of the affine parameters is obtained by taking the total variation of (\ref{eq:parametrizationsint}):
\bea
\delta \lambda &=& \frac{a^2}{a_\calO^2}\,\delta \tilde \lambda + \left(\frac{1}{\ell_\calO^0}\left( \frac{a^2}{a_\calO^2} - 1\right) -\frac{2\dot a_\calO}{a_\calO}\,\lambda\right)\,\delta \eta_\calO 
\nonumber\\
&&+ 
\left( \frac{a^2}{a_\calO^2}\,\tilde\lambda - \lambda\right)\,\frac{\delta \ell_\calO^0}{\ell_\calO^0},
\eea
where $\dot a \equiv \frac{\dd a}{\dd \eta}$.
This is equivalent to
\bea
\delta \tilde\lambda &=& \frac{a_\calO^2}{a^2}\,\delta \lambda + \left(\frac{1}{\ell_\calO^0}\,\left(\frac{a_\calO^2}{a^2}-1\right)+\frac{2\dot a_\calO}{a_\calO}\,\frac{a_\calO^2}{a^2}\,\lambda\right)\,\delta \eta_\calO \nonumber\\ &&+ \left(\frac{a_\calO^2}{a^2}\,\lambda - \tilde\lambda\right)\,\frac{\delta \ell_\calO^0}{\ell_\calO^0}. \label{eq:tildelambdafromlambda}
\eea

We now substitute (\ref{eq:tildelambdafromlambda}) and (\ref{eq:covariantdervrelation}) to (\ref{eq:endpointvariation11null}) and relate the result to (\ref{eq:endpointvariation2null}). 
We obtain  this way
\bea
 0 &=& \left({\cal W}_{XX}{}\UD{\mu}{\nu} - {\tilde \calW}_{XX}{}\UD{\mu}{\nu} - \tilde \calV_{XL}{}\UD{\mu}{i}\,C\UD{i}{\alpha\nu}\,\ell_\calO^\alpha - \tilde\ell_\calE^\mu\,A_\nu \right)\,\delta x_\calO^\nu +
  \nonumber\\ && + \left({\cal V}_{XL}{}\UD{\mu}{i} 
- {\tilde \calV}_{XL}{}\UD{\mu}{i} - \tilde\ell_\calE^\mu\,B_i \right)\,\Delta  \ell_\calO^i \nonumber\\
 &&+ \left(\ell_\calE^\mu - \frac{a_\calO^2}{a^2}\,\tilde\ell_\calE^\mu\right)\,\delta \lambda \label{eq:WviatildeW0}
\eea
with the 1-forms $A_\mu$ and $B_i$ given by complicated expressions. Both 1-forms turn out later to be irrelevant.
Since (\ref{eq:WviatildeW0}) must hold for any admissible variations, we obtain this way general relations between  the operators in the conformal and expanding spacetime:
\bea
 {\cal W}_{XX}{}\UD{\mu}{\nu}  &=& {\tilde \calW}_{XX}{}\UD{\mu}{\nu} + \tilde \calV_{XL}{}\UD{\mu}{i}\,C\UD{i}{\alpha\nu}\,\ell_\calO^\alpha + \tilde\ell_\calE^\mu\,A_\nu \label{eq:WviatildeW1}\\
  {\cal V}_{XL}{}\UD{\mu}{i} &=& {\tilde \calV}_{XL}{}\UD{\mu}{i} + \tilde\ell_\calE^\mu\,B_i  \label{eq:WviatildeW2} \\
 \ell_\calE^\mu &=& \frac{a_\calO^2}{a^2}\,\tilde\ell_\calE^\mu. \nonumber
\eea
The last equation is just a restatement of the relation between the conformal and physical tangent vector. The other two are the relations we have been looking for, i.e. the transformation laws for the BGOs under the conformal rescaling of the metric  by $a(\eta)^2$.

 The relations  (\ref{eq:WviatildeW1})-(\ref{eq:WviatildeW2}) have been derived in the coordinate frame of the $(\eta, \chi, \theta, \varphi)$ coordinates.
 We now need to rewrite them in the conformal SNF $\tilde e_{\bm \mu}$. We begin with (\ref{eq:WviatildeW1}). From (\ref{eq:Cformda}) we see that that $C\UD{\mu}{\alpha\nu}\,\ell_\calO^\nu$ from
 is automatically
orthogonal to $\ell_{\calO\,\mu}$, and therefore from (\ref{eq:VwsW}) we have  $\tilde \calV_{XL}{}\UD{\mu}{i}\,C\UD{i}{\alpha\nu}\,\ell_\calO^\alpha = 
\tilde \calW_{XL}{}\UD{\mu}{\sigma}\,C\UD{\sigma}{\alpha\nu}\,\ell_\calO^\alpha$. Equation (\ref{eq:WviatildeW1}) takes therefore the form
\bea
 {\cal W}_{XX}{}\UD{\bm\mu}{\bm\nu}  &=& {\tilde \calW}_{XX}{}\UD{\bm\mu}{\bm\nu} + \tilde \calW_{XL}{}\UD{\bm\mu}{\bm\sigma}\,C\UD{\bm\sigma}{\bm\alpha\bm\nu}\,\ell_\calO^{\bm\alpha} \nonumber \\ &&+ \tilde\ell_\calE^{\bm\mu}\,A_{\bm\nu} \label{eq:WviatildeW1a}.
 \eea
 We now turn to (\ref{eq:WviatildeW2}). We note that admissible variations $\Delta \ell_\calO$ must have vanishing component $\Delta \ell_\calO^{\bm 0}$ in the SNF. 
 Therefore we get a relation only for the $\bm i = \bm 1, \bm 2, \bm 3$ components of $\calV_{XL}$ and $\tilde\calV_{XL}$, i.e. $\calV_{XL}{}\UD{\bm \mu}{\bm i}$ and  $\tilde\calV_{XL}{}\UD{\bm \mu}{\bm i}$. But in the SNF these components are in turn  equal to the corresponding components of $\calW_{XL}{}\UD{\bm \mu}{\bm i}$ and $\tilde \calW_{XL}{}\UD{\bm \mu}{\bm i}$ respectively, exactly because they correspond to contraction with admissible direction variation vectors, see (\ref{eq:VwsW}). 
Summarizing, we can  rewrite (\ref{eq:WviatildeW2}) as
 \bea
 {\cal V}_{XL}{}\UD{\bm\mu}{\bm i}  = {\cal W}_{XL}{}\UD{\bm\mu}{\bm i} &=& {\tilde \calW}_{XL}{}\UD{\bm\mu}{\bm i} + \tilde\ell_\calE^{\bm\mu}\,B_{\bm i}  \label{eq:WviatildeW2a} .
\eea

Recall that for the Jacobi operator $\calD\UD{\bm A}{\bm B}$ and the emitter-observer asymmetry operator $m\UD{\bm A}{\bm j}$ we only need the transverse components
$\bm 1$ and $\bm 2$ in the upper index $\bm \mu$. Thus the $\tilde\ell_\calE^{\bm\mu}$ terms drop out and the transformation laws simplify to
\bea
 {\cal W}_{XX}{}\UD{\bm A}{\bm\nu}  &=& {\tilde \calW}_{XX}{}\UD{\bm A}{\bm\nu} + \tilde \calW_{XL}{}\UD{\bm A}{\bm\sigma}\,C\UD{\bm\sigma}{\bm\alpha\bm\nu}\,\ell_\calO^{\bm\alpha}\label{eq:WviatildeW1b}. \\
  {\cal V}_{XL}{}\UD{\bm A}{\bm i} &=& {\tilde \calW}_{XL}{}\UD{\bm A}{\bm i} . \label{eq:WviatildeW2b} 
\eea

From (\ref{eq:conformaltangent}) and (\ref{eq:Cformda}) we get
\begin{widetext}
\bea
C\UD{\bm \mu}{\bm \alpha \bm \nu}\,\ell_{\calO}^{\bm \alpha} = \left(
\begin{array}{cccc}
0 & 0 & 0 & 0 \\
0 & -\ell_\calO^0\,\frac{\dot a_\calO}{a_\calO} & 0 & 0 \\
0 & 0 & -\ell_\calO^0\,\frac{\dot a_\calO}{a_\calO} & 0 \\
-\frac{\dot a_\calO}{a_\calO} & 0 & 0 & -2\ell_\calO^0\,\frac{\dot a_\calO}{a_\calO}
\end{array}\right),
\eea
\end{widetext}
where again $\dot a \equiv \frac{\dd a }{\dd \eta}$. Substituting this formula and (\ref{eq:tildeWXX})-(\ref{eq:tildeWXL}) in (\ref{eq:WviatildeW1b})-(\ref{eq:WviatildeW2b}) we obtain 
\bea
 {\cal W}_{XX}{}\UD{\bm A}{\bm\nu}  &=& \left(\begin{array}{cccc}
  0 & C_k + \frac{\dot a_\calO}{a_\calO}\,S_k & 0 & 0 \\
  0& 0 & C_k + \frac{\dot a_\calO}{a_\calO}\,S_k  & 0 
\end{array} 
\right)\\
  {\cal V}_{XL}{}\UD{\bm A}{\bm i} &=& \left(\begin{array}{ccc} - \left(\ell_\calO^0\right)^{-1} \,S_k & 0 & 0 \\
 0 & - \left(\ell_\calO^0\right)^{-1} \,S_k& 0
\end{array}\right)
\eea

\paragraph{Optical operators in the expanding spacetime.} In the final step we need to pass from the conformal frame $\tilde e_{\bm \mu}$ to the physical parallel-transported  SNF $e_{\bm \mu}$ of the expanding metric,  given by
\bea
e_{\bm 0} &=&  \frac{1}{2a}\,\left( \frac{a}{a_\calO} + \frac{a_\calO}{a}\right)\partial_\eta + \frac{1}{2a}\,\left( \frac{a}{a_\calO} - \frac{a_\calO}{a}\right)\partial_\chi  \nonumber\\
&=&
\frac{1}{a_\calO}\,\tilde e_{\bm 0}  - \frac{1}{2a_{\calO}\,{\ell_\calO^0}}\,\left(1 - \frac{a_\calO^2}{a^2}\right)\tilde e_{\bm 3}\label{eq:physicalSNF0}\\
e_{\bm 1} &=&  \left(a\,S_k(\chi)\right)^{-1}\,\partial_\theta =  a^{-1}\,\tilde e_{\bm 1} \label{eq:physicalSNF1} \\
e_{\bm 2} &=& \left(a\,S_k(\chi)\,\sin\theta\right)^{-1}\,\partial_\varphi = a^{-1}\,\tilde e_{\bm 2} \label{eq:physicalSNF2} \\
e_{\bm 3} &=& \frac{a_\calO^2}{a^2}\,\ell_\calO^0\,\left(\partial_\eta - \partial_\chi\right)  =  \frac{a_\calO^2}{a^2}\,\tilde e_{\bm 3}. \label{eq:physicalSNF3}
\eea

The reader may check that this frame is indeed parallel-transported along $\gamma_0$ with respect to the expanding metric $g$ and that $e_{\bm 3}$ coincides with the physical tangent vector to $\gamma_0$, i.e. $\ell^\mu$.
Moreover we see that the transverse vectors $e_{\bm 1}, e_{\bm 2}$ and $e_{\bm 3}$ coincide with $\tilde e_{\bm 1}, \tilde e_{\bm 2}$ and $\tilde e_{\bm 3}$ up to rescalings. Applying
the transformation and remembering that the index $\bm A$ is used for a vector at the emission point $\calE$, while $\bm \nu$ denotes components in $\calO$, we get
\begin{widetext}
\bea
 {\cal W}_{XX}{}\UD{\bm A}{\bm\nu}  &=&   \left(\begin{array}{cccc}0 & \frac{a}{a_\calO}\,C_k + \frac{a\,\dot a_\calO}{a_\calO^2}\,S_k&0&0\\
  0 & 0 & \frac{a}{a_\calO}\,C_k + \frac{a\,\dot a_\calO}{a_\calO^2}\,S_k&0\end{array}\right)  \\
  {\cal V}_{XL}{}\UD{\bm A}{\bm i} &=& \left(\begin{array}{ccc} -\left(\ell_\calO^0 \right)^{-1}\,\frac{a\,S_k}{a_\calO}&0&0\\
  0 & -\left(\ell_\calO^0 \right)^{-1}\,\frac{a\,S_k}{a_\calO}&0\end{array}\right) . 
\eea
\end{widetext}
From this we obtain using (\ref{eq:jacobi}) and (\ref{eq:mfromWXX}):
\bea
 {\cal D}\UD{\bm A}{\bm B}  &=&   -\frac{a\,S_k}{a_\calO\,\ell_\calO^0} \, \delta\UD{\bm A}{\bm B}   \label{eq:Dformulageneral1}  \\
  {m_\perp}{}\UD{\bm A}{\bm B}&=&  \left(\frac{a}{a_\calO}\,C_k + \frac{a\,\dot a_\calO}{a_\calO^2}\,S_k-1\right) \cdot \delta\UD{\bm A}{\bm B} 
  \label{eq:mformulageneral1}\\
  m\UD{\bm A}{\bm 0} &=& m\UD{\bm A}{\bm 3} = 0.
\eea
Note that in our convention $\ell_\calO^0<0$, so the overall sign for the prefactor in (\ref{eq:Dformulageneral1}) is actually positive. We may also simplify (\ref{eq:mformulageneral1}) by noting that $\frac{\dot a_\calO}{a_\calO^2} = H_0$, so 
\bea
 {m_\perp}{}\UD{\bm A}{\bm B}=  \Big(\frac{a}{a_\calO}\,C_k+ a\, H_0\,S_k-1\Big) \cdot \delta\UD{\bm A}{\bm B}   \label{eq:mformulageneral2}.
\eea
Equations (\ref{eq:Dformulageneral1}) and (\ref{eq:mformulageneral2}) agree with the results from \cite{Fleury:2013sna} (equations (4.8) and (4.9)) if we take into account the difference in notation and the parametrizations
assumed there.

As the last step of our derivation we express the coordinate distance $\Delta\chi$, appearing as the argument of $C_k$ and $S_k$, by an integral over the geodesic $\gamma_0$ between $\calO$ and $\calE$.
We begin by calculating 
\bea
\frac{\dd a}{\dd \tilde\lambda}= \frac{\dd a}{\dd t}\cdot\frac{\dd t}{\dd  \eta} \cdot \frac{\dd \eta}{\dd \tilde \lambda} .
\eea
Here $\frac{\dd a}{\dd t} = H(a)\,a$ from the definition of the Hubble parameter, $\frac{\dd t}{\dd  \eta} = a$ from the definition of the conformal time (\ref{eq:conformaltime}) and
$\frac{\dd \eta}{\dd \tilde \lambda} = \ell_\calO^0$ from (\ref{eq:tildegeodesic}). Thus $\frac{\dd a}{\dd \tilde\lambda}= \ell_\calO^0\,H(a)\,a^2$ or 
\bea
\frac{\dd \tilde\lambda}{\dd a} =  \frac{1}{\ell_\calO^0\,H(a)\,a^{2}}
\eea
valid along the null geodesic $\gamma_0$.
Integrating this relation from $\calO$, where $\tilde\lambda = 0$, up to the emission point $\calE$ we obtain
\bea
-\ell_\calO^0\,\tilde\lambda &=& \Delta \chi= - \int_{a_\calO}^{a} \frac{\dd \hat a}{H(\hat a)\,\hat a^2} \nonumber \\ &=& \int_{a}^{a_\calO} \frac{\dd \hat a}{H(\hat a)\,\hat a^2}.
\eea
 We may also change the integration variable to the redshift $\hat z = \frac{a_\calO}{\hat a} - 1$:
\bea\label{eq:chiapp}
\Delta\chi= \frac{1}{a_\calO} \int_{0}^{z} \frac{\dd \hat z}{H(\hat z)}.
\eea
Using the first Friedmann equation in the form of
\bea
 H(z)^2 &=& H_0^2\,\left( \Omegamzero(1+z)^3 + \Omegakzero(1+z)^2 \nonumber \right.\\
 &&\left. + \OmegaLambdazero\right)
\eea
(see \cite{peeblesbook, peacock_1998_2, Hogg:1999ad}) the integral in \eqref{eq:chiapp} can be recast in the following form:
\bea
\Delta\chi &=& \frac{1}{a_\calO\,H_0} \int_{0}^{z} \dd \hat z\,\left(\Omegamzero(1+\hat z)^3 + \Omegakzero(1+\hat z)^2 \right. \nonumber \\
 &&\left. + \OmegaLambdazero\right)^{-1/2}.
\eea
The integral above is related to the total line-of-sight comoving distance $D_C$ between $\calE$ and $\calO$ evaluated at the observation moment \cite{Hogg:1999ad}, namely we have
$\Delta \chi =  \frac{1}{a_\calO}\,D_C$. 
We may now impose the standard convention, in which at the observation moment we have $a_\calO = 1$, the 
observation point is located at the origin, i.e. $\chi_\calO = 0$, and the null vector $\ell_\calO$ is normalized so that $\ell_\calO^0 = -1$, as it is assumed in Section \ref{sec:cosmo}. In this case we have simply 
$D_C = \Delta\chi$ for the standard comoving  distance and $\chi = \Delta\chi = \int_{0}^{z} \frac{\dd \hat z}{H(\hat z)}$. Applying these relations to (\ref{eq:Dformulageneral1}) and
(\ref{eq:mformulageneral2}) we obtain:
\bea
 {\cal D}\UD{\bm A}{\bm B}  &=&   a\,S_k\left(\chi\right)\, \delta\UD{\bm A}{\bm B}   \label{eq:Dformulageneral3}  \\
 {m_\perp}{}\UD{\bm A}{\bm B}&=&  \left(a\,C_k(\chi ) + 
 a\, H_0\,S_k(\chi )-1\right) \, \delta\UD{\bm A}{\bm B} \label{eq:mformulageneral3}. 
\eea

With these results we may evaluate the distance slip using the relation (\ref{eq:defofmu2}): 
\bea
 \mu = 1 - a^2\left(C_k(\chi ) + 
  H_0\,S_k(\chi )\right)^2.
\eea
Noting that $a = (1+z)^{-1}$ for comoving sources we recover equation (\ref{eq:muFLRW}) in the main text.

\bibliography{ms}

\begin{thebibliography}{52}%
\makeatletter
\providecommand \@ifxundefined [1]{%
 \@ifx{#1\undefined}
}%
\providecommand \@ifnum [1]{%
 \ifnum #1\expandafter \@firstoftwo
 \else \expandafter \@secondoftwo
 \fi
}%
\providecommand \@ifx [1]{%
 \ifx #1\expandafter \@firstoftwo
 \else \expandafter \@secondoftwo
 \fi
}%
\providecommand \natexlab [1]{#1}%
\providecommand \enquote  [1]{``#1''}%
\providecommand \bibnamefont  [1]{#1}%
\providecommand \bibfnamefont [1]{#1}%
\providecommand \citenamefont [1]{#1}%
\providecommand \href@noop [0]{\@secondoftwo}%
\providecommand \href [0]{\begingroup \@sanitize@url \@href}%
\providecommand \@href[1]{\@@startlink{#1}\@@href}%
\providecommand \@@href[1]{\endgroup#1\@@endlink}%
\providecommand \@sanitize@url [0]{\catcode `\\12\catcode `\$12\catcode
  `\&12\catcode `\#12\catcode `\^12\catcode `\_12\catcode `\%12\relax}%
\providecommand \@@startlink[1]{}%
\providecommand \@@endlink[0]{}%
\providecommand \url  [0]{\begingroup\@sanitize@url \@url }%
\providecommand \@url [1]{\endgroup\@href {#1}{\urlprefix }}%
\providecommand \urlprefix  [0]{URL }%
\providecommand \Eprint [0]{\href }%
\providecommand \doibase [0]{http://dx.doi.org/}%
\providecommand \selectlanguage [0]{\@gobble}%
\providecommand \bibinfo  [0]{\@secondoftwo}%
\providecommand \bibfield  [0]{\@secondoftwo}%
\providecommand \translation [1]{[#1]}%
\providecommand \BibitemOpen [0]{}%
\providecommand \bibitemStop [0]{}%
\providecommand \bibitemNoStop [0]{.\EOS\space}%
\providecommand \EOS [0]{\spacefactor3000\relax}%
\providecommand \BibitemShut  [1]{\csname bibitem#1\endcsname}%
\let\auto@bib@innerbib\@empty
\bibitem [{\citenamefont {Grasso}\ \emph {et~al.}(2019)\citenamefont {Grasso},
  \citenamefont {Korzy\'nski},\ and\ \citenamefont
  {Serbenta}}]{Grasso:2018mei}%
  \BibitemOpen
  \bibfield  {author} {\bibinfo {author} {\bibfnamefont {M.}~\bibnamefont
  {Grasso}}, \bibinfo {author} {\bibfnamefont {M.}~\bibnamefont {Korzy\'nski}},
  \ and\ \bibinfo {author} {\bibfnamefont {J.}~\bibnamefont {Serbenta}},\
  }\href {\doibase 10.1103/PhysRevD.99.064038} {\bibfield  {journal} {\bibinfo
  {journal} {Phys. Rev.}\ }\textbf {\bibinfo {volume} {D99}},\ \bibinfo {pages}
  {064038} (\bibinfo {year} {2019})},\ \Eprint
  {http://arxiv.org/abs/1811.10284} {arXiv:1811.10284 [gr-qc]} \BibitemShut
  {NoStop}%
\bibitem [{\citenamefont {{Sachs}}(1961)}]{sachs}%
  \BibitemOpen
  \bibfield  {author} {\bibinfo {author} {\bibfnamefont {R.}~\bibnamefont
  {{Sachs}}},\ }\href {\doibase 10.1098/rspa.1961.0202} {\bibfield  {journal}
  {\bibinfo  {journal} {Proceedings of the Royal Society of London Series A}\
  }\textbf {\bibinfo {volume} {264}},\ \bibinfo {pages} {309} (\bibinfo {year}
  {1961})}\BibitemShut {NoStop}%
\bibitem [{\citenamefont {{Seitz}}\ \emph {et~al.}(1994)\citenamefont
  {{Seitz}}, \citenamefont {{Schneider}},\ and\ \citenamefont
  {{Ehlers}}}]{seitzschneiderehlers}%
  \BibitemOpen
  \bibfield  {author} {\bibinfo {author} {\bibfnamefont {S.}~\bibnamefont
  {{Seitz}}}, \bibinfo {author} {\bibfnamefont {P.}~\bibnamefont
  {{Schneider}}}, \ and\ \bibinfo {author} {\bibfnamefont {J.}~\bibnamefont
  {{Ehlers}}},\ }\href {\doibase 10.1088/0264-9381/11/9/016} {\bibfield
  {journal} {\bibinfo  {journal} {Classical and Quantum Gravity}\ }\textbf
  {\bibinfo {volume} {11}},\ \bibinfo {pages} {2345} (\bibinfo {year}
  {1994})},\ \Eprint {http://arxiv.org/abs/astro-ph/9403056} {astro-ph/9403056}
  \BibitemShut {NoStop}%
\bibitem [{\citenamefont {Perlick}(2004)}]{perlick-lrr}%
  \BibitemOpen
  \bibfield  {author} {\bibinfo {author} {\bibfnamefont {V.}~\bibnamefont
  {Perlick}},\ }\href {\doibase 10.12942/lrr-2004-9} {\bibfield  {journal}
  {\bibinfo  {journal} {Living Reviews in Relativity}\ }\textbf {\bibinfo
  {volume} {7}},\ \bibinfo {pages} {9} (\bibinfo {year} {2004})}\BibitemShut
  {NoStop}%
\bibitem [{\citenamefont {Ehlers}\ \emph {et~al.}(1961)\citenamefont {Ehlers},
  \citenamefont {Jordan},\ and\ \citenamefont {Sachs}}]{ehlers-jordan-sachs}%
  \BibitemOpen
  \bibfield  {author} {\bibinfo {author} {\bibfnamefont {J.}~\bibnamefont
  {Ehlers}}, \bibinfo {author} {\bibfnamefont {P.}~\bibnamefont {Jordan}}, \
  and\ \bibinfo {author} {\bibfnamefont {R.~K.}\ \bibnamefont {Sachs}},\
  }\href@noop {} {\emph {\bibinfo {title} {{Beitr\"age zur Theorie der reinen
  Gravitationsstrahlung}}}},\ \bibinfo {series} {{Abhandlungen der
  Mathematisch-Naturwissenschaftlichen Klasse}}, Vol.~\bibinfo {volume} {1}\
  (\bibinfo  {publisher} {Verlag der Akademie der Wissenschaften und der
  Literatur in Mainz},\ \bibinfo {address} {Wiesbaden, Germany},\ \bibinfo
  {year} {1961})\BibitemShut {NoStop}%
\bibitem [{\citenamefont {Jordan}\ \emph {et~al.}(2013)\citenamefont {Jordan},
  \citenamefont {Ehlers},\ and\ \citenamefont {Sachs}}]{Jordan2013reprint}%
  \BibitemOpen
  \bibfield  {author} {\bibinfo {author} {\bibfnamefont {P.}~\bibnamefont
  {Jordan}}, \bibinfo {author} {\bibfnamefont {J.}~\bibnamefont {Ehlers}}, \
  and\ \bibinfo {author} {\bibfnamefont {R.~K.}\ \bibnamefont {Sachs}},\ }\href
  {\doibase 10.1007/s10714-013-1590-1} {\bibfield  {journal} {\bibinfo
  {journal} {General Relativity and Gravitation}\ }\textbf {\bibinfo {volume}
  {45}},\ \bibinfo {pages} {2691} (\bibinfo {year} {2013})}\BibitemShut
  {NoStop}%
\bibitem [{\citenamefont {Quercellini}\ \emph {et~al.}(2012)\citenamefont
  {Quercellini}, \citenamefont {Amendola}, \citenamefont {Balbi}, \citenamefont
  {Cabella},\ and\ \citenamefont {Quartin}}]{Quercellini:2010zr}%
  \BibitemOpen
  \bibfield  {author} {\bibinfo {author} {\bibfnamefont {C.}~\bibnamefont
  {Quercellini}}, \bibinfo {author} {\bibfnamefont {L.}~\bibnamefont
  {Amendola}}, \bibinfo {author} {\bibfnamefont {A.}~\bibnamefont {Balbi}},
  \bibinfo {author} {\bibfnamefont {P.}~\bibnamefont {Cabella}}, \ and\
  \bibinfo {author} {\bibfnamefont {M.}~\bibnamefont {Quartin}},\ }\href
  {\doibase 10.1016/j.physrep.2012.09.002} {\bibfield  {journal} {\bibinfo
  {journal} {Phys. Rept.}\ }\textbf {\bibinfo {volume} {521}},\ \bibinfo
  {pages} {95} (\bibinfo {year} {2012})},\ \Eprint
  {http://arxiv.org/abs/1011.2646} {arXiv:1011.2646 [astro-ph.CO]} \BibitemShut
  {NoStop}%
\bibitem [{\citenamefont {Korzy\'nski}\ and\ \citenamefont
  {Kopi\'nski}(2018)}]{Korzynski:2017nas}%
  \BibitemOpen
  \bibfield  {author} {\bibinfo {author} {\bibfnamefont {M.}~\bibnamefont
  {Korzy\'nski}}\ and\ \bibinfo {author} {\bibfnamefont {J.}~\bibnamefont
  {Kopi\'nski}},\ }\href {\doibase 10.1088/1475-7516/2018/03/012} {\bibfield
  {journal} {\bibinfo  {journal} {JCAP}\ }\textbf {\bibinfo {volume} {1803}},\
  \bibinfo {pages} {012} (\bibinfo {year} {2018})},\ \Eprint
  {http://arxiv.org/abs/1711.00584} {arXiv:1711.00584 [gr-qc]} \BibitemShut
  {NoStop}%
\bibitem [{\citenamefont {Marcori}\ \emph {et~al.}(2018)\citenamefont
  {Marcori}, \citenamefont {Pitrou}, \citenamefont {Uzan},\ and\ \citenamefont
  {Pereira}}]{Marcori:2018cwn}%
  \BibitemOpen
  \bibfield  {author} {\bibinfo {author} {\bibfnamefont {O.~H.}\ \bibnamefont
  {Marcori}}, \bibinfo {author} {\bibfnamefont {C.}~\bibnamefont {Pitrou}},
  \bibinfo {author} {\bibfnamefont {J.-P.}\ \bibnamefont {Uzan}}, \ and\
  \bibinfo {author} {\bibfnamefont {T.~S.}\ \bibnamefont {Pereira}},\ }\href
  {\doibase 10.1103/PhysRevD.98.023517} {\bibfield  {journal} {\bibinfo
  {journal} {Phys. Rev.}\ }\textbf {\bibinfo {volume} {D98}},\ \bibinfo {pages}
  {023517} (\bibinfo {year} {2018})},\ \Eprint
  {http://arxiv.org/abs/1805.12121} {arXiv:1805.12121 [astro-ph.CO]}
  \BibitemShut {NoStop}%
\bibitem [{\citenamefont {Hellaby}\ and\ \citenamefont
  {Walters}(2018)}]{Hellaby:2017soj}%
  \BibitemOpen
  \bibfield  {author} {\bibinfo {author} {\bibfnamefont {C.}~\bibnamefont
  {Hellaby}}\ and\ \bibinfo {author} {\bibfnamefont {A.}~\bibnamefont
  {Walters}},\ }\href {\doibase 10.1088/1475-7516/2018/02/015} {\bibfield
  {journal} {\bibinfo  {journal} {JCAP}\ }\textbf {\bibinfo {volume} {1802}},\
  \bibinfo {pages} {015} (\bibinfo {year} {2018})},\ \Eprint
  {http://arxiv.org/abs/1708.01031} {arXiv:1708.01031 [gr-qc]} \BibitemShut
  {NoStop}%
\bibitem [{\citenamefont {Yoo}\ and\ \citenamefont
  {Scaccabarozzi}(2016)}]{Yoo:2016vne}%
  \BibitemOpen
  \bibfield  {author} {\bibinfo {author} {\bibfnamefont {J.}~\bibnamefont
  {Yoo}}\ and\ \bibinfo {author} {\bibfnamefont {F.}~\bibnamefont
  {Scaccabarozzi}},\ }\href {\doibase 10.1088/1475-7516/2016/09/046} {\bibfield
   {journal} {\bibinfo  {journal} {JCAP}\ }\textbf {\bibinfo {volume} {1609}},\
  \bibinfo {pages} {046} (\bibinfo {year} {2016})},\ \Eprint
  {http://arxiv.org/abs/1606.08453} {arXiv:1606.08453 [astro-ph.CO]}
  \BibitemShut {NoStop}%
\bibitem [{\citenamefont {Fleury}\ \emph {et~al.}(2013)\citenamefont {Fleury},
  \citenamefont {Dupuy},\ and\ \citenamefont {Uzan}}]{Fleury:2013sna}%
  \BibitemOpen
  \bibfield  {author} {\bibinfo {author} {\bibfnamefont {P.}~\bibnamefont
  {Fleury}}, \bibinfo {author} {\bibfnamefont {H.}~\bibnamefont {Dupuy}}, \
  and\ \bibinfo {author} {\bibfnamefont {J.-P.}\ \bibnamefont {Uzan}},\ }\href
  {\doibase 10.1103/PhysRevD.87.123526} {\bibfield  {journal} {\bibinfo
  {journal} {Phys. Rev.}\ }\textbf {\bibinfo {volume} {D87}},\ \bibinfo {pages}
  {123526} (\bibinfo {year} {2013})},\ \Eprint {http://arxiv.org/abs/1302.5308}
  {arXiv:1302.5308 [astro-ph.CO]} \BibitemShut {NoStop}%
\bibitem [{\citenamefont {Räsänen}\ \emph {et~al.}(2015)\citenamefont
  {Räsänen}, \citenamefont {Bolejko},\ and\ \citenamefont
  {Finoguenov}}]{Rasanen:2014mca}%
  \BibitemOpen
  \bibfield  {author} {\bibinfo {author} {\bibfnamefont {S.}~\bibnamefont
  {Räsänen}}, \bibinfo {author} {\bibfnamefont {K.}~\bibnamefont {Bolejko}},
  \ and\ \bibinfo {author} {\bibfnamefont {A.}~\bibnamefont {Finoguenov}},\
  }\href {\doibase 10.1103/PhysRevLett.115.101301} {\bibfield  {journal}
  {\bibinfo  {journal} {Phys. Rev. Lett.}\ }\textbf {\bibinfo {volume} {115}},\
  \bibinfo {pages} {101301} (\bibinfo {year} {2015})},\ \Eprint
  {http://arxiv.org/abs/1412.4976} {arXiv:1412.4976 [astro-ph.CO]} \BibitemShut
  {NoStop}%
\bibitem [{\citenamefont {Sturm}\ \emph {et~al.}(2018)\citenamefont {Sturm}
  \emph {et~al.}}]{2018Natur.563..657G}%
  \BibitemOpen
  \bibfield  {author} {\bibinfo {author} {\bibfnamefont {E.}~\bibnamefont
  {Sturm}} \emph {et~al.} (\bibinfo {collaboration} {Gravity collaboration}),\
  }\href {\doibase 10.1038/s41586-018-0731-9} {\bibfield  {journal} {\bibinfo
  {journal} {Nature}\ }\textbf {\bibinfo {volume} {563}},\ \bibinfo {pages}
  {657} (\bibinfo {year} {2018})},\ \Eprint {http://arxiv.org/abs/1811.11195}
  {arXiv:1811.11195 [astro-ph.GA]} \BibitemShut {NoStop}%
\bibitem [{\citenamefont {Risaliti}\ and\ \citenamefont
  {Lusso}(2019)}]{Risaliti:2018reu}%
  \BibitemOpen
  \bibfield  {author} {\bibinfo {author} {\bibfnamefont {G.}~\bibnamefont
  {Risaliti}}\ and\ \bibinfo {author} {\bibfnamefont {E.}~\bibnamefont
  {Lusso}},\ }\href {\doibase 10.1038/s41550-018-0657-z} {\bibfield  {journal}
  {\bibinfo  {journal} {Nat. Astron.}\ }\textbf {\bibinfo {volume} {3}},\
  \bibinfo {pages} {272} (\bibinfo {year} {2019})},\ \Eprint
  {http://arxiv.org/abs/1811.02590} {arXiv:1811.02590 [astro-ph.CO]}
  \BibitemShut {NoStop}%
\bibitem [{\citenamefont {Uzun}(2018)}]{Uzun:2018yes}%
  \BibitemOpen
  \bibfield  {author} {\bibinfo {author} {\bibfnamefont {N.}~\bibnamefont
  {Uzun}},\ }\href@noop {} {\  (\bibinfo {year} {2018})},\ \Eprint
  {http://arxiv.org/abs/1811.10917} {arXiv:1811.10917 [gr-qc]} \BibitemShut
  {NoStop}%
\bibitem [{\citenamefont {R\"as\"anen}(2014)}]{rasanen}%
  \BibitemOpen
  \bibfield  {author} {\bibinfo {author} {\bibfnamefont {S.}~\bibnamefont
  {R\"as\"anen}},\ }\href {http://stacks.iop.org/1475-7516/2014/i=03/a=035}
  {\bibfield  {journal} {\bibinfo  {journal} {Journal of Cosmology and
  Astroparticle Physics}\ }\textbf {\bibinfo {volume} {2014}},\ \bibinfo
  {pages} {035} (\bibinfo {year} {2014})}\BibitemShut {NoStop}%
\bibitem [{\citenamefont {Etherington}(1933)}]{etherington}%
  \BibitemOpen
  \bibfield  {author} {\bibinfo {author} {\bibfnamefont {I.}~\bibnamefont
  {Etherington}},\ }\href {\doibase 10.1080/14786443309462220} {\bibfield
  {journal} {\bibinfo  {journal} {The London, Edinburgh, and Dublin
  Philosophical Magazine and Journal of Science}\ }\textbf {\bibinfo {volume}
  {15}},\ \bibinfo {pages} {761} (\bibinfo {year} {1933})},\ \Eprint
  {http://arxiv.org/abs/http://dx.doi.org/10.1080/14786443309462220}
  {http://dx.doi.org/10.1080/14786443309462220} \BibitemShut {NoStop}%
\bibitem [{\citenamefont {Etherington}(2007)}]{etherington2}%
  \BibitemOpen
  \bibfield  {author} {\bibinfo {author} {\bibfnamefont {I.~M.~H.}\
  \bibnamefont {Etherington}},\ }\href {\doibase 10.1007/s10714-007-0447-x}
  {\bibfield  {journal} {\bibinfo  {journal} {General Relativity and
  Gravitation}\ }\textbf {\bibinfo {volume} {39}},\ \bibinfo {pages} {1055}
  (\bibinfo {year} {2007})}\BibitemShut {NoStop}%
\bibitem [{\citenamefont {{Klioner}}(2003)}]{klioner2003}%
  \BibitemOpen
  \bibfield  {author} {\bibinfo {author} {\bibfnamefont {S.~A.}\ \bibnamefont
  {{Klioner}}},\ }\href {\doibase 10.1086/367593} {\bibfield  {journal}
  {\bibinfo  {journal} {The Astronomical Journal}\ }\textbf {\bibinfo {volume}
  {125}},\ \bibinfo {pages} {1580} (\bibinfo {year} {2003})}\BibitemShut
  {NoStop}%
\bibitem [{\citenamefont {{Sanna}}\ \emph {et~al.}(2017)\citenamefont
  {{Sanna}}, \citenamefont {{Reid}}, \citenamefont {{Dame}}, \citenamefont
  {{Menten}},\ and\ \citenamefont {{Brunthaler}}}]{2017Sci...358..227S}%
  \BibitemOpen
  \bibfield  {author} {\bibinfo {author} {\bibfnamefont {A.}~\bibnamefont
  {{Sanna}}}, \bibinfo {author} {\bibfnamefont {M.~J.}\ \bibnamefont {{Reid}}},
  \bibinfo {author} {\bibfnamefont {T.~M.}\ \bibnamefont {{Dame}}}, \bibinfo
  {author} {\bibfnamefont {K.~M.}\ \bibnamefont {{Menten}}}, \ and\ \bibinfo
  {author} {\bibfnamefont {A.}~\bibnamefont {{Brunthaler}}},\ }\href {\doibase
  10.1126/science.aan5452} {\bibfield  {journal} {\bibinfo  {journal}
  {Science}\ }\textbf {\bibinfo {volume} {358}},\ \bibinfo {pages} {227}
  (\bibinfo {year} {2017})},\ \Eprint {http://arxiv.org/abs/1710.06489}
  {arXiv:1710.06489 [astro-ph.GA]} \BibitemShut {NoStop}%
\bibitem [{\citenamefont {Mignard}\ \emph {et~al.}(2018)\citenamefont {Mignard}
  \emph {et~al.}}]{2018A&A...616A..14G}%
  \BibitemOpen
  \bibfield  {author} {\bibinfo {author} {\bibfnamefont {F.}~\bibnamefont
  {Mignard}} \emph {et~al.} (\bibinfo {collaboration} {Gaia collaboration}),\
  }\href {\doibase 10.1051/0004-6361/201832916} {\bibfield  {journal} {\bibinfo
   {journal} {Astronomy and Astrophysics}\ }\textbf {\bibinfo {volume} {616}},\
  \bibinfo {eid} {A14} (\bibinfo {year} {2018})}\BibitemShut {NoStop}%
\bibitem [{\citenamefont {Humphreys}\ \emph {et~al.}(2013)\citenamefont
  {Humphreys}, \citenamefont {Reid}, \citenamefont {Moran}, \citenamefont
  {Greenhill},\ and\ \citenamefont {Argon}}]{Humphreys_2013}%
  \BibitemOpen
  \bibfield  {author} {\bibinfo {author} {\bibfnamefont {E.~M.~L.}\
  \bibnamefont {Humphreys}}, \bibinfo {author} {\bibfnamefont {M.~J.}\
  \bibnamefont {Reid}}, \bibinfo {author} {\bibfnamefont {J.~M.}\ \bibnamefont
  {Moran}}, \bibinfo {author} {\bibfnamefont {L.~J.}\ \bibnamefont
  {Greenhill}}, \ and\ \bibinfo {author} {\bibfnamefont {A.~L.}\ \bibnamefont
  {Argon}},\ }\href {\doibase 10.1088/0004-637x/775/1/13} {\bibfield  {journal}
  {\bibinfo  {journal} {The Astrophysical Journal}\ }\textbf {\bibinfo {volume}
  {775}},\ \bibinfo {pages} {13} (\bibinfo {year} {2013})}\BibitemShut
  {NoStop}%
\bibitem [{\citenamefont {{Pietrzy{\'n}ski}}\ \emph {et~al.}(2013)\citenamefont
  {{Pietrzy{\'n}ski}}, \citenamefont {{Graczyk}}, \citenamefont {{Gieren}},
  \citenamefont {{Thompson}}, \citenamefont {{Pilecki}}, \citenamefont
  {{Udalski}}, \citenamefont {{Soszy{\'n}ski}}, \citenamefont {{Koz{\l}owski}},
  \citenamefont {{Konorski}}, \citenamefont {{Suchomska}}, \citenamefont
  {{Bono}}, \citenamefont {{Moroni}}, \citenamefont {{Villanova}},
  \citenamefont {{Nardetto}}, \citenamefont {{Bresolin}}, \citenamefont
  {{Kudritzki}}, \citenamefont {{Storm}}, \citenamefont {{Gallenne}},
  \citenamefont {{Smolec}}, \citenamefont {{Minniti}}, \citenamefont
  {{Kubiak}}, \citenamefont {{Szyma{\'n}ski}}, \citenamefont {{Poleski}},
  \citenamefont {{Wyrzykowski}}, \citenamefont {{Ulaczyk}}, \citenamefont
  {{Pietrukowicz}}, \citenamefont {{G{\'o}rski}},\ and\ \citenamefont
  {{Karczmarek}}}]{2013Nature}%
  \BibitemOpen
  \bibfield  {author} {\bibinfo {author} {\bibfnamefont {G.}~\bibnamefont
  {{Pietrzy{\'n}ski}}}, \bibinfo {author} {\bibfnamefont {D.}~\bibnamefont
  {{Graczyk}}}, \bibinfo {author} {\bibfnamefont {W.}~\bibnamefont {{Gieren}}},
  \bibinfo {author} {\bibfnamefont {I.~B.}\ \bibnamefont {{Thompson}}},
  \bibinfo {author} {\bibfnamefont {B.}~\bibnamefont {{Pilecki}}}, \bibinfo
  {author} {\bibfnamefont {A.}~\bibnamefont {{Udalski}}}, \bibinfo {author}
  {\bibfnamefont {I.}~\bibnamefont {{Soszy{\'n}ski}}}, \bibinfo {author}
  {\bibfnamefont {S.}~\bibnamefont {{Koz{\l}owski}}}, \bibinfo {author}
  {\bibfnamefont {P.}~\bibnamefont {{Konorski}}}, \bibinfo {author}
  {\bibfnamefont {K.}~\bibnamefont {{Suchomska}}}, \bibinfo {author}
  {\bibfnamefont {G.}~\bibnamefont {{Bono}}}, \bibinfo {author} {\bibfnamefont
  {P.~G.~P.}\ \bibnamefont {{Moroni}}}, \bibinfo {author} {\bibfnamefont
  {S.}~\bibnamefont {{Villanova}}}, \bibinfo {author} {\bibfnamefont
  {N.}~\bibnamefont {{Nardetto}}}, \bibinfo {author} {\bibfnamefont
  {F.}~\bibnamefont {{Bresolin}}}, \bibinfo {author} {\bibfnamefont {R.~P.}\
  \bibnamefont {{Kudritzki}}}, \bibinfo {author} {\bibfnamefont
  {J.}~\bibnamefont {{Storm}}}, \bibinfo {author} {\bibfnamefont
  {A.}~\bibnamefont {{Gallenne}}}, \bibinfo {author} {\bibfnamefont
  {R.}~\bibnamefont {{Smolec}}}, \bibinfo {author} {\bibfnamefont
  {D.}~\bibnamefont {{Minniti}}}, \bibinfo {author} {\bibfnamefont
  {M.}~\bibnamefont {{Kubiak}}}, \bibinfo {author} {\bibfnamefont {M.~K.}\
  \bibnamefont {{Szyma{\'n}ski}}}, \bibinfo {author} {\bibfnamefont
  {R.}~\bibnamefont {{Poleski}}}, \bibinfo {author} {\bibfnamefont
  {{\L}.}~\bibnamefont {{Wyrzykowski}}}, \bibinfo {author} {\bibfnamefont
  {K.}~\bibnamefont {{Ulaczyk}}}, \bibinfo {author} {\bibfnamefont
  {P.}~\bibnamefont {{Pietrukowicz}}}, \bibinfo {author} {\bibfnamefont
  {M.}~\bibnamefont {{G{\'o}rski}}}, \ and\ \bibinfo {author} {\bibfnamefont
  {P.}~\bibnamefont {{Karczmarek}}},\ }\href {\doibase 10.1038/nature11878}
  {\bibfield  {journal} {\bibinfo  {journal} {Nature}\ }\textbf {\bibinfo
  {volume} {495}},\ \bibinfo {pages} {76} (\bibinfo {year} {2013})},\ \Eprint
  {http://arxiv.org/abs/1303.2063} {arXiv:1303.2063 [astro-ph.GA]} \BibitemShut
  {NoStop}%
\bibitem [{\citenamefont {{Benedict}}\ \emph {et~al.}(2007)\citenamefont
  {{Benedict}}, \citenamefont {{McArthur}}, \citenamefont {{Feast}},
  \citenamefont {{Barnes}}, \citenamefont {{Harrison}}, \citenamefont
  {{Patterson}}, \citenamefont {{Menzies}}, \citenamefont {{Bean}},\ and\
  \citenamefont {{Freedman}}}]{2007AJ....133.1810B}%
  \BibitemOpen
  \bibfield  {author} {\bibinfo {author} {\bibfnamefont {G.~F.}\ \bibnamefont
  {{Benedict}}}, \bibinfo {author} {\bibfnamefont {B.~E.}\ \bibnamefont
  {{McArthur}}}, \bibinfo {author} {\bibfnamefont {M.~W.}\ \bibnamefont
  {{Feast}}}, \bibinfo {author} {\bibfnamefont {T.~G.}\ \bibnamefont
  {{Barnes}}}, \bibinfo {author} {\bibfnamefont {T.~E.}\ \bibnamefont
  {{Harrison}}}, \bibinfo {author} {\bibfnamefont {R.~J.}\ \bibnamefont
  {{Patterson}}}, \bibinfo {author} {\bibfnamefont {J.~W.}\ \bibnamefont
  {{Menzies}}}, \bibinfo {author} {\bibfnamefont {J.~L.}\ \bibnamefont
  {{Bean}}}, \ and\ \bibinfo {author} {\bibfnamefont {W.~L.}\ \bibnamefont
  {{Freedman}}},\ }\href {\doibase 10.1086/511980} {\bibfield  {journal}
  {\bibinfo  {journal} {The Astrophysical Journal}\ }\textbf {\bibinfo {volume}
  {133}},\ \bibinfo {pages} {1810} (\bibinfo {year} {2007})},\ \Eprint
  {http://arxiv.org/abs/astro-ph/0612465} {arXiv:astro-ph/0612465 [astro-ph]}
  \BibitemShut {NoStop}%
\bibitem [{\citenamefont {Van~Leeuwen}\ \emph {et~al.}(2007)\citenamefont
  {Van~Leeuwen}, \citenamefont {Feast}, \citenamefont {Whitelock},\ and\
  \citenamefont {Laney}}]{10.1111/j.1365-2966.2007.11972.x}%
  \BibitemOpen
  \bibfield  {author} {\bibinfo {author} {\bibfnamefont {F.}~\bibnamefont
  {Van~Leeuwen}}, \bibinfo {author} {\bibfnamefont {M.~W.}\ \bibnamefont
  {Feast}}, \bibinfo {author} {\bibfnamefont {P.~A.}\ \bibnamefont
  {Whitelock}}, \ and\ \bibinfo {author} {\bibfnamefont {C.~D.}\ \bibnamefont
  {Laney}},\ }\href {\doibase 10.1111/j.1365-2966.2007.11972.x} {\bibfield
  {journal} {\bibinfo  {journal} {Monthly Notices of the Royal Astronomical
  Society}\ }\textbf {\bibinfo {volume} {379}},\ \bibinfo {pages} {723}
  (\bibinfo {year} {2007})},\ \Eprint
  {http://arxiv.org/abs/http://oup.prod.sis.lan/mnras/article-pdf/379/2/723/3398063/mnras0379-0723.pdf}
  {http://oup.prod.sis.lan/mnras/article-pdf/379/2/723/3398063/mnras0379-0723.pdf}
  \BibitemShut {NoStop}%
\bibitem [{\citenamefont {Casertano}\ \emph {et~al.}(2016)\citenamefont
  {Casertano}, \citenamefont {Riess}, \citenamefont {Anderson}, \citenamefont
  {Anderson}, \citenamefont {Bowers}, \citenamefont {Clubb}, \citenamefont
  {Cukierman}, \citenamefont {Filippenko}, \citenamefont {Graham},
  \citenamefont {MacKenty}, \citenamefont {Melis}, \citenamefont {Tucker},\
  and\ \citenamefont {Upadhya}}]{Casertano_2016}%
  \BibitemOpen
  \bibfield  {author} {\bibinfo {author} {\bibfnamefont {S.}~\bibnamefont
  {Casertano}}, \bibinfo {author} {\bibfnamefont {A.~G.}\ \bibnamefont
  {Riess}}, \bibinfo {author} {\bibfnamefont {J.}~\bibnamefont {Anderson}},
  \bibinfo {author} {\bibfnamefont {R.~I.}\ \bibnamefont {Anderson}}, \bibinfo
  {author} {\bibfnamefont {J.~B.}\ \bibnamefont {Bowers}}, \bibinfo {author}
  {\bibfnamefont {K.~I.}\ \bibnamefont {Clubb}}, \bibinfo {author}
  {\bibfnamefont {A.~R.}\ \bibnamefont {Cukierman}}, \bibinfo {author}
  {\bibfnamefont {A.~V.}\ \bibnamefont {Filippenko}}, \bibinfo {author}
  {\bibfnamefont {M.~L.}\ \bibnamefont {Graham}}, \bibinfo {author}
  {\bibfnamefont {J.~W.}\ \bibnamefont {MacKenty}}, \bibinfo {author}
  {\bibfnamefont {C.}~\bibnamefont {Melis}}, \bibinfo {author} {\bibfnamefont
  {B.~E.}\ \bibnamefont {Tucker}}, \ and\ \bibinfo {author} {\bibfnamefont
  {G.}~\bibnamefont {Upadhya}},\ }\href {\doibase 10.3847/0004-637x/825/1/11}
  {\bibfield  {journal} {\bibinfo  {journal} {The Astrophysical Journal}\
  }\textbf {\bibinfo {volume} {825}},\ \bibinfo {pages} {11} (\bibinfo {year}
  {2016})}\BibitemShut {NoStop}%
\bibitem [{\citenamefont {{Lindegren, L.}}\ \emph {et~al.}(2016)\citenamefont
  {{Lindegren, L.}}, \citenamefont {{Lammers, U.}}, \citenamefont {{Bastian,
  U.}}, \citenamefont {{Hern\'andez, J.}}, \citenamefont {{Klioner, S.}},
  \citenamefont {{Hobbs, D.}}, \citenamefont {{Bombrun, A.}}, \citenamefont
  {{Michalik, D.}}, \citenamefont {{Ramos-Lerate, M.}}, \citenamefont
  {{Butkevich, A.}}, \citenamefont {{Comoretto, G.}}, \citenamefont {{Joliet,
  E.}}, \citenamefont {{Holl, B.}}, \citenamefont {{Hutton, A.}}, \citenamefont
  {{Parsons, P.}}, \citenamefont {{Steidelm\"uller, H.}}, \citenamefont
  {{Abbas, U.}}, \citenamefont {{Altmann, M.}}, \citenamefont {{Andrei, A.}},
  \citenamefont {{Anton, S.}}, \citenamefont {{Bach, N.}}, \citenamefont
  {{Barache, C.}}, \citenamefont {{Becciani, U.}}, \citenamefont {{Berthier,
  J.}}, \citenamefont {{Bianchi, L.}}, \citenamefont {{Biermann, M.}},
  \citenamefont {{Bouquillon, S.}}, \citenamefont {{Bourda, G.}}, \citenamefont
  {{Br\"usemeister, T.}}, \citenamefont {{Bucciarelli, B.}}, \citenamefont
  {{Busonero, D.}}, \citenamefont {{Carlucci, T.}}, \citenamefont
  {{Casta\~neda, J.}}, \citenamefont {{Charlot, P.}}, \citenamefont {{Clotet,
  M.}}, \citenamefont {{Crosta, M.}}, \citenamefont {{Davidson, M.}},
  \citenamefont {{de Felice, F.}}, \citenamefont {{Drimmel, R.}}, \citenamefont
  {{Fabricius, C.}}, \citenamefont {{Fienga, A.}}, \citenamefont {{Figueras,
  F.}}, \citenamefont {{Fraile, E.}}, \citenamefont {{Gai, M.}}, \citenamefont
  {{Garralda, N.}}, \citenamefont {{Geyer, R.}}, \citenamefont
  {{Gonz\'alez-Vidal, J. J.}}, \citenamefont {{Guerra, R.}}, \citenamefont
  {{Hambly, N. C.}}, \citenamefont {{Hauser, M.}}, \citenamefont {{Jordan,
  S.}}, \citenamefont {{Lattanzi, M. G.}}, \citenamefont {{Lenhardt, H.}},
  \citenamefont {{Liao, S.}}, \citenamefont {{L\"offler, W.}}, \citenamefont
  {{McMillan, P. J.}}, \citenamefont {{Mignard, F.}}, \citenamefont {{Mora,
  A.}}, \citenamefont {{Morbidelli, R.}}, \citenamefont {{Portell, J.}},
  \citenamefont {{Riva, A.}}, \citenamefont {{Sarasso, M.}}, \citenamefont
  {{Serraller, I.}}, \citenamefont {{Siddiqui, H.}}, \citenamefont {{Smart,
  R.}}, \citenamefont {{Spagna, A.}}, \citenamefont {{Stampa, U.}},
  \citenamefont {{Steele, I.}}, \citenamefont {{Taris, F.}}, \citenamefont
  {{Torra, J.}}, \citenamefont {{van Reeven, W.}}, \citenamefont {{Vecchiato,
  A.}}, \citenamefont {{Zschocke, S.}}, \citenamefont {{de Bruijne, J.}},
  \citenamefont {{Gracia, G.}}, \citenamefont {{Raison, F.}}, \citenamefont
  {{Lister, T.}}, \citenamefont {{Marchant, J.}}, \citenamefont {{Messineo,
  R.}}, \citenamefont {{Soffel, M.}}, \citenamefont {{Osorio, J.}},
  \citenamefont {{de Torres, A.}},\ and\ \citenamefont {{O\'{}Mullane,
  W.}}}]{refId0}%
  \BibitemOpen
  \bibfield  {author} {\bibinfo {author} {\bibnamefont {{Lindegren, L.}}},
  \bibinfo {author} {\bibnamefont {{Lammers, U.}}}, \bibinfo {author}
  {\bibnamefont {{Bastian, U.}}}, \bibinfo {author} {\bibnamefont
  {{Hern\'andez, J.}}}, \bibinfo {author} {\bibnamefont {{Klioner, S.}}},
  \bibinfo {author} {\bibnamefont {{Hobbs, D.}}}, \bibinfo {author}
  {\bibnamefont {{Bombrun, A.}}}, \bibinfo {author} {\bibnamefont {{Michalik,
  D.}}}, \bibinfo {author} {\bibnamefont {{Ramos-Lerate, M.}}}, \bibinfo
  {author} {\bibnamefont {{Butkevich, A.}}}, \bibinfo {author} {\bibnamefont
  {{Comoretto, G.}}}, \bibinfo {author} {\bibnamefont {{Joliet, E.}}}, \bibinfo
  {author} {\bibnamefont {{Holl, B.}}}, \bibinfo {author} {\bibnamefont
  {{Hutton, A.}}}, \bibinfo {author} {\bibnamefont {{Parsons, P.}}}, \bibinfo
  {author} {\bibnamefont {{Steidelm\"uller, H.}}}, \bibinfo {author}
  {\bibnamefont {{Abbas, U.}}}, \bibinfo {author} {\bibnamefont {{Altmann,
  M.}}}, \bibinfo {author} {\bibnamefont {{Andrei, A.}}}, \bibinfo {author}
  {\bibnamefont {{Anton, S.}}}, \bibinfo {author} {\bibnamefont {{Bach, N.}}},
  \bibinfo {author} {\bibnamefont {{Barache, C.}}}, \bibinfo {author}
  {\bibnamefont {{Becciani, U.}}}, \bibinfo {author} {\bibnamefont {{Berthier,
  J.}}}, \bibinfo {author} {\bibnamefont {{Bianchi, L.}}}, \bibinfo {author}
  {\bibnamefont {{Biermann, M.}}}, \bibinfo {author} {\bibnamefont
  {{Bouquillon, S.}}}, \bibinfo {author} {\bibnamefont {{Bourda, G.}}},
  \bibinfo {author} {\bibnamefont {{Br\"usemeister, T.}}}, \bibinfo {author}
  {\bibnamefont {{Bucciarelli, B.}}}, \bibinfo {author} {\bibnamefont
  {{Busonero, D.}}}, \bibinfo {author} {\bibnamefont {{Carlucci, T.}}},
  \bibinfo {author} {\bibnamefont {{Casta\~neda, J.}}}, \bibinfo {author}
  {\bibnamefont {{Charlot, P.}}}, \bibinfo {author} {\bibnamefont {{Clotet,
  M.}}}, \bibinfo {author} {\bibnamefont {{Crosta, M.}}}, \bibinfo {author}
  {\bibnamefont {{Davidson, M.}}}, \bibinfo {author} {\bibnamefont {{de Felice,
  F.}}}, \bibinfo {author} {\bibnamefont {{Drimmel, R.}}}, \bibinfo {author}
  {\bibnamefont {{Fabricius, C.}}}, \bibinfo {author} {\bibnamefont {{Fienga,
  A.}}}, \bibinfo {author} {\bibnamefont {{Figueras, F.}}}, \bibinfo {author}
  {\bibnamefont {{Fraile, E.}}}, \bibinfo {author} {\bibnamefont {{Gai, M.}}},
  \bibinfo {author} {\bibnamefont {{Garralda, N.}}}, \bibinfo {author}
  {\bibnamefont {{Geyer, R.}}}, \bibinfo {author} {\bibnamefont
  {{Gonz\'alez-Vidal, J. J.}}}, \bibinfo {author} {\bibnamefont {{Guerra,
  R.}}}, \bibinfo {author} {\bibnamefont {{Hambly, N. C.}}}, \bibinfo {author}
  {\bibnamefont {{Hauser, M.}}}, \bibinfo {author} {\bibnamefont {{Jordan,
  S.}}}, \bibinfo {author} {\bibnamefont {{Lattanzi, M. G.}}}, \bibinfo
  {author} {\bibnamefont {{Lenhardt, H.}}}, \bibinfo {author} {\bibnamefont
  {{Liao, S.}}}, \bibinfo {author} {\bibnamefont {{L\"offler, W.}}}, \bibinfo
  {author} {\bibnamefont {{McMillan, P. J.}}}, \bibinfo {author} {\bibnamefont
  {{Mignard, F.}}}, \bibinfo {author} {\bibnamefont {{Mora, A.}}}, \bibinfo
  {author} {\bibnamefont {{Morbidelli, R.}}}, \bibinfo {author} {\bibnamefont
  {{Portell, J.}}}, \bibinfo {author} {\bibnamefont {{Riva, A.}}}, \bibinfo
  {author} {\bibnamefont {{Sarasso, M.}}}, \bibinfo {author} {\bibnamefont
  {{Serraller, I.}}}, \bibinfo {author} {\bibnamefont {{Siddiqui, H.}}},
  \bibinfo {author} {\bibnamefont {{Smart, R.}}}, \bibinfo {author}
  {\bibnamefont {{Spagna, A.}}}, \bibinfo {author} {\bibnamefont {{Stampa,
  U.}}}, \bibinfo {author} {\bibnamefont {{Steele, I.}}}, \bibinfo {author}
  {\bibnamefont {{Taris, F.}}}, \bibinfo {author} {\bibnamefont {{Torra, J.}}},
  \bibinfo {author} {\bibnamefont {{van Reeven, W.}}}, \bibinfo {author}
  {\bibnamefont {{Vecchiato, A.}}}, \bibinfo {author} {\bibnamefont {{Zschocke,
  S.}}}, \bibinfo {author} {\bibnamefont {{de Bruijne, J.}}}, \bibinfo {author}
  {\bibnamefont {{Gracia, G.}}}, \bibinfo {author} {\bibnamefont {{Raison,
  F.}}}, \bibinfo {author} {\bibnamefont {{Lister, T.}}}, \bibinfo {author}
  {\bibnamefont {{Marchant, J.}}}, \bibinfo {author} {\bibnamefont {{Messineo,
  R.}}}, \bibinfo {author} {\bibnamefont {{Soffel, M.}}}, \bibinfo {author}
  {\bibnamefont {{Osorio, J.}}}, \bibinfo {author} {\bibnamefont {{de Torres,
  A.}}}, \ and\ \bibinfo {author} {\bibnamefont {{O\'{}Mullane, W.}}},\ }\href
  {\doibase 10.1051/0004-6361/201628714} {\bibfield  {journal} {\bibinfo
  {journal} {A\&A}\ }\textbf {\bibinfo {volume} {595}},\ \bibinfo {pages} {A4}
  (\bibinfo {year} {2016})}\BibitemShut {NoStop}%
\bibitem [{\citenamefont {{Riess}}\ \emph {et~al.}(2018)\citenamefont
  {{Riess}}, \citenamefont {{Casertano}}, \citenamefont {{Yuan}}, \citenamefont
  {{Macri}}, \citenamefont {{Anderson}}, \citenamefont {{MacKenty}},
  \citenamefont {{Bowers}}, \citenamefont {{Clubb}}, \citenamefont
  {{Filippenko}}, \citenamefont {{Jones}},\ and\ \citenamefont
  {{Tucker}}}]{2018ApJ...855..136R}%
  \BibitemOpen
  \bibfield  {author} {\bibinfo {author} {\bibfnamefont {A.~G.}\ \bibnamefont
  {{Riess}}}, \bibinfo {author} {\bibfnamefont {S.}~\bibnamefont
  {{Casertano}}}, \bibinfo {author} {\bibfnamefont {W.}~\bibnamefont {{Yuan}}},
  \bibinfo {author} {\bibfnamefont {L.}~\bibnamefont {{Macri}}}, \bibinfo
  {author} {\bibfnamefont {J.}~\bibnamefont {{Anderson}}}, \bibinfo {author}
  {\bibfnamefont {J.~W.}\ \bibnamefont {{MacKenty}}}, \bibinfo {author}
  {\bibfnamefont {J.~B.}\ \bibnamefont {{Bowers}}}, \bibinfo {author}
  {\bibfnamefont {K.~I.}\ \bibnamefont {{Clubb}}}, \bibinfo {author}
  {\bibfnamefont {A.~V.}\ \bibnamefont {{Filippenko}}}, \bibinfo {author}
  {\bibfnamefont {D.~O.}\ \bibnamefont {{Jones}}}, \ and\ \bibinfo {author}
  {\bibfnamefont {B.~E.}\ \bibnamefont {{Tucker}}},\ }\href {\doibase
  10.3847/1538-4357/aaadb7} {\bibfield  {journal} {\bibinfo  {journal} {The
  Astrophysical Journal}\ }\textbf {\bibinfo {volume} {855}},\ \bibinfo {eid}
  {136} (\bibinfo {year} {2018})},\ \Eprint {http://arxiv.org/abs/1801.01120}
  {arXiv:1801.01120 [astro-ph.SR]} \BibitemShut {NoStop}%
\bibitem [{\citenamefont {Kardashev}(1986)}]{kardashev}%
  \BibitemOpen
  \bibfield  {author} {\bibinfo {author} {\bibfnamefont {N.~S.}\ \bibnamefont
  {Kardashev}},\ }\href@noop {} {\bibfield  {journal} {\bibinfo  {journal}
  {Soviet astronomy}\ }\textbf {\bibinfo {volume} {30}} (\bibinfo {year}
  {1986})}\BibitemShut {NoStop}%
\bibitem [{\citenamefont {Bel}\ and\ \citenamefont
  {Marinoni}(2018)}]{PhysRevLett.121.021101}%
  \BibitemOpen
  \bibfield  {author} {\bibinfo {author} {\bibfnamefont {J.}~\bibnamefont
  {Bel}}\ and\ \bibinfo {author} {\bibfnamefont {C.}~\bibnamefont {Marinoni}},\
  }\href {\doibase 10.1103/PhysRevLett.121.021101} {\bibfield  {journal}
  {\bibinfo  {journal} {Phys. Rev. Lett.}\ }\textbf {\bibinfo {volume} {121}},\
  \bibinfo {pages} {021101} (\bibinfo {year} {2018})}\BibitemShut {NoStop}%
\bibitem [{\citenamefont {Ding}\ and\ \citenamefont
  {Croft}(2009)}]{Ding:2009xs}%
  \BibitemOpen
  \bibfield  {author} {\bibinfo {author} {\bibfnamefont {F.}~\bibnamefont
  {Ding}}\ and\ \bibinfo {author} {\bibfnamefont {R.~A.~C.}\ \bibnamefont
  {Croft}},\ }\href {\doibase 10.1111/j.1365-2966.2009.15111.x} {\bibfield
  {journal} {\bibinfo  {journal} {Mon. Not. Roy. Astron. Soc.}\ }\textbf
  {\bibinfo {volume} {397}},\ \bibinfo {pages} {1739} (\bibinfo {year}
  {2009})},\ \Eprint {http://arxiv.org/abs/0903.3402} {arXiv:0903.3402
  [astro-ph.CO]} \BibitemShut {NoStop}%
\bibitem [{\citenamefont {Quartin}\ and\ \citenamefont
  {Amendola}(2010)}]{Quartin:2009xr}%
  \BibitemOpen
  \bibfield  {author} {\bibinfo {author} {\bibfnamefont {M.}~\bibnamefont
  {Quartin}}\ and\ \bibinfo {author} {\bibfnamefont {L.}~\bibnamefont
  {Amendola}},\ }\href {\doibase 10.1103/PhysRevD.81.043522} {\bibfield
  {journal} {\bibinfo  {journal} {Phys. Rev.}\ }\textbf {\bibinfo {volume}
  {D81}},\ \bibinfo {pages} {043522} (\bibinfo {year} {2010})},\ \Eprint
  {http://arxiv.org/abs/0909.4954} {arXiv:0909.4954 [astro-ph.CO]} \BibitemShut
  {NoStop}%
\bibitem [{\citenamefont {{Riess}}\ \emph {et~al.}(1996)\citenamefont
  {{Riess}}, \citenamefont {{Press}},\ and\ \citenamefont
  {{Kirshner}}}]{1996ApJ...473...88R}%
  \BibitemOpen
  \bibfield  {author} {\bibinfo {author} {\bibfnamefont {A.~G.}\ \bibnamefont
  {{Riess}}}, \bibinfo {author} {\bibfnamefont {W.~H.}\ \bibnamefont
  {{Press}}}, \ and\ \bibinfo {author} {\bibfnamefont {R.~P.}\ \bibnamefont
  {{Kirshner}}},\ }\href {\doibase 10.1086/178129} {\bibfield  {journal}
  {\bibinfo  {journal} {The Astrophysical Journal}\ }\textbf {\bibinfo {volume}
  {473}},\ \bibinfo {pages} {88} (\bibinfo {year} {1996})},\ \Eprint
  {http://arxiv.org/abs/astro-ph/9604143} {arXiv:astro-ph/9604143 [astro-ph]}
  \BibitemShut {NoStop}%
\bibitem [{\citenamefont {Riess}\ \emph {et~al.}(1998)\citenamefont {Riess},
  \citenamefont {Filippenko}, \citenamefont {Challis}, \citenamefont
  {Clocchiatti}, \citenamefont {Diercks}, \citenamefont {Garnavich},
  \citenamefont {Gilliland}, \citenamefont {Hogan}, \citenamefont {Jha},
  \citenamefont {Kirshner}, \citenamefont {Leibundgut}, \citenamefont
  {Phillips}, \citenamefont {Reiss}, \citenamefont {Schmidt}, \citenamefont
  {Schommer}, \citenamefont {Smith}, \citenamefont {Spyromilio}, \citenamefont
  {Stubbs}, \citenamefont {Suntzeff},\ and\ \citenamefont
  {Tonry}}]{Riess_1998}%
  \BibitemOpen
  \bibfield  {author} {\bibinfo {author} {\bibfnamefont {A.~G.}\ \bibnamefont
  {Riess}}, \bibinfo {author} {\bibfnamefont {A.~V.}\ \bibnamefont
  {Filippenko}}, \bibinfo {author} {\bibfnamefont {P.}~\bibnamefont {Challis}},
  \bibinfo {author} {\bibfnamefont {A.}~\bibnamefont {Clocchiatti}}, \bibinfo
  {author} {\bibfnamefont {A.}~\bibnamefont {Diercks}}, \bibinfo {author}
  {\bibfnamefont {P.~M.}\ \bibnamefont {Garnavich}}, \bibinfo {author}
  {\bibfnamefont {R.~L.}\ \bibnamefont {Gilliland}}, \bibinfo {author}
  {\bibfnamefont {C.~J.}\ \bibnamefont {Hogan}}, \bibinfo {author}
  {\bibfnamefont {S.}~\bibnamefont {Jha}}, \bibinfo {author} {\bibfnamefont
  {R.~P.}\ \bibnamefont {Kirshner}}, \bibinfo {author} {\bibfnamefont
  {B.}~\bibnamefont {Leibundgut}}, \bibinfo {author} {\bibfnamefont {M.~M.}\
  \bibnamefont {Phillips}}, \bibinfo {author} {\bibfnamefont {D.}~\bibnamefont
  {Reiss}}, \bibinfo {author} {\bibfnamefont {B.~P.}\ \bibnamefont {Schmidt}},
  \bibinfo {author} {\bibfnamefont {R.~A.}\ \bibnamefont {Schommer}}, \bibinfo
  {author} {\bibfnamefont {R.~C.}\ \bibnamefont {Smith}}, \bibinfo {author}
  {\bibfnamefont {J.}~\bibnamefont {Spyromilio}}, \bibinfo {author}
  {\bibfnamefont {C.}~\bibnamefont {Stubbs}}, \bibinfo {author} {\bibfnamefont
  {N.~B.}\ \bibnamefont {Suntzeff}}, \ and\ \bibinfo {author} {\bibfnamefont
  {J.}~\bibnamefont {Tonry}},\ }\href {\doibase 10.1086/300499} {\bibfield
  {journal} {\bibinfo  {journal} {The Astronomical Journal}\ }\textbf {\bibinfo
  {volume} {116}},\ \bibinfo {pages} {1009} (\bibinfo {year}
  {1998})}\BibitemShut {NoStop}%
\bibitem [{\citenamefont {{Perlmutter}}\ \emph {et~al.}(1999)\citenamefont
  {{Perlmutter}}, \citenamefont {{Aldering}}, \citenamefont {{Goldhaber}},
  \citenamefont {{Knop}}, \citenamefont {{Nugent}}, \citenamefont {{Castro}},
  \citenamefont {{Deustua}}, \citenamefont {{Fabbro}}, \citenamefont
  {{Goobar}}, \citenamefont {{Groom}}, \citenamefont {{Hook}}, \citenamefont
  {{Kim}}, \citenamefont {{Kim}}, \citenamefont {{Lee}}, \citenamefont
  {{Nunes}}, \citenamefont {{Pain}}, \citenamefont {{Pennypacker}},
  \citenamefont {{Quimby}}, \citenamefont {{Lidman}}, \citenamefont {{Ellis}},
  \citenamefont {{Irwin}}, \citenamefont {{McMahon}}, \citenamefont
  {{Ruiz-Lapuente}}, \citenamefont {{Walton}}, \citenamefont {{Schaefer}},
  \citenamefont {{Boyle}}, \citenamefont {{Filippenko}}, \citenamefont
  {{Matheson}}, \citenamefont {{Fruchter}}, \citenamefont {{Panagia}},
  \citenamefont {{Newberg}}, \citenamefont {{Couch}},\ and\ \citenamefont
  {{Project}}}]{1999ApJ...517..565P}%
  \BibitemOpen
  \bibfield  {author} {\bibinfo {author} {\bibfnamefont {S.}~\bibnamefont
  {{Perlmutter}}}, \bibinfo {author} {\bibfnamefont {G.}~\bibnamefont
  {{Aldering}}}, \bibinfo {author} {\bibfnamefont {G.}~\bibnamefont
  {{Goldhaber}}}, \bibinfo {author} {\bibfnamefont {R.~A.}\ \bibnamefont
  {{Knop}}}, \bibinfo {author} {\bibfnamefont {P.}~\bibnamefont {{Nugent}}},
  \bibinfo {author} {\bibfnamefont {P.~G.}\ \bibnamefont {{Castro}}}, \bibinfo
  {author} {\bibfnamefont {S.}~\bibnamefont {{Deustua}}}, \bibinfo {author}
  {\bibfnamefont {S.}~\bibnamefont {{Fabbro}}}, \bibinfo {author}
  {\bibfnamefont {A.}~\bibnamefont {{Goobar}}}, \bibinfo {author}
  {\bibfnamefont {D.~E.}\ \bibnamefont {{Groom}}}, \bibinfo {author}
  {\bibfnamefont {I.~M.}\ \bibnamefont {{Hook}}}, \bibinfo {author}
  {\bibfnamefont {A.~G.}\ \bibnamefont {{Kim}}}, \bibinfo {author}
  {\bibfnamefont {M.~Y.}\ \bibnamefont {{Kim}}}, \bibinfo {author}
  {\bibfnamefont {J.~C.}\ \bibnamefont {{Lee}}}, \bibinfo {author}
  {\bibfnamefont {N.~J.}\ \bibnamefont {{Nunes}}}, \bibinfo {author}
  {\bibfnamefont {R.}~\bibnamefont {{Pain}}}, \bibinfo {author} {\bibfnamefont
  {C.~R.}\ \bibnamefont {{Pennypacker}}}, \bibinfo {author} {\bibfnamefont
  {R.}~\bibnamefont {{Quimby}}}, \bibinfo {author} {\bibfnamefont
  {C.}~\bibnamefont {{Lidman}}}, \bibinfo {author} {\bibfnamefont {R.~S.}\
  \bibnamefont {{Ellis}}}, \bibinfo {author} {\bibfnamefont {M.}~\bibnamefont
  {{Irwin}}}, \bibinfo {author} {\bibfnamefont {R.~G.}\ \bibnamefont
  {{McMahon}}}, \bibinfo {author} {\bibfnamefont {P.}~\bibnamefont
  {{Ruiz-Lapuente}}}, \bibinfo {author} {\bibfnamefont {N.}~\bibnamefont
  {{Walton}}}, \bibinfo {author} {\bibfnamefont {B.}~\bibnamefont
  {{Schaefer}}}, \bibinfo {author} {\bibfnamefont {B.~J.}\ \bibnamefont
  {{Boyle}}}, \bibinfo {author} {\bibfnamefont {A.~V.}\ \bibnamefont
  {{Filippenko}}}, \bibinfo {author} {\bibfnamefont {T.}~\bibnamefont
  {{Matheson}}}, \bibinfo {author} {\bibfnamefont {A.~S.}\ \bibnamefont
  {{Fruchter}}}, \bibinfo {author} {\bibfnamefont {N.}~\bibnamefont
  {{Panagia}}}, \bibinfo {author} {\bibfnamefont {H.~J.~M.}\ \bibnamefont
  {{Newberg}}}, \bibinfo {author} {\bibfnamefont {W.~J.}\ \bibnamefont
  {{Couch}}}, \ and\ \bibinfo {author} {\bibfnamefont {T.~S.~C.}\ \bibnamefont
  {{Project}}},\ }\href {\doibase 10.1086/307221} {\bibfield  {journal}
  {\bibinfo  {journal} {The Astrophysical Journal}\ }\textbf {\bibinfo {volume}
  {517}},\ \bibinfo {pages} {565} (\bibinfo {year} {1999})},\ \Eprint
  {http://arxiv.org/abs/astro-ph/9812133} {arXiv:astro-ph/9812133 [astro-ph]}
  \BibitemShut {NoStop}%
\bibitem [{\citenamefont {{Schutz}}(1986)}]{1986Natur.323..310S}%
  \BibitemOpen
  \bibfield  {author} {\bibinfo {author} {\bibfnamefont {B.~F.}\ \bibnamefont
  {{Schutz}}},\ }\href {\doibase 10.1038/323310a0} {\bibfield  {journal}
  {\bibinfo  {journal} {Nature}\ }\textbf {\bibinfo {volume} {323}},\ \bibinfo
  {pages} {310} (\bibinfo {year} {1986})}\BibitemShut {NoStop}%
\bibitem [{\citenamefont {{Holz}}\ and\ \citenamefont
  {{Hughes}}(2005)}]{2005ApJ...629...15H}%
  \BibitemOpen
  \bibfield  {author} {\bibinfo {author} {\bibfnamefont {D.~E.}\ \bibnamefont
  {{Holz}}}\ and\ \bibinfo {author} {\bibfnamefont {S.~A.}\ \bibnamefont
  {{Hughes}}},\ }\href {\doibase 10.1086/431341} {\bibfield  {journal}
  {\bibinfo  {journal} {The Astrophysical Journal}\ }\textbf {\bibinfo {volume}
  {629}},\ \bibinfo {pages} {15} (\bibinfo {year} {2005})},\ \Eprint
  {http://arxiv.org/abs/astro-ph/0504616} {arXiv:astro-ph/0504616 [astro-ph]}
  \BibitemShut {NoStop}%
\bibitem [{\citenamefont {Elvis}\ and\ \citenamefont
  {Karovska}(2002)}]{Elvis:2002ja}%
  \BibitemOpen
  \bibfield  {author} {\bibinfo {author} {\bibfnamefont {M.}~\bibnamefont
  {Elvis}}\ and\ \bibinfo {author} {\bibfnamefont {M.}~\bibnamefont
  {Karovska}},\ }\href {\doibase 10.1086/346015} {\bibfield  {journal}
  {\bibinfo  {journal} {Astrophys. J.}\ }\textbf {\bibinfo {volume} {581}},\
  \bibinfo {pages} {L67} (\bibinfo {year} {2002})},\ \Eprint
  {http://arxiv.org/abs/astro-ph/0211385} {arXiv:astro-ph/0211385 [astro-ph]}
  \BibitemShut {NoStop}%
\bibitem [{\citenamefont {Panda}\ \emph {et~al.}(2019)\citenamefont {Panda},
  \citenamefont {Martínez-Aldama}, \citenamefont {Zajaček},\ and\
  \citenamefont {Czerny}}]{Panda:2019cvs}%
  \BibitemOpen
  \bibfield  {author} {\bibinfo {author} {\bibfnamefont {S.}~\bibnamefont
  {Panda}}, \bibinfo {author} {\bibfnamefont {M.~L.}\ \bibnamefont
  {Martínez-Aldama}}, \bibinfo {author} {\bibfnamefont {M.}~\bibnamefont
  {Zajaček}}, \ and\ \bibinfo {author} {\bibfnamefont {B.}~\bibnamefont
  {Czerny}} (\bibinfo {collaboration} {LSST AGN Science}),\ }\href@noop {} {\
  (\bibinfo {year} {2019})},\ \Eprint {http://arxiv.org/abs/1909.05572}
  {arXiv:1909.05572 [astro-ph.HE]} \BibitemShut {NoStop}%
\bibitem [{\citenamefont {Di~Dio}\ \emph {et~al.}(2016)\citenamefont {Di~Dio},
  \citenamefont {Montanari}, \citenamefont {Raccanelli}, \citenamefont
  {Durrer}, \citenamefont {Kamionkowski},\ and\ \citenamefont
  {Lesgourgues}}]{DiDio:2016ykq}%
  \BibitemOpen
  \bibfield  {author} {\bibinfo {author} {\bibfnamefont {E.}~\bibnamefont
  {Di~Dio}}, \bibinfo {author} {\bibfnamefont {F.}~\bibnamefont {Montanari}},
  \bibinfo {author} {\bibfnamefont {A.}~\bibnamefont {Raccanelli}}, \bibinfo
  {author} {\bibfnamefont {R.}~\bibnamefont {Durrer}}, \bibinfo {author}
  {\bibfnamefont {M.}~\bibnamefont {Kamionkowski}}, \ and\ \bibinfo {author}
  {\bibfnamefont {J.}~\bibnamefont {Lesgourgues}},\ }\href {\doibase
  10.1088/1475-7516/2016/06/013} {\bibfield  {journal} {\bibinfo  {journal}
  {JCAP}\ }\textbf {\bibinfo {volume} {1606}},\ \bibinfo {pages} {013}
  (\bibinfo {year} {2016})},\ \Eprint {http://arxiv.org/abs/1603.09073}
  {arXiv:1603.09073 [astro-ph.CO]} \BibitemShut {NoStop}%
\bibitem [{\citenamefont {Aghanim}\ \emph {et~al.}(2018)\citenamefont {Aghanim}
  \emph {et~al.}}]{Aghanim:2018eyx}%
  \BibitemOpen
  \bibfield  {author} {\bibinfo {author} {\bibfnamefont {N.}~\bibnamefont
  {Aghanim}} \emph {et~al.} (\bibinfo {collaboration} {Planck collaboration}),\
  }\href@noop {} {\  (\bibinfo {year} {2018})},\ \Eprint
  {http://arxiv.org/abs/1807.06209} {arXiv:1807.06209 [astro-ph.CO]}
  \BibitemShut {NoStop}%
\bibitem [{\citenamefont {Chevallier}\ and\ \citenamefont
  {Polarski}(2001)}]{Chevallier:2000qy}%
  \BibitemOpen
  \bibfield  {author} {\bibinfo {author} {\bibfnamefont {M.}~\bibnamefont
  {Chevallier}}\ and\ \bibinfo {author} {\bibfnamefont {D.}~\bibnamefont
  {Polarski}},\ }\href {\doibase 10.1142/S0218271801000822} {\bibfield
  {journal} {\bibinfo  {journal} {Int. J. Mod. Phys.}\ }\textbf {\bibinfo
  {volume} {D10}},\ \bibinfo {pages} {213} (\bibinfo {year} {2001})},\ \Eprint
  {http://arxiv.org/abs/gr-qc/0009008} {arXiv:gr-qc/0009008 [gr-qc]}
  \BibitemShut {NoStop}%
\bibitem [{\citenamefont {Linder}(2003)}]{Linder:2002et}%
  \BibitemOpen
  \bibfield  {author} {\bibinfo {author} {\bibfnamefont {E.~V.}\ \bibnamefont
  {Linder}},\ }\href {\doibase 10.1103/PhysRevLett.90.091301} {\bibfield
  {journal} {\bibinfo  {journal} {Phys. Rev. Lett.}\ }\textbf {\bibinfo
  {volume} {90}},\ \bibinfo {pages} {091301} (\bibinfo {year} {2003})},\
  \Eprint {http://arxiv.org/abs/astro-ph/0208512} {arXiv:astro-ph/0208512
  [astro-ph]} \BibitemShut {NoStop}%
\bibitem [{\citenamefont {Kermack}\ \emph {et~al.}(1934)\citenamefont
  {Kermack}, \citenamefont {McCrea},\ and\ \citenamefont
  {Whittaker}}]{kermack_mccrea_whittaker_1934}%
  \BibitemOpen
  \bibfield  {author} {\bibinfo {author} {\bibfnamefont {W.~O.}\ \bibnamefont
  {Kermack}}, \bibinfo {author} {\bibfnamefont {W.~H.}\ \bibnamefont {McCrea}},
  \ and\ \bibinfo {author} {\bibfnamefont {E.~T.}\ \bibnamefont {Whittaker}},\
  }\href {\doibase 10.1017/S0370164600015479} {\bibfield  {journal} {\bibinfo
  {journal} {Proceedings of the Royal Society of Edinburgh}\ }\textbf {\bibinfo
  {volume} {53}},\ \bibinfo {pages} {31–47} (\bibinfo {year}
  {1934})}\BibitemShut {NoStop}%
\bibitem [{\citenamefont {Weinberg}(1970)}]{weinberg-letter}%
  \BibitemOpen
  \bibfield  {author} {\bibinfo {author} {\bibfnamefont {S.}~\bibnamefont
  {Weinberg}},\ }\href@noop {} {\bibfield  {journal} {\bibinfo  {journal} {The
  Astrophisical Journal}\ }\textbf {\bibinfo {volume} {161}},\ \bibinfo {pages}
  {L233} (\bibinfo {year} {1970})}\BibitemShut {NoStop}%
\bibitem [{\citenamefont {Kasai}(1988)}]{kasai}%
  \BibitemOpen
  \bibfield  {author} {\bibinfo {author} {\bibfnamefont {M.}~\bibnamefont
  {Kasai}},\ }\href {http://dx.doi.org/10.1143/PTP.79.777} {\bibfield
  {journal} {\bibinfo  {journal} {Prog. Theor. Phys.}\ }\textbf {\bibinfo
  {volume} {79}},\ \bibinfo {pages} {777} (\bibinfo {year} {1988})}\BibitemShut
  {NoStop}%
\bibitem [{\citenamefont {Rosquist}(1988)}]{rosquist}%
  \BibitemOpen
  \bibfield  {author} {\bibinfo {author} {\bibfnamefont {K.}~\bibnamefont
  {Rosquist}},\ }\href {http://dx.doi.org/10.1086/166588} {\bibfield  {journal}
  {\bibinfo  {journal} {Astrophys. J.}\ }\textbf {\bibinfo {volume} {331}},\
  \bibinfo {pages} {648} (\bibinfo {year} {1988})}\BibitemShut {NoStop}%
\bibitem [{\citenamefont {Korzy\'nski}\ and\ \citenamefont
  {Villa}()}]{KorzynskiVilla2}%
  \BibitemOpen
  \bibfield  {author} {\bibinfo {author} {\bibfnamefont {M.}~\bibnamefont
  {Korzy\'nski}}\ and\ \bibinfo {author} {\bibfnamefont {E.}~\bibnamefont
  {Villa}},\ }\href@noop {} {\enquote {\bibinfo {title} {Parallax, drift
  effects and distance slip and in a perturbed {FLRW} spacetime},}\ }\bibinfo
  {note} {In preparation}\BibitemShut {NoStop}%
\bibitem [{\citenamefont {{Peebles}}(1993)}]{peeblesbook}%
  \BibitemOpen
  \bibfield  {author} {\bibinfo {author} {\bibfnamefont {P.~J.~E.}\
  \bibnamefont {{Peebles}}},\ }\href@noop {} {\emph {\bibinfo {title}
  {{Principles of Physical Cosmology}}}}\ (\bibinfo {year} {1993})\BibitemShut
  {NoStop}%
\bibitem [{\citenamefont {Peacock}(1998)}]{peacock_1998_2}%
  \BibitemOpen
  \bibfield  {author} {\bibinfo {author} {\bibfnamefont {J.~A.}\ \bibnamefont
  {Peacock}},\ }\href {\doibase 10.1017/CBO9780511804533} {\emph {\bibinfo
  {title} {Cosmological Physics}}}\ (\bibinfo  {publisher} {Cambridge
  University Press},\ \bibinfo {year} {1998})\BibitemShut {NoStop}%
\bibitem [{\citenamefont {{Hogg}}(1999)}]{Hogg:1999ad}%
  \BibitemOpen
  \bibfield  {author} {\bibinfo {author} {\bibfnamefont {D.~W.}\ \bibnamefont
  {{Hogg}}},\ }\href@noop {} {\bibfield  {journal} {\bibinfo  {journal} {arXiv
  e-prints}\ ,\ \bibinfo {eid} {astro-ph/9905116}} (\bibinfo {year} {1999})},\
  \Eprint {http://arxiv.org/abs/astro-ph/9905116} {arXiv:astro-ph/9905116
  [astro-ph]} \BibitemShut {NoStop}%
\end{thebibliography}%

\end{document}